\documentclass[a4paper,12pt,twoside,openany]{memoir}
\usepackage{graphicx,eso-pic}
\usepackage{geometry}
\setlrmarginsandblock{2.6cm}{*}{1}
\setulmarginsandblock{3.5cm}{*}{1}
\checkandfixthelayout

\usepackage{stix}

\usepackage{amsmath,mathtools,tensor}
\DeclarePairedDelimiter\abs{\lvert}{\rvert}%
\newcommand{\metdens}{\sqrt{\abs{g}}}
\newcommand{\bmetdens}{\sqrt{\abs{\bar g}}}
\newcommand{\ee}{\mathrm{e}}
\newcommand{\ii}{\mathrm{i}}
\newcommand{\dd}{\mathrm{d}}

\newcommand{\dx}{\mathrm{d}^dx\,}
\newcommand{\qqquad}{\qquad\qquad}
\numberwithin{equation}{section}

\usepackage{tikz}
\usetikzlibrary{decorations.pathmorphing}
\usetikzlibrary{decorations.markings}
\usetikzlibrary{positioning, shapes, arrows}
\tikzset{
	graviton/.style={line width=.8pt, -latex,decorate, decoration={snake, segment length=4pt,amplitude=1.8pt, pre length=.1cm, post length=.25cm}},
	worldline/.style={gray, line width=1pt},
	worldlineBold/.style={black, line width=.6pt},
	zUndirected/.style={line width=1pt},
	zParticle/.style={line width=1pt,postaction={decorate},decoration={markings,mark=at position .6 with {\arrow[#1]{latex}}}},
	zParticleF/.style={line width=1pt,postaction={decorate}},
	cscalar/.style={line width=1pt,postaction={decorate},decoration={markings,mark=at position .6 with {\arrow[#1]{latex}}}},
	cscalar2/.style={line width=1pt,postaction={decorate},decoration={markings,mark=at position .8 with {\arrow[#1]{latex}}}},
	photon/.style={line width=.8pt, decorate, decoration={snake, segment length=4pt, amplitude=1.8pt,  pre length=.1cm, post length=.1cm}},
    photon2/.style={line width=2pt, darkgray, decorate, decoration={snake, segment length=5pt, amplitude=1.8pt,  pre length=.1cm, post length=.1cm,}}
}

\usepackage[numbers,sort&compress]{natbib}

\renewenvironment{thebibliography}[1]{%
\begin{oldthebibliography}{#1}%
\small%
\raggedright%
\setlength{\itemsep}{5pt}%
}%
{%
\end{oldthebibliography}%
}

\usepackage{lipsum}

\usepackage{hyperref}
\hypersetup{
	colorlinks=true,
	linktoc=page,
	citecolor=cadmiumgreen,
	linkcolor=cadmiumgreen,
	urlcolor=cadmiumgreen} 
\definecolor{cadmiumgreen}{rgb}{0.0, 0.42, 0.24}
\definecolor{kured}{RGB}{147, 26, 30}

\nouppercaseheads

\makechapterstyle{frontmatter}{
  \chapterstyle{default}
}
\makechapterstyle{mainmatter}{
  \chapterstyle{default}
}
\makechapterstyle{backmatter}{
  \chapterstyle{frontmatter}
}
\let\oldfrontmatter\frontmatter
\renewcommand{\frontmatter}{\chapterstyle{frontmatter}\oldfrontmatter}
\let\oldmainmatter\mainmatter
\renewcommand{\mainmatter}{\chapterstyle{mainmatter}\oldmainmatter}
\let\oldbackmatter\backmatter
\renewcommand{\backmatter}{\chapterstyle{backmatter}\oldbackmatter}

\settocdepth{subsection}

\begin{document}
\begin{titlingpage}
    \newgeometry{left=5cm,right=2.2cm,bottom=3cm}
    \AddToShipoutPictureBG*{
        \AtPageLowerLeft{
            \includegraphics[width=\paperwidth,height=\paperheight]{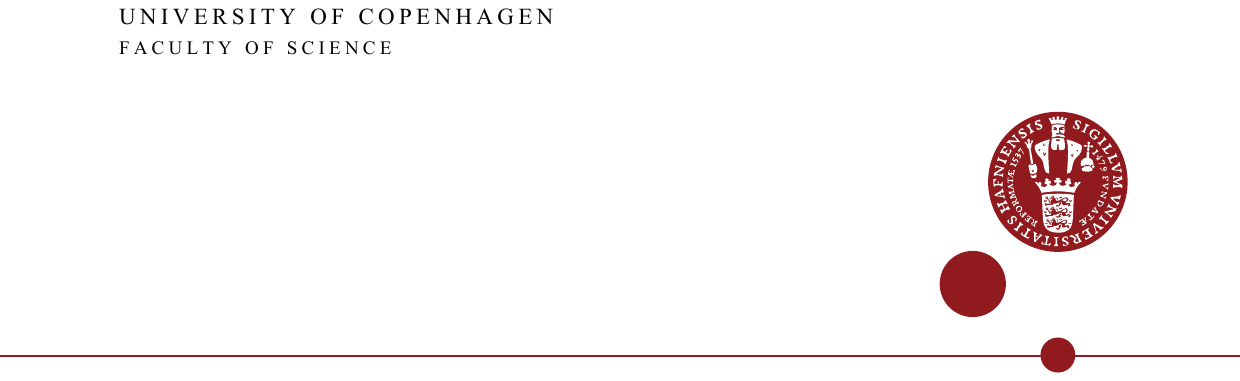}
        }
    }
    \vspace*{8cm}
    \noindent\textcolor{kured}{\textsf{\large Master's thesis}}
    
    \vspace*{1cm}
    \noindent\textcolor{kured}{\textsf{\huge Gravitational scattering amplitudes \\[.1cm] from curved space}}
    
    \vspace*{6cm}
    \noindent\textsf{\large \textbf{Carl Jordan Eriksen}}
    
    \vspace*{1cm}
    \noindent\textsf{\large Advisor \\ N. Emil J. Bjerrum-Bohr}

    \vspace*{.5cm}
    \noindent\textsf{\large Co-advisor \\ Gang Chen}

    \vspace*{1cm}
    \noindent\textsf{\large May 20, 2025. Niels Bohr Institute}
\end{titlingpage}
\restoregeometry
\frontmatter
\chapter{Abstract}
Motivated by the study of extreme mass-ratio binary systems, recent work has explored the use of curved backgrounds in computations of classical gravitational amplitudes \cite{Kosmopoulos:2023bwc,Cheung:2023lnj,Cheung:2024byb}. While these investigations concern the self-force expansion in the ratio of masses of the binaries, the use of curved backgrounds is interesting in its own right. In this thesis, I examine how gravitational computations can be done in a curved background. After having reviewed aspects of general relativity and the $d$-dimensional metric generated by a point mass (known as the Schwarz\-schild--Tangherlini solution), I quantize general relativity on an arbitrary background and compute Feynman rules for gravity in two cases: when the background is flat, and when it is a Schwarz\-schild--Tangherlini background. I then outline worldline quantum field theory. Using this newly-developed perturbation theory for the partition function of a worldline coupled to gravity in a curved background, I reformulate the perturbative expansion of the Compton amplitude, which describes the scattering of a graviton off a compact object. Having established this framework, I compute the first and second post-Minkowskian contributions to the Compton amplitude. Both are shown to match the results obtained from a flat-space computation. In addition, the second-order amplitude displays the expected infrared behavior and agrees with earlier results on massless gravitational scattering.
\chapter{Acknowledgments}
I would first of all like to express my gratitude to my supervisors, Emil Bjerrum-Bohr and Gang Chen, for pointing me toward this interesting problem and granting me a large degree of freedom in how I approached it. I have valued our relaxed discussions about the project and your help and support through the PhD application process.

I would like to thank Nabha Shah for her collaboration on this project and our frequent discussions about Compton amplitudes and related topics.

Then, I would thank my office mates: Pablo, Kostas, Filip, and Max-Emil for their company in the past year and for our frequent lively conversations on varied subjects, within and outside physics.

Finally, I am indebted to my family and friends for their unwavering support.

\chapter{Declaration}
Parts of this thesis have since appeared in print \cite{Bjerrum-Bohr:2025bqg}.
\chapter{Notations and conventions}\label{sec:notation_and_convention}
\paragraph{Units.} I use natural units throughout this thesis. This means that both the speed of light $c$ and Planck's reduced constant $\hbar$ are set to unity.

\paragraph{Tensors.} In most of the thesis, I take the dimension of spacetime $d$ to be arbitrary. I use a mostly-minus convention for the metric tensor $g_{\mu\nu}$, meaning that the sign of the determinant of the metric is $\operatorname{sgn}\det{g} = (-1)^{d-1}$. Due to the choice of sign convention, I denote the metric density $\sqrt{\abs{\det g}}$ by $\metdens$. The summation convention is used throughout, meaning that, e.g., the dot product of two vectors $V^\mu$ and $W^\mu$ is written as interchangeably as
\begin{equation*}
    V\cdot W = g_{\mu\nu}V^\mu W^\nu,
\end{equation*}
with the summation over $\mu$ and $\nu$ being implied. If $W^\mu = V^\mu$ I write $V\cdot V = V^2$, and I denote the length of $V$ by
\begin{equation*}
    \abs{V} = \sqrt{\abs{V^2}}.
\end{equation*}

\paragraph{Fourier transforms.}
I define the Fourier transform and its inverse of a field $\phi(x)$ defined on Minkowski spacetime to be
\begin{equation*}
    \phi(x) = \int\frac{\mathrm{d}^dk}{(2\pi)^d}\ee^{-\ii k\cdot x}\phi(k), \qquad \phi(k) = \int\mathrm{d}^dx\,\ee^{\ii k\cdot x}\phi(x).
\end{equation*}
Almost always, I will use the convenient notation
\begin{equation*}
    \int_k = \int\frac{\mathrm{d}^dk}{(2\pi)^d}.
\end{equation*}
In conjunction with the above, it is convenient to define the hatted delta function
\begin{equation*}
    \hat\delta(k) = (2\pi)^d\delta^{(d)}(k),
\end{equation*}
such that
\begin{equation*}
    \int_k\hat\delta(k) = 1.
\end{equation*}
For a field $f$ defined on a worldline embedded in spacetime parameterized by $\tau$, I define the Fourier transform and its inverse in analogy to the above, i.e.,
\begin{equation*}
    f(\tau) = \int\frac{\mathrm{d}\omega}{2\pi}\ee^{-\ii\omega\tau}f(\omega), \qquad f(\omega) = \int\mathrm{d}\tau\,\ee^{\ii\omega\tau}f(\tau).
\end{equation*}
Again, I will almost always use
\begin{equation*}
    \int_\omega = \int\frac{\mathrm{d}\omega}{2\pi},
\end{equation*}
along with
\begin{equation*}
    \hat\delta(\omega) = 2\pi\delta(\omega),
\end{equation*}
such that
\begin{equation*}
    \int_\omega\hat\delta(\omega) = 1.
\end{equation*}
I symmetrize and anti-symmetrize with unit weight, meaning that
\begin{align*}
    T^{(\mu\nu)} = \frac{T^{\mu\nu} + T^{\nu\mu}}{2}, \\
    T^{[\mu\nu]} = \frac{T^{\mu\nu} - T^{\nu\mu}}{2}.
\label{eq:symmetrization_convention}\end{align*}
\newpage
\begin{KeepFromToc}
    \tableofcontents
\end{KeepFromToc}
\mainmatter
\setsecnumdepth{subsection}
\chapter{Introduction}
Of the four fundamental interactions, the force of gravity is the one whose effects are most readily discernible at the scale of everyday experience. The ease with which its influence is perceived, and the magnificence of the night sky it governs, are chief among the reasons why gravity was the first force to be subjected to rigorous mathematical study. In antiquity, Apollonius of Perga ventured that the movement of the planets, then known as \emph{wandering stars}, against the stellar background could be explained by a theory of epicycles and deferents---a theory which was later developed into a quantitative, predictive framework by Hipparchus and Ptolemy \cite{Evans:1998hpa}. Centuries later, in the wake of the Copernican revolution, Kepler proposed that the planets, including Earth, moved on ellipses around a stationary Sun. Additionally, Galileo, who was a contemporary of Kepler, developed his law of inertia and observed that the acceleration due to gravity is independent of mass \cite{Swerdlow:2013ohh}. These advances set the stage for Newton to perform the first great unification in physics: the unified description of terrestrial and celestial mechanics provided by his laws of motion and gravitation \cite{Newton:1687eqk}. Applying these laws to the binary problem---the determination of the dynamics of two bodies interacting gravitationally---leads exactly to Kepler's elliptical orbits in the bound case, and more generally to motion along conic sections. Following Newton, it is difficult to overstate the importance of Bernoulli's introduction of the gravitational potential \cite{Bernoulli:1738ab}, as it represented a stepping stone to field theory, and led Lagrange and others to formulate physics in the language of action principles \cite{Caparrini:2013ohh}.

The power of field theory showed itself in Maxwell's unification of electric and magnetic phenomena \cite{Maxwell:1865zz}. Maxwell's theory implies that the speed of light is constant in all reference frames, as the theory is Lorentz invariant. As is well known, this fact was the chief motivation for Einstein when he abolished the notion that space and time are essentially disparate concepts \cite{Einstein:1905ve}. This insight spurred an important innovation due to Minkowski: the \emph{spacetime continuum}, in which space and time are considered a unified whole. Building on this work, Einstein extended the Galilean equivalence principle with his postulate that gravitational forces are locally equivalent to inertial forces (such as Coriolis forces), and this finally lead him to the insight that gravitational phenomena are manifestations of the curvature of spacetime, described by the metric field. This is the celebrated theory of general relativity \cite{Einstein:1915rt}.

Since its introduction, general relativity has consistently shown itself to be an exceptionally precise theory of gravity, unsurpassed in its agreement with a wealth of rigorously obtained experimental data. It also makes (among others) two startling predictions: the existence of black holes and gravitational waves. With the first experimental detection of the gravitational wave signal from the merger of two black holes by the LIGO/Virgo collaboration \cite{LIGOScientific:2016aoc} and the first image having been taken of the supermassive black hole at the center of the galaxy M87 by the Event Horizon Telescope \cite{EventHorizonTelescope:2019dse}, these predictions have been definitively confirmed, once again bringing gravitational physics into a new and exciting era. Neutron star mergers as well as mergers between neutron stars and black holes are now regularly observed \cite{LIGOScientific:2017vwq,LIGOScientific:2021qlt}.

While the binary problem admitted an exact, stable solution in Newtonian theory, this is no longer the case in general relativity, as gravitational waves are constantly leaking energy from the system. Binary systems of compact objects such as black holes or neutron stars as described in general relativity thus have a finite lifetime: a generally long-lived \emph{inspiral} stage is followed by a violent \emph{merger}, after which the system settles down through a \emph{ringdown} stage. To study binary (and more general) dynamics, one must resort to a perturbative scheme. For the inspiral stage, the most popular approximation method has traditionally been the \emph{post-Newtonian approximation}, where a joint expansion in velocities and Newton's constant is performed \cite{Blanchet:2013haa}. In recent years, however, surprising new tools for tackling gravitational dynamics have emerged from quantum theory.

In the early 20th century, an overwhelming body of evidence necessitated the introduction of quantum mechanics. The reconciliation between quantum mechanics and special relativity gave rise to quantum field theory, where particles emerge as excitations of quantum fields. The quantum field theory that is believed to describe the real world, the standard model, has enjoyed innumerable successes as a predictive physical theory \cite{Schwartz:2014sze,Peskin:1995ev}.

The marriage of general relativity to quantum field theory is infamously troubled. As was shown by Donoghue, however, general relativity admits a perfectly sensible description as a low-energy effective quantum field theory \cite{Donoghue:1993eb,Donoghue:1994dn}. This has yielded new insights into the low-energy quantum corrections to gravity \cite{Bjerrum-Bohr:2002gqz,BjerrumBohr:2002ks}, but also made the toolbox normally associated with scattering amplitudes available to run-of-the-mill, classical gravitational computations. The first foray into this exciting application was made by Goldberger and Rothstein, who developed a point particle effective field theory approach to the binary problem utilizing the post-Newtonian approximation \cite{Goldberger:2004jt,Goldberger:2006bd,Goldberger:2009qd}. More recently, many approaches have exploited that quantum field theories are naturally expanded in their coupling constant, which in the case of gravity corresponds to a \emph{post-Minkowskian expansion}---an expansion only in Newton's constant. Progress in this direction has been quick, and the state of the art is now fifth post-Minkowskian order (5PM) (restricted to first self-force order, see below) \cite{Driesse:2024feo}.

In the quantum field theory approach to gravity, the role of the non-spinning compact object is played by a scalar, massive field. At some point, one must take a classical limit, which is somewhat subtle. There exists an alternative approach, worldline quantum field theory, where this limit is taken at the level of the action \cite{Mogull:2020sak, Jakobsen:2021zvh, Jakobsen:2021lvp, Jakobsen:2022fcj, Jakobsen:2023ndj,Driesse:2024feo}.\footnote{The state of the art for post-Minkowskian observables was in fact computed with worldline quantum field theory \cite{Driesse:2024feo}.} Worldline quantum field theory, which naturally uses the post-Minkowskian expansion, is related to the aforementioned point particle effective field theory.

The planned launch of the space-based laser interferometer detector antenna (LISA) has also spurred new developments. LISA is planned to be sensitive to much lower frequency ranges, and should thus be able to probe extreme-mass ratio systems currently out of reach. These systems are uniquely suited for the \emph{self-force approximation}, which is an expansion in the mass ratio of the two bodies comprising the binary system \cite{Barack:2018yvs}, which is exact in Newton's constant at each order in the ratio. This has also been investigated from the quantum field theory side \cite{Kosmopoulos:2023bwc,Cheung:2023lnj,Cheung:2024byb}, where the self-force expansion entails an expansion around a curved background.

In this thesis, I focus on the computation of the classical Compton amplitude using both a weak-field expansion and an expansion around curved space. The Compton amplitude is the amplitude for a graviton scattering off a massive compact object, such as a black hole.\footnote{Throughout the thesis, I will use compact object, point particle and black hole interchangeably.} I carry out the computation using worldline quantum field theory. Specifically, I compute the first (1PM) and second post-Minkowskian (2PM) Compton amplitude. The 1PM amplitude has appeared many places in the literature, e.g., \cite{Brandhuber:2021eyq}. Certain pieces of the 2PM amplitude were presented in \cite{Akpinar:2025huz}, however, to my knowledge, the full result is new. The 1PM Compton amplitude is usually used as a building block in the approach to amplitude computations known as \emph{generalized unitarity} \cite{Bern:2019crd}. The discontinuity (or cut) of an amplitude in a certain kinematic channel can be computed by gluing lower-order, on-shell amplitudes together \cite{Bren:2011bgu}. Additionally, amplitudes are always expressible as sums of Feynman diagrams. These two facts allow one to determine the amplitude by computing cuts. However, indications that the classical Compton amplitude is closely tied with tidal effects have appeared recently \cite{Correia:2024jgr,Caron-Huot:2025tlq}.

I give here an outline of the thesis: In chapter \ref{ch:general relativity}, I provide a review of the aspects of general relativity that I need for the subsequent developments. The action formulation of general relativity and the action for a point particle is covered. In addition, I discuss a derivation of the Schwarzschild--Tangherlini solution to Einstein's equation, which is the $d$-dimensional generalization of the Schwarzschild solution. Chapter \ref{ch:perturbative quantization of gravity} is devoted to the quantum field-theoretic description of gravity. I cover the treatment of gravity as a gauge theory, after which I quantize gravity around an arbitrary background. Feynman rules are derived both in the weak-field expansion and around the Schwarzschild--Tangherlini background. In this chapter, I also give brief reviews of path integral quantization, gravity as an effective field theory, and the classical limit. In chapter \ref{ch:worldline formalism}, I discuss the alternative approach to the classical limit provided by worldline quantum field theory. The Feynman rules in this formalism are derived both in the weak-field and Schwarzschild--Tangherlini cases. It is using the worldline formalism that I shall carry out my computations of the Compton amplitude, which are presented in detail in chapter \ref{ch:compton amplitude}. In this chapter, I first discuss the kinematics of the scattering, after which I compute the 1PM Compton amplitude using both the flat and curved Feynman rules. Then follows the computation of the second 2PM Compton amplitude, which necessitates a discussion of integration-by-parts identities and integral evaluation. The result is then presented, and consistency checks are performed. Lastly, I provide concluding remarks in chapter \ref{ch:conclusion}.

\chapter{General relativity}\label{ch:general relativity}
This chapter provides an exposition of the necessary theoretical background in general relativity. I begin in section \ref{sec:actionformulation} with a review of the action formulation of general relativity. In section \ref{sec:action for a point particle}, I discuss the treatment of point particles in this framework. Finally, I provide in section \ref{sec:schwarzschild--tangherlini} a derivation of the Schwarzschild--Tangherlini solution to Einstein's equation and a conversion of this metric into the form needed for the later developments in the thesis.

\section{Action formulation}\label{sec:actionformulation}
The action for general relativity, known as the Einstein--Hilbert action, was first discovered by Hilbert \cite{Hilbert:1915dgd}. It is given by
\begin{equation}
    S_\text{EH}[g] = -\frac{1}{16\pi G}\int\dd^4x\sqrt{\abs{\det g_{\mu\nu}}} R,
\label{eq:eh4d}\end{equation}
where $g_{\mu\nu}$ is the metric field; it is a dimensionless, symmetric rank-2 tensor with components $g_{\mu\nu}$, where $\mu,\nu = 0,\ldots,3$. The prefactor of the action contains the four-dimensional Newton's constant $G$, which has mass dimension $[G] = [\text{M}]^{-2}$. The Ricci scalar $R$ is defined in terms of the Ricci tensor,
\begin{equation}
    R = g^{\mu\nu}R_{\mu\nu},
\end{equation}
which in turn is a contraction of the Riemann tensor,
\begin{equation}
    R_{\mu\nu} = R\indices{^\lambda_{\mu\lambda\nu}}.
\end{equation}
The Riemann tensor is given by\footnote{The definitions of the Ricci and Riemann tensors given here coincide with the ones in \cite{Carroll:2004st}. The extra minus in the Einstein--Hilbert action is due to the fact that the Ricci scalar contains an odd number of metrics and hence changes sign between metric signatures.}\footnote{I always symmetrize and anti-symmetrize with unit weight.}
\begin{equation}
    R\indices{^\mu_{\nu\rho\sigma}} = 2\partial_{[\rho}\Gamma\indices{^\mu_{\sigma]\nu}} + 2
    \Gamma\indices{^\lambda_{\nu[\sigma}}\Gamma\indices{^\mu_{\rho]\lambda}},
\end{equation}
where
\begin{equation}
    \Gamma\indices{^\lambda_{\mu\nu}} = \frac12g^{\lambda\sigma}(2\partial_{(\mu}g_{\nu)\lambda} - \partial_\lambda g_{\mu\nu})
\label{eq:christoffel}\end{equation}
is the Christoffel symbol, also known as the connection. It is sometimes convenient to have an expression for the Riemann tensor with all indices down. Using the compatibility of the covariant derivative associated with the Christoffel symbol,
\begin{equation}
    \nabla_{\rho}g_{\mu\nu} = \partial_\rho g_{\mu\nu} - 2\Gamma_{(\mu\nu)\rho} = 0,
\end{equation}
it is easy to see that
\begin{equation}
    g_{\kappa\mu}\partial_{\rho}\Gamma\indices{^\mu_{\nu\sigma}} = \partial_\rho\Gamma_{\kappa\nu\sigma} - 2\Gamma\indices{^\mu_{\nu\sigma}}\Gamma_{(\mu\kappa)\rho}
\end{equation}
which implies that
\begin{equation}
    R_{\mu\nu\rho\sigma} = 2\partial_{[\nu\vert}\Gamma_{\sigma\rho\vert\mu]} + 2\Gamma_{\lambda\sigma[\mu}\Gamma\indices{^\lambda_{\nu]\rho}}.
\end{equation}
The determinant of the metric appears almost exclusively as part of the metric density $\sqrt{\abs{\det g_{\mu\nu}}}$, so from now on, I will abbreviate the determinant of the metric $\det g_{\mu\nu}$ simply as $g$. It will be obvious from the context if $g$ means the metric or its determinant.

Later in the thesis, it will be necessary to regulate expressions which are formally infinite in four dimensions. The modern procedure for doing this is called \emph{dimensional regularization}, and it entails working in a $d$-dimensional setup, necessitating a $d$-dimensional generalization of the Einstein--Hilbert action. This generalization is
\begin{equation}
    S_\text{EH}[g] = -\frac{1}{16\pi G_d}\int\dx\metdens R.
\end{equation}
In $d$ dimensions, the components of the metric are $g_{\mu\nu}$ where $\mu,\nu = 0,\ldots,d-1$. The $G_d$ appearing here is the $d$-dimensional Newton's constant with mass dimension $[G] = 2 - d$. In dimensional regularization, once the $d\to4$ limit has been taken, any piece originating purely from $d\neq4$ is unphysical and must not appear in any physical observable. Hence, it is convenient to parametrize this piece of $G_d$ by introducing the arbitrary mass scale $\mu$,
\begin{equation}
    G_d = G\tilde\mu^{4-d}, \qqquad \tilde\mu^2 = \frac{\ee^{\gamma_\text{E}}\mu^2}{4\pi},
\label{eq:ddimnewtonconstant}\end{equation}
where the additional factors involving $\pi$ and the Euler--Mascheroni constant $\gamma_\text{E}$ are conventional. I will often work with the gravitational coupling constant $\kappa$ defined by
\begin{equation}
    \kappa^2 = 32\pi G_d,
\end{equation}
in terms of which the Einstein--Hilbert action reads
\begin{equation}
    S_\text{EH}[g] = -\frac{2}{\kappa^2}\int\dx\metdens R.
\label{eq:ehaction}\end{equation}

The Einstein--Hilbert action (in any dimension) has the important property of being invariant under diffeomorphisms, which I now briefly discuss. A diffeomorphism given by
\begin{equation}
    x^\mu \to x^{\mu'}(x)
\end{equation}
will have an associated Jacobian
\begin{equation}
    J\indices{^{\mu'}_\mu} = \frac{\partial x^{\mu'}}{\partial x^\mu}, \qqquad (J^{-1})\indices{^\mu_{\mu'}} = J\indices{_{\mu'}^\mu} = \frac{\partial x^\mu}{\partial x^{\mu'}}, \qqquad \det J = J.
\label{eq:jacobiannotation}\end{equation}
Under this transformation, the old and new basis vectors and basis one-forms become
\begin{equation}
    \partial_\mu \to \partial_{\mu'} = J\indices{_{\mu'}^\mu}\partial_\mu, \qqquad \dd x^\mu \to \dd x^{\mu'} = J\indices{^{\mu'}_\mu}\dd x^\mu.
\label{eq:basistransforms}\end{equation}
Any scalar, including the Ricci scalar, will hence remain invariant, as it contains an equal number of upper and lower indices. From the one-form transformation rule it is possible to derive that the measure from eq. \eqref{eq:ehaction} acquires a factor of the Jacobian determinant,
\begin{equation}
    \dd^dx \to \dd^dx' = \abs{J}\dd^dx.
\label{eq:measure-transform}\end{equation}
The components of the metric acquire the opposite factor to the basis one-forms, meaning that the metric density acquires a compensating factor of the Jacobian determinant,\footnote{This property of the metric density means that it is a tensor density of weight -1.}
\begin{equation}
    \metdens \to \sqrt{\abs{g'}} = \abs{J}^{-1}\metdens,
\label{eq:metdenstransform}\end{equation}
meaning that the combination $\dd^4x\metdens$ is invariant. This implies the invariance that was claimed.

For now, I will let any matter fields $\phi$ present in the theory be described by the (diffeomorphism invariant) action $S_\text{matter}[g, \phi]$. Then, by the action principle, Einstein's equation is obtained by varying the total action,
\begin{equation}
    \delta S_\text{EH} + \delta S_\text{matter} = -\frac{2}{\kappa^2}\int\dx\Big(\metdens\,\delta R + R\,\delta\metdens\Big) + \int\dx\frac{\delta S_\text{matter}}{\delta g_{\mu\nu}}\delta g_{\mu\nu},
\end{equation}
and requiring that it vanish. The variation of the metric density is
\begin{equation}
    \delta\metdens = \frac{1}{2\metdens}\operatorname{sgn}g\,\delta g,
\end{equation}
and the variation of the determinant of the metric $\delta g$ can be obtained directly from the Leibnitz formula.\footnote{It is also possible to calculate this variation from the matrix identity $\det{A} = \ee^{\operatorname{tr}\log{A}}$.} By direct computation,
\begin{align}
    \delta g &= \frac{\delta g}{\delta g_{\mu\nu}}\delta g_{\mu\nu} \notag\\
    &= \frac{1}{d!}\varepsilon^{\alpha_0\cdots\alpha_{d-1}}\varepsilon^{\beta_0\cdots\beta_{d-1}}\frac{\delta}{\delta g_{\mu\nu}}(g_{\alpha_0\beta_0}\cdots g_{\alpha_{d-1}\beta_{d-1}})\delta g_{\mu\nu} \notag\\
    &= \frac{1}{d!}\varepsilon^{\alpha_0\cdots\alpha_{d-1}}\varepsilon^{\beta_0\cdots\beta_{d-1}}\big(\delta\indices*{^{(\mu}_{\alpha_0}}\delta\indices*{^{\nu)}_{\beta_0}}\cdots g_{\alpha_{d-1}\beta_{d-1}} + \cdots + g_{\alpha_0\beta_0}\cdots\delta\indices*{^{(\mu}_{\alpha_{d-1}}}\delta\indices*{^{\nu)}_{\beta_{d-1}}}\big)\delta g_{\mu\nu}.
\end{align}
Here, $\varepsilon^{\alpha_0\cdots\alpha_{d-1}}$ is the $d$-dimensional Levi-Civita symbol. The quantity $\delta g_{\mu\nu}$ is the arbitrary variation of the metric which will later play the role of the graviton. The index symmetry of $\delta g_{\mu\nu}$ implies that one can forget about the index symmetrization inside the parenthesis. By taking the Levi-Civita symbols inside the parenthesis and renaming indices, it is possible to bring all the terms to the form of the first. Thus,
\begin{align}
    \delta g &= \frac{1}{(d - 1)!}\varepsilon^{\mu\alpha_1\cdots\alpha_{d-1}}\varepsilon^{\nu\beta_1\cdots\beta_{d-1}}g_{\alpha_1\beta_1}\cdots g_{\alpha_{d-1}\beta_{d-1}}\delta g_{\mu\nu} \notag\\
    &= M^{\mu\nu}\delta g_{\mu\nu}
\end{align}
The result now follows by showing that $M^{\mu\nu}$ is proportional to the inverse metric $g^{\mu\nu}$ with proportionality factor $g$. Consider the contraction
\begin{equation}
    M^{\mu\beta_0}g_{\beta_0\nu} = \frac{1}{(d-1)!}\varepsilon^{\mu\alpha_1\cdots\alpha_{d-1}}\varepsilon^{\beta_0\beta_1\cdots\beta_{d-1}}g_{\alpha_1\beta_1}\cdots g_{\alpha_{d-1}\beta_{d-1}}g_{\beta_0\nu}.
\end{equation}
Potentially non-zero terms must have $\alpha_1, \ldots, \alpha_{d-1} \neq \mu$. However, assuming $\mu \neq \nu$ forces there to be one $\alpha_i = \nu$ for some $i = 1, \ldots, d-1$. This ensures that $g_{\alpha_1\beta_1}\cdots g_{\alpha_{d-1}\beta_{d-1}}g_{\beta_0\nu}$ is symmetric in $\beta_i$ and $\beta_0$, causing the contraction with $\varepsilon^{\beta_0\beta_1\cdots\beta_{d-1}}$ to vanish identically. Contrariwise, with $\mu = \nu$, $(d - 1)!$ terms of the form
\begin{equation}
    \varepsilon^{\mu\alpha_1\cdots\alpha_{d-1}}\varepsilon^{\beta_0\beta_1\cdots\beta_{d-1}}g_{\mu\beta_0}g_{\alpha_1\beta_1}\cdots g_{\alpha_{d-1}\beta_{d-1}}, \quad \text{(No implied sum over $\mu, \alpha_1, \ldots, \alpha_{d-1}$.)}
\end{equation}
are obtained. This can be more lucidly expressed as
\begin{equation}
    \operatorname{sgn}{\sigma}\,\varepsilon^{\beta_0\beta_1\cdots\beta_{d-1}}g_{\sigma(0)\beta_0}g_{\sigma(1)\beta_1}\cdots g_{\sigma(d-1)\beta_{d-1}}
\end{equation}
where $\sigma$ is an element of the symmetric group of degree $d$. The advantage is that relabeling indices immediately gives
\begin{equation}
    \operatorname{sgn}{\sigma}\,\varepsilon^{\beta_{\sigma(0)}\beta_{\sigma(1)}\cdots\beta_{\sigma(d-1)}}g_{\sigma(0)\beta_{\sigma(0)}}g_{\sigma(1)\beta_{\sigma(1)}}\cdots g_{\sigma(d-1)\beta_{\sigma(d-1)}},
\end{equation}
which by use of the inverse permutation $\sigma^{-1}$ and the fact that a permutation and its inverse have equal parity can be reorganized to
\begin{equation}
    \varepsilon^{\beta_0\beta_1\cdots\beta_{d-1}}g_{0\beta_0}g_{1\beta_1}\cdots g_{(d-1)\beta_{d-1}},
\end{equation}
which is simply the determinant. This shows that $M^{\mu\beta_0}g_{\beta_0\nu} = g\delta^\mu_\nu$ which implies that $M^{\mu\nu} = gg^{\mu\nu}$ and hence that $\delta g = gg^{\mu\nu}\delta g_{\mu\nu}$. Finally, I obtain
\begin{equation}
    \delta\metdens = \frac12\metdens g^{\mu\nu}\delta g_{\mu\nu}.
\label{eq:metdens-variation}\end{equation}
To compute the variation of the Ricci scalar, the variation of the Riemann tensor is needed, making it necessary to first vary the Christoffel symbol. From the definition,
\begin{align}
    \delta \Gamma\indices{^\rho_{\mu\nu}} = \frac12g^{\rho\lambda}(2\partial_{(\mu}\delta g_{\nu)\lambda} - \partial_\lambda\delta g_{\mu\nu}) + \frac12\delta (g^{\rho\lambda})(2\partial_{(\mu}g_{\nu)\lambda} - \partial_\lambda g_{\mu\nu}).
\end{align}
Using that $\delta(g^{\mu\rho}g_{\rho\nu}) = 0$ to determine the variation of the inverse metric,
\begin{equation}
    \delta \Gamma\indices{^\rho_{\mu\nu}} = \frac12g^{\rho\lambda}(2\partial_{(\mu}\delta g_{\nu)\lambda} - \partial_\lambda \delta g_{\mu\nu}) - g^{\rho\lambda}\Gamma\indices{^\sigma_{\mu\nu}}\delta g_{\sigma\lambda}.
\end{equation}
It is prudent at this point to write $g^{\rho\lambda}\Gamma\indices{^\sigma_{\mu\nu}}\delta g_{\sigma\lambda} = g^{\rho\lambda}\Gamma\indices{^\sigma_{(\mu\nu)}}\delta g_{\sigma\lambda}$ and to add and subtract $\lambda g^{\rho\lambda}\Gamma\indices{^\sigma_{\lambda(\mu}}h\indices{_{\nu)\sigma}}$ from the expression, as this enables the following simplification:
\begin{align}
    \delta \Gamma\indices{^\rho_{\mu\nu}} &= \frac12g^{\rho\lambda}(2\partial_{(\mu}\delta g_{\nu)\lambda} - \partial_\lambda \delta g_{\mu\nu}) - g^{\rho\lambda}\Gamma\indices{^\sigma_{(\mu\nu)}}\delta g_{\sigma\lambda} + g^{\rho\lambda}\Gamma\indices{^\sigma_{\lambda(\mu}}\delta g\indices{_{\nu)\sigma}} - g^{\rho\lambda}\Gamma\indices{^\sigma_{\lambda(\mu}}\delta g\indices{_{\nu)\sigma}} \notag\\
    &= \frac12g^{\rho\lambda}\Big(2\partial_{(\mu}\delta g_{\nu)\lambda} - 2\Gamma\indices{^\sigma_{(\mu\nu)}}\delta g_{\sigma\lambda} - 2\Gamma\indices{^\sigma_{\lambda(\mu}}\delta g\indices{_{\nu)\sigma}} - \partial_\lambda \delta g_{\mu\nu} + 2\Gamma\indices{^\sigma_{\lambda(\mu}}\delta g\indices{_{\nu)\sigma}}\Big) \notag\\
    &= \frac12\big(2\nabla_{(\mu}\delta g\indices{_{\nu)}^\rho} - \nabla^\rho \delta g_{\mu\nu}\big).
\end{align}
It is important to note that the variation of the metric with raised indices is defined as
\begin{equation}
    \delta g^{\mu\nu} = g^{\mu\rho}g^{\nu\sigma}\delta g_{\rho\sigma}.
\end{equation}
The result for the Christoffel symbol is not surprising; the variation of the connection is the difference between the connection of the perturbed and unperturbed geometry, and it is a dictum of general relativity that the difference between two connections is a tensor, necessitating that each partial derivative, being non-tensorial, be compensated with pieces non-tensorial in exactly the opposite way. With the above result, the variation of the Riemann tensor can be calculated. From the definition,
\begin{align}
    \delta R\indices{^\mu_{\nu\rho\sigma}} &= \delta\big(2\partial_{[\rho}\Gamma\indices{^\mu_{\sigma]\nu}} + 2
    \Gamma\indices{^\lambda_{\nu[\sigma}}\Gamma\indices{^\mu_{\rho]\lambda}}\big) \notag\\
    &= 2\partial_{[\rho}\delta\Gamma\indices{^\mu_{\sigma]\nu}} + 2
    \delta\Gamma\indices{^\lambda_{\nu[\sigma}}\Gamma\indices{^\mu_{\rho]\lambda}} + 2\Gamma\indices{^\lambda_{\nu[\sigma}}\delta\Gamma\indices{^\mu_{\rho]\lambda}}.
\label{eq:riemann-variation-start}\end{align}
By exchanging $\rho$ and $\sigma$ in the third term and adding and subtracting $\Gamma\indices{^\lambda_{\rho\sigma}}\delta\Gamma\indices{^\mu_{\nu\lambda}}$, the terms combine into covariant derivatives, and the expression simplifies into the so-called Palatini identity
\begin{equation}
    \delta R\indices{^\mu_{\nu\rho\sigma}} = 2\nabla\indices{_{[\rho}}\delta\Gamma\indices{^\mu_{\sigma]\nu}}.
\end{equation}
Again, note that this identity was implied merely by $\delta R\indices{^\mu_{\nu\rho\sigma}}$ being a tensor and the appearance of the two partial derivatives in eq. \eqref{eq:riemann-variation-start}. To get the result in terms of $\delta g_{\mu\nu}$, it remains only to insert the variation of the Christoffel symbol. This gives
\begin{equation}
    \delta R\indices{^\mu_{\nu\rho\sigma}} = \nabla_\rho\nabla\indices{_{(\sigma}}\delta g\indices{_{\nu)}^\mu} - \nabla_\sigma\nabla\indices{_{(\rho}}\delta g\indices{_{\nu)}^\mu} + \nabla_{[\sigma}\nabla^\mu \delta g_{\rho]\nu}.
\end{equation}
Contracting two indices in this expression gives
\begin{align}
    \delta R_{\mu\nu} &=  \delta R\indices{^\lambda_{\mu\lambda\nu}} \notag\\
    &= \nabla_\lambda\nabla\indices{_{(\mu}}\delta g\indices{_{\nu)}^\lambda} - \frac12\nabla_\mu\nabla_\nu \delta g - \frac12\nabla^2\delta g_{\mu\nu},
\end{align}
where I defined $\delta g = g^{\mu\nu}\delta g_{\mu\nu}$. Finally, the variation of the Ricci scalar is
\begin{align}
    \delta R &= \delta(g^{\mu\nu}R_{\mu\nu}) \notag\\
    &= -\delta g_{\mu\nu}R^{\mu\nu} + g^{\mu\nu}\delta R_{\mu\nu} \notag\\
    &= -\delta g_{\mu\nu}R^{\mu\nu} + \nabla^\mu\nabla^\nu\delta g_{\mu\nu} - \nabla^2\delta g.
\label{eq:ricci-variation}\end{align}
The motivation for these calculations was to obtain Einstein's equation from the action principle, and the variation of the the action can now be determined using eqs. \eqref{eq:metdens-variation} and \eqref{eq:ricci-variation}. I obtain
\begin{equation}
    \delta S_\text{EH} + \delta S_\text{matter} = \frac{2}{\kappa^2}\int\dx\metdens\big(G^{\mu\nu}\delta g_{\mu\nu} + \nabla_\rho V^{\rho\mu\nu}\delta g_{\mu\nu}\big) + \int\dx\frac{\delta S_\text{matter}}{\delta g_{\mu\nu}}\delta g_{\mu\nu},
\label{eq:eh-variation}\end{equation}
where
\begin{equation}
    G_{\mu\nu} = R_{\mu\nu} - \frac12Rg_{\mu\nu}
\end{equation}
is the Einstein tensor, and
\begin{equation}
    V^{\rho\mu\nu} = \nabla^\rho g^{\mu\nu} - \nabla^\mu g^{\rho\nu}.
\end{equation}
The second term in eq. \eqref{eq:eh-variation} is a total derivative originating from the last two terms in eq. \eqref{eq:ricci-variation}. By Stokes' theorem, it can be written as an integral over the boundary at temporal and spatial infinity, which can be made to vanish by assuming that $\delta g_{\mu\nu}$ vanish there.\footnote{If one is working with an Einstein--Hilbert action defined on a manifold with a boundary, or if one is for other reasons concerned with what happens at temporal and spatial infinity, it is necessary to include in the action the so-called Gibbons--Hawking--York term.} Requiring that eq. \eqref{eq:eh-variation} vanish and remembering that $\delta g_{\mu\nu}$ is arbitrary, Einstein's equation is obtained
\begin{equation}
    G_{\mu\nu} = \frac{\kappa^2}{4}T_{\mu\nu},
\label{eq:einsteins-eq}\end{equation}
where I defined the energy-momentum tensor
\begin{equation}
    T^{\mu\nu} = \frac{-2}{\metdens}\frac{\delta S_\text{matter}}{\delta g_{\mu\nu}}.
\label{eq:emtensordef}\end{equation}
Through the trace of eq. \eqref{eq:einsteins-eq}, the Ricci scalar can be related to the trace of the energy-momentum tensor $T$. This, in turn, can be used to transform Einstein's equation into the useful form
\begin{equation}
    R_{\mu\nu} = \frac{\kappa^2}{4}\Big(T_{\mu\nu} - \frac{1}{d-2}Tg_{\mu\nu}\Big).
\end{equation}
From this one concludes that if there is no matter in the theory, the metric satisfies
\begin{equation}
    R_{\mu\nu} = 0.
\label{eq:einsteins-eq-ricci}\end{equation}
Spacetimes satisfying this are called Ricci flat.

\section{Action for a point particle}\label{sec:action for a point particle}
Before discussing the Schwarzschild--Tangherlini solution, it is necessary to review the action for a point particle. The motion of a point particle of mass $M$ traces out a worldline in spacetime described by the field $x^\mu(\tau)$, where $\tau$ is a parameter along the worldline. This is the trajectory of the particle. The parameter $\tau$ is arbitrary in the sense that if I transform $\tau \to \tau'(\tau)$, where $\tau'$ is monotonic, and
\begin{equation}
    x^\mu(\tau) \to x'^\mu(\tau'(\tau)) = x^\mu(\tau),
\end{equation}
then $x'^\mu$ and $x^\mu$ describe the same worldline. In general relativity, point particles follow geodesics, which maximize the proper time felt by the particle. This means the action has to be
\begin{equation}
    S_\text{wl}[x,g] = -M\int_\text{wl}\dd\tau,
\end{equation}
where the mass is needed to make the dimensions work out. To calculate $\dd\tau$, the line element,
\begin{equation}
    \dd\tau^2 = g_{\mu\nu}\dd x^\mu\dd x^\nu,
\end{equation}
must be pulled back to the worldline using the trajectory. The pullback of the basis one-forms is
\begin{equation}
    \dd x^\mu = \dot x^\mu \dd\tau,
\end{equation}
where $\dot x^\mu = \dd x^\mu/\dd\tau$, so the action becomes
\begin{align}
    S_\text{wl}[x,g] = -M\int\dd\tau\,\sqrt{g_{\mu\nu}(x)\dot x^\mu\dot x^\nu}.
\label{eq:wlactionnambugoto}\end{align}
In fact, this version of the point particle action is not nice to work with due to nonlinearity of the coupling between $x^\mu$ and $g_{\mu\nu}$. This issue is cured by defining the alternative action
\begin{equation}
    S_\text{wl}[x,g,e] = -\frac{M}{2}\int\dd\tau\,\big(e^{-1}g_{\mu\nu}(x)\dot x^\mu\dot x^\nu + e\big),
\label{eq:wlactionnotgaugefixed}\end{equation}
where $e = e(\tau)$ is a new worldline field known as the einbein. Its purpose in life is to guarantee the action be invariant under reparameterizations $\tau \to \tau'(\tau)$. To do this, it must transform under such a transformation as
\begin{equation}
    e(\tau) \to e'(\tau'(\tau)) = \frac{\dd \tau'}{\dd \tau}e(\tau)
\end{equation}
such as to cancel the factor coming from the measure. The equation of motion for the einbein, obtained from varying the action with respect to it, is
\begin{equation}
    e^2 = g_{\mu\nu}\dot x^\mu\dot x^\nu.
\label{eq:einbeineom}\end{equation}
Inserting this into eq. \eqref{eq:wlactionnotgaugefixed} yields eq. \eqref{eq:wlactionnambugoto}. The advantage of eq. \eqref{eq:wlactionnotgaugefixed}, is that it exposes better than eq. \eqref{eq:wlactionnambugoto} the gauge invariance of the description that reparameterization invariance represents, which enables a smart choice of gauge to be made. In proper time gauge, where $e = 1$, the parameter $\tau$ is forced to be the proper time as can be seen from eq. \eqref{eq:einbeineom}. In this gauge, the action is simply
\begin{equation}
    S_\text{wl}[x,g] = -\frac{M}{2}\int\dd\tau\,\big(g_{\mu\nu}(x)\dot x^\mu\dot x^\nu + 1\big),
\label{eq:wlaction}\end{equation}
and this is the gauge I shall work in. Using eq. \eqref{eq:emtensordef}, the energy-momentum tensor for a point particle can now be determined to be
\begin{equation}
    T_{\mu\nu}(x) = M\int\dd\tau\,\frac{\delta^{(d)}(x - x(\tau))}{\metdens}\dot x_\mu\dot x_\nu.
\label{eq:pp-emt}\end{equation}
For completeness, I also briefly review how to vary the worldline action. The variation is
\begin{equation}
    \delta S_\text{wl} = -\frac{M}{2}\int\dd\tau\,\big(\dot x^\mu\dot x^\nu\delta g_{\mu\nu}(x) + 2g_{\mu\nu}(x)\dot x^\mu\delta\dot x^\nu\big)
\end{equation}
Considering the endpoints of the trajectory fixed, total derivatives can freely be discarded from the above. Thus, I can integrate the first term by parts and use that $\delta g_{\mu\nu}(x) = \delta x^\rho\partial_\rho g_{\mu\nu}(x)$ to get
\begin{equation}
    \delta S_\text{wl} = M\int\dd\tau\,\Big(\ddot x^\rho + \dot x^\mu \dot g_{\mu\rho}(x) - \frac12\dot x^\mu\dot x^\nu g^{\rho\lambda}\partial_\lambda g_{\mu\nu}(x)\Big)\delta x_\rho.
\end{equation}
Using that $\dot g_{\mu\rho}(x) = \dot x^\nu\partial_\nu g_{\mu\rho}$, the last two terms combine into a Christoffel symbol,
\begin{equation}
    \delta S_\text{wl} = M\int\dd\tau\,\big(\ddot x^\rho + \Gamma\indices{^\rho_{\mu\nu}}(x)\dot x^\mu\dot x^\nu\big)\delta x_\rho,
\end{equation}
and the geodesic equation results if the variation is set to zero.

\section{Derivation of the Schwarzschild--Tangherlini solution}\label{sec:schwarzschild--tangherlini}
Having covered the energy-momentum tensor of a point particle, it is now possible to review the derivation of the geometry it will source. This solution to eq. \eqref{eq:einsteins-eq} is known as the Schwarzschild solution and was derived by Schwarzschild in 1915 \cite{Schwarzschild:1916uq}.\footnote{My derivation follows in broad strokes the four-dimensional derivation in \cite{Bartelmann:2019gr}. See also appendix J in \cite{Carroll:2004st} for a lucid introduction to non-coordinate bases.} Schwarzschild derived his solution in four dimensions; I shall need it in $d$ dimensions. The extension to this case was first worked out by Tangherlini and is known as the Schwarzschild--Tangherlini solution \cite{Tangherlini:1963bw}.

Consider a $d$-dimensional, static spacetime containing a single point particle of mass $M$ at rest. This spacetime possesses spherical symmetry around the particle, so a reasonable set of coordinates to introduce are a time coordinate $t$, a radial coordinate $r$ such that the location of the particle is at $r = 0$, and a set of angles $\chi^i$, where letters from the middle of the Latin alphabet run over $2,\ldots,d-1$.

Outside the origin, the line element can be expressed using the spherically symmetric ansatz
\begin{equation}
    \dd s^2 = \ee^{2A(r)}\dd t^2 - \ee^{2B(r)}\dd r^2 - r^2\dd\Omega^2_{d-2}, \qqquad \dd\Omega_{d-2}^2 = \gamma_{ij}\dd\chi^i\dd\chi^j,
\label{eq:scwh-ansatz}\end{equation}
where $\dd\Omega_{d-2}^2$ is the line element of the unit ($d-2$)-sphere. The metric must be asymptotically flat, implying that $A$ and $B$ must vanish asymptotically at infinity. To determine $A$ and $B$, the Riemann tensor must first be computed. This somewhat laborious task is made easier by introducing the set of basis one-forms
\begin{equation}
    e^{t} = \ee^A\dd t, \qqquad e^r = \ee^B\dd r, \qqquad e^i = r\varepsilon^i,
\end{equation}
and the dual set of basis vectors
\begin{equation}
    e_t = \ee^{-A}\partial_t, \qqquad e_r = \ee^{-B}\partial_r, \qqquad e_i = \frac{1}{r}\varepsilon_i.
\label{eq:vielbeinvecs}\end{equation}
The $\varepsilon^i$ and $\varepsilon_i$ are chosen to satisfy
\begin{equation}
    \eta_{ij}\varepsilon^i\varepsilon^j = \eta^{ij}\varepsilon_i\varepsilon_j =  -\gamma_{ij}\dd\chi^i\dd\chi^j.
\end{equation}
This basis is orthonormal, $e^a(e_b) = \delta^a_b$, and is an example of a vielbein. Expressed in this basis, the line element is simply
\begin{equation}
    \dd s^2 = \eta_{ab}e^a e^b,
\end{equation}
where letters from the beginning of the Latin alphabet run over $t, r, 2,\ldots, d-1$. The exterior derivative of the basis one-forms satisfies
\begin{equation}
    \dd e^a = -\omega\indices{^a_b}\wedge e^b, 
\label{eq:cartan1}\end{equation}
which is known as Cartan's first structure equation. The quantity $\omega\indices{^a_b}$ is known as the spin connection, and it is a one-form taking values in the space of tensors with one upper and one lower index,
\begin{equation}
    \omega\indices{^a_b} = \omega\indices{_c^a_b}e^c.
\end{equation}
Likewise, the Riemann tensor can be seen as a two-form taking values in the space of tensors with one upper and one lower index,
\begin{equation}
    R\indices{^a_b} = R\indices{^a_{bcd}}e^c\wedge e^d.
\end{equation}
Expressed in this way, the Riemann tensor satisfies the equation
\begin{equation}
    R\indices{^a_b} = \dd\omega\indices{^a_b} + \omega\indices{^a_c}\wedge\omega\indices{^c_b},
\label{eq:cartan2}\end{equation}
which is known as Cartan's second structure equation. Thus, one can determine the Riemann tensor through eq. \eqref{eq:cartan2} by first determining the spin connection through eq. \eqref{eq:cartan1}.

The first step is to calculate the exterior derivatives of the vielbeins,
\begin{subequations}
\begin{align}
    \dd e^t &= A'\ee^{-B}e^r\wedge e^t, \\
    \dd e^r &= 0, \\
    \dd e^i &= \frac{\ee^{-B}}{r}e^r\wedge e^i + r\dd\varepsilon^i.
\end{align}
\end{subequations}
I denote radial derivatives by primes. Using these, a system of equations for the spin connection can be set up with eq. \eqref{eq:cartan1}. I find
\begin{subequations}
\begin{align}
    A'\ee^{-B}e^r\wedge e^t &= -\omega\indices{^t_r}\wedge e^r -\omega\indices{^t_i}\wedge e^i, \label{eq:cartan1sys1}\\
    0 &= -\omega\indices{^r_t}\wedge e^t -\omega\indices{^r_i}\wedge e^i, \label{eq:cartan1sys2}\\
    \frac{\ee^{-B}}{r}e^r\wedge e^i + r\dd\varepsilon^i &= -\omega\indices{^i_t}\wedge e^t -\omega\indices{^i_r}\wedge e^r -r\omega\indices{^i_j}\wedge \varepsilon^j. \label{eq:cartan1sys3}
\end{align}
\end{subequations}
Eq. \eqref{eq:cartan1sys1} fixes
\begin{equation}
    \omega\indices{^t_r} = \omega\indices{^r_t} = A'\ee^{-B}e^t, \qqquad \omega\indices{^t_i} = \omega\indices{^i_t} = 0,
\end{equation}
which upon insertion into eq. \eqref{eq:cartan1sys2} implies
\begin{equation}
    \omega\indices{^r_i} = -\omega\indices{^i_r} = Ce^i
\end{equation}
for some function $C$. Finally, eq. \eqref{eq:cartan1sys3} implies that
\begin{equation}
    C = -\frac{e^{-B}}{r}, \qqquad \dd\varepsilon^i = -\omega\indices{^i_j}\wedge \varepsilon^j.
\label{eq:schwcrucialstep}\end{equation}
With the spin connection in hand, the Riemann tensor can now be found using eq. \eqref{eq:cartan2}. The components involving $t$ are simple to obtain:
\begin{subequations}
\begin{alignat}{2}
    R\indices{^t_r} &= R\indices{^r_t} &&= (A'' - A'B' + A'^2)\ee^{-2B}e^r\wedge e^t, \\
    R\indices{^t_i} &= R\indices{^i_t} &&= -\frac{A'\ee^{-2B}}{r}e^t\wedge e^i.
\end{alignat}
\end{subequations}
The purely spatial parts are slightly more involved:
\begin{align}
    R\indices{^r_i} &= \dd\omega\indices{^r_i} + \omega\indices{^r_j}\wedge\omega\indices{^j_i} \notag\\
    &= \bigg(\frac{B'\ee^{-2B}}{r} + \frac{\ee^{-2B}}{r^2}\bigg)e^r\wedge e^i - \frac{\ee^{-2B}}{r^2}e^r\wedge e^i - \ee^{-B}\dd\varepsilon^i - \ee^{-B}\varepsilon^j\wedge\omega\indices{^j_i} \notag\\
    &= \frac{B'\ee^{-2B}}{r}e^r\wedge e^i - \ee^{-B}\dd\varepsilon^i - \ee^{-B}\varepsilon^j\wedge\omega\indices{^j_i}.
\end{align}
A sum over $j$ should still be understood in the above. The anti-symmetry of the spin connection along with eq. \eqref{eq:schwcrucialstep} imply that the last two terms cancel, leaving
\begin{equation}
    R\indices{^r_i} = -R\indices{^i_r} = \frac{B'\ee^{-2B}}{r}e^r\wedge e^i.
\end{equation}
Finally, the spherical pieces become
\begin{align}
    R\indices{^i_j} &= \dd\omega\indices{^i_j} + \omega\indices{^i_k}\wedge\omega\indices{^k_j} + \omega\indices{^i_r}\wedge\omega\indices{^r_j} \notag\\
    &= \hat R\indices{^i_j} - \frac{\ee^{-2B}}{r^2}e^i\wedge e^j.
\end{align}
Here, $\hat R\indices{^i_j}$ is the Riemann tensor for the $(d-2)$-sphere with radius $r$. This space has constant sectional curvature $1/r^2$, so it takes the simple form
\begin{equation}
    \hat R\indices{^i_j} = \frac{1}{r^2}e^i\wedge e^j,
\end{equation}
from which it follows that
\begin{equation}
    R\indices{^i_j} = \frac{1-\ee^{-2B}}{r^2}e^i\wedge e^j.
\end{equation}
The Ricci tensor in the vielbein basis can now simply be obtained by evaluating the Riemann tensor on the basis vectors introduced in eq. \eqref{eq:vielbeinvecs}. For this it is necessary to know that
\begin{equation}
    (e^a\wedge e^b)(e_c,e_d) = 2\delta^{[a}_c\delta^{b]}_d, \qqquad e^i(e_i) = d-2.
\end{equation}
The non-zero components are
\begin{subequations}
\begin{alignat}{2}
    R_{tt} &= R\indices{^a_t}(e_a, e_t) &&= (A'' - A'B' + A'^2)\ee^{-2B} + (d-2)\frac{A'\ee^{-2B}}{r}, \\
    R_{rr} &= R\indices{^a_r}(e_a, e_r) &&= -(A'' - A'B' + A'^2)\ee^{-2B} + (d-2)\frac{B'\ee^{-2B}}{r}, \\
    R_{ij} &= R\indices{^a_i}(e_a, e_j) &&= \frac{B' - A'}{r}\ee^{-2B}\delta^i_j + (d-3)\frac{1-\ee^{-2B}}{r^2}\delta^i_j.
\end{alignat}
\end{subequations}
With this, the Ricci scalar can be determined to be
\begin{align}
    R &= \eta^{ab}R_{ab} \notag \\
    &= 2(A'' - A'B' + A'^2)\ee^{-2B} - 2(d-2)\frac{B' - A'}{r}\ee^{-2B} - (d-2)(d-3)\frac{1-\ee^{-2B}}{r^2}.
\label{eq:schw-ricci}\end{align}
Finally, the non-zero components of the Einstein tensor in the vielbein basis turn out to be
\begin{subequations}
\begin{alignat}{2}
    G_{tt} &= (d-2)\Bigg[\frac{d-3}{2r^2} + \ee^{-2B}\Big(\frac{B'}{r} - \frac{d-3}{2r^2}\Big)\Bigg], \label{eq:scwh-einstein1}\\
    G_{rr} &= (d-2)\Bigg[-\frac{d-3}{r^2} + \ee^{-2B}\Big(\frac{A'}{r} + \frac{d-3}{2r^2}\Big)\Bigg], \label{eq:scwh-einstein2}\\
    G_{ii} &= (A''-A'B'+A'^2)\ee^{-2B} - (d-3)\Bigg[\ee^{-2B}\frac{B'-A'}{r} + \frac{d}{2}\frac{1-\ee^{-2B}}{r}\Bigg]. \label{eq:scwh-einstein3}
\end{alignat}
\end{subequations}
Away from the origin, the metric satisfies $G_{ab} = 0$. By adding eqs. \eqref{eq:scwh-einstein1} and \eqref{eq:scwh-einstein2}, this can be seen to imply
\begin{equation}
    A' + B' = 0.
\label{eq:AB-middlestep}\end{equation}
It was required that $A$ and $B$ vanish at infinity. Hence, integrating eq. \eqref{eq:AB-middlestep} from infinity to any finite radius implies
\begin{equation}
    A = -B.
\end{equation}
Now, $G_{tt} = 0$ implies that
\begin{equation}
    \ee^{-2B}\Big(rB'-\frac{d-3}{2}\Big) = \frac{d-3}{2}.
\end{equation}
By defining $f(r) = \ee^{-2B}$, this can be turned into
\begin{equation}
    rf'(r) + (d-3)(f(r) - 1) = 0.
\end{equation}
Then, writing $f' = \dd f/\dd r$ and rearranging, this can be straightforwardly integrated,
\begin{equation}
    \int\frac{\dd f}{f-1} = -(d-3)\int\frac{\dd r}{r},
\end{equation}
yielding the familiar solution
\begin{equation}
    f(r) = 1 - \frac{\Lambda_d}{r^{d-3}},
\end{equation}
where $\Lambda_d$ is a potentially dimensionally dependent integration constant. The metric functions can be determined explicitly to be
\begin{equation}
    \ee^{2A} = f(r), \qqquad \ee^{2B} = \frac{1}{f(r)},
\end{equation}
implying that the line element is
\begin{equation}
    \dd s^2 = f(r)\dd t^2 - \frac{1}{f(r)}\dd r^2 - r^2\dd\Omega_{d-2}^2.
\end{equation}
The remaining task is to fix the integration constant. I gave the energy-momentum tensor in eq. \eqref{eq:pp-emt}. As the particle is at rest, $x = (\tau,0,\ldots)$, so the integral over $\tau$ simply enforces $\tau = t$. Using this, the energy-momentum tensor simplifies to
\begin{equation}
    T_{\mu\nu}(\mathbf{x}) = M\frac{\delta^{(d-1)}(\mathbf{x})}{\metdens}\delta_\mu^t\delta_\nu^t,
\end{equation}
where $\mathbf{x} = (r,\chi^i)$. The components of the Einstein tensor were calculated in the vielbein basis, so the energy-momentum tensor must also be expressed in this basis. This is done by computing
\begin{align}
    T_{ab} &= e\indices{_a^\mu}e\indices{_b^\nu}T_{\mu\nu} \notag\\
    &= \frac{M}{f(r)}\frac{\delta^{(d-1)}(\mathbf{x})}{\metdens}\delta_a^t\delta_b^t,
\end{align}
where I used that $e\indices{_t^t} = f^{-\frac12}(r)$. Then, Einstein's equation including the origin is
\begin{equation}
    G_{tt} = \frac{1}{f(r)}\frac{\kappa^2M}{4}\frac{\delta^{(d-1)}(\mathbf{x})}{\metdens}.
\label{eq:ttcompregularize}\end{equation}
On the face of it, this equation appears quite ill-posed. By construction, $G_{tt}$ vanishes and $f$ is divergent at $r=0$. These facts mean a regularization procedure is required before proper sense can be made of eq. \eqref{eq:ttcompregularize}. One such procedure is to introduce
\begin{equation}
    f_\alpha(r) = 1 - h_\alpha(r)\frac{\Lambda_d}{r^{d-3}},
\end{equation}
where $\alpha$ is a regulator that should be taken to zero at the end of the computation \cite{Hayman:2024aev}. The regulating function must satisfy
\begin{subequations}
\begin{align}
    \lim_{\alpha\to0}h_\alpha(r) = 1, \\
    \lim_{r\to0}\frac{h_\alpha(r)}{r^{d-3}} = 0;
\end{align}
\end{subequations}
an example of a function satisfying these conditions is $\ee^{-\alpha/r}$. Now eq. \eqref{eq:ttcompregularize} can be integrated over a ball of radius $R$ centered at the origin. The right-hand side works out to
\begin{equation}
    \frac{\kappa^2M}{4}\int_{r\leq R}\dd^{d-1}\mathbf{x}\,\metdens\frac{1}{f_\alpha(r)}\frac{\delta^{(d-1)}(\mathbf{x})}{\metdens} = \frac{\kappa^2M}{4}.
\label{eq:lambda_d-rhs}\end{equation}
The left-hand side is determined by inserting $f_\alpha(r)$ into eq. \eqref{eq:scwh-einstein1}. This yields the finite
\begin{equation}
    G_{tt} = \frac{(d-2)\Lambda_d}{2r^{d-2}}h'_\alpha(r),
\end{equation}
which upon integration results in
\begin{align}
    \int_{r\leq R}\dd^{d-1}\mathbf{x}\,\metdens G_{tt} &= \frac{(d-2)\Omega_{d-2}\Lambda_d}{2}\int_0^R\dd r\, h'_\alpha(r) \notag\\
    &= \frac{(d-2)\Omega_{d-2}\Lambda_dh_\alpha(R)}{2},
\label{eq:lambda_d-lhs}\end{align}
where
\begin{equation}
    \Omega_{d-2} = \frac{2\pi^\frac{d-1}{2}}{\Gamma\big(\frac{d-1}{2}\big)}
\label{eq:sphere-surface-area}\end{equation}
is the surface area of the $(d-2)$-sphere. I also used that
\begin{equation}
    \metdens = r^{d-2}\sqrt{\abs{\gamma}},
\end{equation}
where $\gamma$ was defined in eq. \eqref{eq:scwh-ansatz}. By taking $\alpha\to0$ and identifying eqs. \eqref{eq:lambda_d-rhs} and \eqref{eq:lambda_d-lhs}, the integration constant is determined to be
\begin{equation}
    \Lambda_d = \frac{\kappa^2M}{2(d-2)\Omega_{d-2}}.
\label{eq:lambdaddef}\end{equation}
This concludes the derivation of the Schwarzschild--Tangherlini solution.

For amplitude computations, the Schwarzschild coordinates employed in the above are not very convenient. It is in fact nicer to work in an isotropic coordinate system, where every spatial direction is on an equal footing. This is computationally very convenient, as will become obvious later in the thesis. The line element in isotropic coordinates is
\begin{equation}
    \dd s^2 = \alpha(r')\dd t^2 - \beta(r')\dd\Sigma_{d-1}^2.
\label{eq:isotropic-def}\end{equation}
Here, $r'$ is some new radial coordinate related to the Schwarzschild one, and $\dd\Sigma_{d-2}^2$ is the line element of $(d-1)$-dimensional Euclidean space,
\begin{equation}
    \dd\Sigma_{d-1}^2 = \dd r'^2 + r'^2\dd\Omega_{d-2}^2,
\label{eq:euclid-line-element-def}\end{equation}
To derive $\alpha$ and $\beta$, one needs only to derive the new radial coordinate $r'$, which can be done by equating the spatial part of the metric in the old and new coordinates,
\begin{equation}
    \frac{1}{f(r)}\dd r^2 + r^2\dd\Omega_{d-2}^2 = \beta(r')\big(\dd r'^2 + r'^2\dd \Omega_{d-2}^2\big).
\end{equation}
This provides two equations for $r'$, namely
\begin{subequations}
\begin{align}
    r^2 &= \beta(r')r'^2, \\
    \frac{\dd r^2}{f(r)} &= \beta(r')\dd r'^2.
\end{align}
\end{subequations}
Inserting the first in the second and taking the square root yields
\begin{equation}
    \frac{\dd r'}{r'} = \pm\frac{\dd r}{r\sqrt{f(r)}},
\end{equation}
which can be integrated to give
\begin{equation}
    r'(r) = r'_0\Bigg[\frac{2r^{d-3}-\Lambda_d\pm2\sqrt{r^{d-3}(r^{d-3}-\Lambda_d)}}{\Lambda_d}\Bigg]^\frac{1}{d-3},
\label{eq:isotropic-radial-coord}\end{equation}
where $r'_0$ is an integration constant. The sign and the constant can be determined by requiring that the line element in eq. \eqref{eq:isotropic-def} asymptote to the Minkowski one, which can only occur if $r'$ is asymptotically equal to $r$. The asymptotic expansion of eq. \eqref{eq:isotropic-radial-coord} is
\begin{equation}
    r'(r) \sim r'_0\bigg(\frac{2\pm 2}{\Lambda_d}\bigg)^\frac{1}{d-3}r \qquad\text{as}\qquad r\to\infty,
\end{equation}
which implies that the positive sign must be chosen and that
\begin{equation}
    r'_0 = \bigg(\frac{\Lambda_d}{4}\bigg)^\frac{1}{d-3}.
\end{equation}
Inserting this into eq. \eqref{eq:isotropic-radial-coord} and inverting yields the old radial coordinate in terms of the new one,
\begin{equation}
    r(r') = r'\bigg(1 + \frac{\Lambda_d}{4r'^{d-3}}\bigg)^\frac{2}{d-3}.
\end{equation}
The coordinate transformation can now be carried out, and the resulting line element is
\begin{equation}
    \dd s^2 = \frac{\Big(1-\frac{\Lambda_d}{4r'^{d-3}}\Big)^2}{\Big(1+\frac{\Lambda_d}{4r'^{d-3}}\Big)^2}\dd t^2 - \bigg(1 + \frac{\Lambda_d}{4r'^{d-3}}\bigg)^\frac{4}{d-3}\dd\Sigma_{d-1}^2.
\label{eq:isotropic-line-element}\end{equation}
The line element in eq. \eqref{eq:isotropic-line-element} describes the field generated by a point particle at rest. In an amplitude computation, one is interested in maintaining manifest Lorentz covariance. For this reason, it is convenient already at this point to generalize eq. \eqref{eq:isotropic-line-element} to the case where the metric is sourced by a point mass moving with constant velocity $v^\mu$ with respect to an observer at asymptotic infinity. To do this, it is useful first to write $\dd\Sigma_{d-1}^2$ in cartesian coordinates,
\begin{equation}
    \dd\Sigma_{d-1}^2 = (\dd x^1)^2 + \cdots +(\dd x^{d-1})^2,
\end{equation}
which upon comparison with eq. \eqref{eq:euclid-line-element-def} yields the relation
\begin{equation}
    r'^2 = (x^1)^2 + \cdots +(x^{d-1})^2,
\end{equation}
I will from now on let $x^0 = t$. In the rest frame of the particle, $v^\mu = (1,0,\ldots)$. This means the coordinate basis one-forms can be written
\begin{equation}
    (\dd x^0)^2 = \eta_{\parallel\mu\nu}\dd x^\mu\dd x^\nu, \qqquad \dd\Sigma_{d-1}^2 = -\eta_{\perp\mu\nu}\dd x^\mu\dd x^\nu,
\end{equation}
where I defined the restricted metrics
\begin{equation}
    \eta_{\parallel\mu\nu} = v_\mu v_\nu, \qqquad \eta_{\perp\mu\nu} = \eta_{\mu\nu} - v_\mu v_\nu,
\end{equation}
that act as projection operators onto, respectively, the one-dimensional, timelike subspace collinear with the point particle velocity and the $(d-1)$-dimensional, spacelike subspace orthogonal to it. The radius can also be written in this way as
\begin{equation}
    r'^2 = -\eta_{\perp\mu\nu}x^\mu x^\nu = \abs{x_\perp}^2,
\end{equation}
where I used the notation
\begin{subequations}
\begin{align}
    x_\perp^\mu &= \eta_{\perp\nu}^\mu x^\nu, \\
    \abs{x_\perp} &= \sqrt{\abs{x_\perp^2}}.
\end{align}
\end{subequations}
As these quantities are Lorentz scalars, they are valid in any (asymptotically) inertial frame. These final definitions and observations land me on the line element I will need for the following developments:
\begin{equation}
    \dd s^2 = g_{\mu\nu}\dd x^\mu \dd x^\nu,
\label{eq:metric-final}\end{equation}
where the metric is
\begin{equation}
    g_{\mu\nu} = \frac{\Big(1-\frac{\Lambda_d}{4\abs{x_\perp}^{d-3}}\Big)^2}{\Big(1+\frac{\Lambda_d}{4\abs{x_\perp}^{d-3}}\Big)^2}\eta_{\parallel\mu\nu} + \bigg(1 + \frac{\Lambda_d}{4\abs{x_\perp}^{d-3}}\bigg)^\frac{4}{d-3}\eta_{\perp\mu\nu}.
\label{eq:schwarzschild-tangherlini}\end{equation}

\chapter{Quantization of gravity}\label{ch:perturbative quantization of gravity}
Having covered the classical, geometric formulation of gravitation, I now turn to the quantum field-theoretic treatment. A non-geometric point of view on gravitation was famously promulgated by Feynman in \cite{Feynman:2003kb}, where he by a recursive argument based on weak-field gravity proved the remarkable result that general relativity is the only self-consistent theory of an interacting massless spin-two field. Another quantum field-theoretic point of view on gravity expounded in e.g. \cite{Donoghue:2017pgk} is based on the dramatic parallels that exist between gauge theories and gravity. Specifically, it is possible to obtain the Einstein--Hilbert action in the same way one obtains scalar quantum electrodynamics by gauging the global U(1) symmetry. I discuss this in section \ref{sec:gravityasgauge}. Then I turn to the problem of quantizing this theory. In section \ref{sec:pathintegralquant}, I will provide a brief review of path integral quantization and the relevant quantities involved. Then, in section \ref{sec:quantgeneralback}, I carry out the quantization by expanding the metric about an arbitrary background. In section \ref{sec:feynman rules in the weak-field expansion}, I discuss the Feynman rules when this background is taken to be the Minkowski metric. Then, in section \ref{sec:feynmanrulesgeneralback}, I generalize the discussion of Feynman rules to an arbitrary background and specialize afterwards to the Schwarzschild--Tangherlini metric. It is at this point appropriate to discuss the difficulties that have been encountered in the quantization of gravity, and, crucially, how to circumvent them by treating the quantum theory of general relativity as a low-energy effective field theory. I discuss this in section \ref{sec:gravityaseft}. Finally, in section \ref{sec:classicalphysics}, I discuss how one may extract  classical quantities from the quantum theory of gravity.

\section{Gravity as a gauge theory}\label{sec:gravityasgauge}
Before discussing gravity, I will briefly review how abelian Yang-Mills theory arises from promoting the global gauge symmetry associated with a conserved current to a local one. In scalar quantum electrodynamics in flat space, one has a complex scalar field $\Phi$ described by the action,
\begin{equation}
    S_\text{matter}[\Phi] = \int\dx\Big[(\partial_\mu\Phi)(\partial^\mu\Phi)^* - m^2\abs{\Phi}^2\Big],
\label{eq:sqedlagrangian}\end{equation}
where $\Phi^*$ denotes the complex conjugate of $\Phi$. This action is invariant under the global gauge transformation
\begin{equation}
    \Phi\to\ee^{\ii\theta}\Phi,
\label{eq:sqedgaugetrans}\end{equation}
which, from Noether's theorem, gives rise to the current
\begin{equation}
    j^\mu = \ii\big(\Phi\partial^\mu\Phi^* - \Phi^*\partial^\mu\Phi\big),
\end{equation}
that is conserved, $\partial_\mu j^\mu = 0$, when $\Phi$ satisfies its equation of motion. To promote the gauge symmetry of \eqref{eq:sqedlagrangian} from a global one to a local one,
\begin{equation}
    \Phi\to\ee^{\ii\theta(x)}\Phi,
\end{equation}
one introduces a gauge field $A_\mu$ and a covariant derivative
\begin{equation}
    D_\mu = \partial_\mu - \ii eA_\mu,
\end{equation}
where $e$ characterizes the strength of the coupling of the gauge field to the scalar field. This gauge field transforms under the gauge transformation in eq. \eqref{eq:sqedgaugetrans} according to
\begin{equation}
    A_\mu \to A_\mu - e^{-1}\partial_\mu\theta.
\label{eq:photongaugetrans}\end{equation}
One can now make the theory gauge invariant by replacing the partial derivatives in the action in eq. \eqref{eq:sqedlagrangian} with covariant ones. Expanding, one finds
\begin{equation}
    S_\text{matter}[\Phi, A] = \int\dx\Big[(\partial_\mu\Phi)(\partial^\mu\Phi)^* - m^2\abs{\Phi}^2 - ej^\mu A_\mu + e^2A_\mu A^\mu\abs{\Phi}^2\Big].
\label{eq:invariantscalarqed}\end{equation}
As promised, the gauge field couples with strength $e$ to the conserved current of the ungauged theory $j^\mu$ with coupling $e$. For this theory, a quartic term also appears. To include the dynamics of the gauge field in the theory, one must construct a kinetic term that is bilinear in derivatives of $A_\mu$, a Lorentz scalar, and gauge-invariant. This can be constructed by considering the curvature,
\begin{equation}
    [D_\mu,D_\nu] = -\ii eF_{\mu\nu},
\end{equation}
where
\begin{equation}
    F_{\mu\nu} = \partial_\mu A_\nu - \partial_\nu A_\mu
\end{equation}
is the field strength tensor. The kinetic term can be uniquely chosen as 
\begin{equation}
    S_\text{gauge}[A] = -\frac14\int\dx F_{\mu\nu}F^{\mu\nu},
\end{equation}
where the prefactor is included such that the kinetic term for the physical degrees of freedom in $A_\mu$ has the canonical prefactor of $1/2$.

To repeat this analysis for gravity, I first introduce a real, massive scalar field $\varphi$ described by the action
\begin{equation}
    S_\text{matter}[\varphi] = \int\dx\bigg[\frac12\partial_\mu\varphi\partial^\mu\varphi - \frac12M^2\varphi\bigg].
\label{eq:phiaction}\end{equation}
It is not a recent observation that gravity couples to mass \cite{Newton:1687eqk}. Somewhat more recent is the observation that mass is equivalent to energy \cite{Einstein:1905nhh}. This and other arguments (see \cite{Feynman:2003kb, Donoghue:2017pgk}) implies that gravity must be mediated by a spin-two field coupling to the energy-momentum tensor $T_{\mu\nu}$. In certain restricted cases, such as for the action in eq. \eqref{eq:phiaction}, $T_{\mu\nu}$ can be seen as the conserved current associated with the invariance of the theory under spacetime translations
\begin{equation}
    x^\mu \to x^\mu + \xi^\mu,
\end{equation}
where $\xi^\mu$ is a constant vector.

In general, the current one obtains by applying the Noether procedure to spacetime translations, known as the \emph{canonical energy-momentum tensor}, does not coincide with the definition in eq. \eqref{eq:emtensordef}. On the contrary, it possesses some unnerving qualities, such as being neither symmetric nor gauge-invariant. Various \textit{ad hoc} approaches to ``improving'' the canonical energy-momentum have existed for many years, among whom perhaps the most notable is the Belinfante-Rosenfeld tensor. For discussions of other approaches to the solution of this problem, see \cite{Gotay:1992sem, Forger:2003ut}. As mentioned, the canonical energy-momentum tensor for scalar matter does coincide with eq. \eqref{eq:emtensordef}, so no improvement is needed in this case.

Thus, I proceed with gauging spacetime translations by allowing $\xi^\mu$ to be position-dependent
\begin{equation}
    x^\mu \to x^\mu + \xi^\mu(x).
\end{equation}
Without loss of generality, one can take $\xi^\mu(x) = -x^\mu + x^{\mu'}(x)$, implying that the new class of transformations
\begin{equation}
    x^\mu \to x^{\mu'}(x),
\label{eq:diffeo}\end{equation}
is just the class of diffeomorphisms, which have an associated Jacobian as defined in eq. \eqref{eq:jacobiannotation}. The first task is now to make eq. \eqref{eq:phiaction} invariant under diffeomorphisms by analogy with the steps that lead from eq. \eqref{eq:sqedlagrangian} to eq. \eqref{eq:invariantscalarqed}. The theory started out enjoying Poincaré invariance, whence the line element in the ungauged theory was
\begin{equation}
    \dd s^2 = \eta_{\mu\nu}\dd x^\mu\dd x^\nu.
\label{eq:linelementpregauge}\end{equation}
This is not invariant under diffeomorphisms, as can easily be seen by using the transformation properties of the basis vectors and basis one-forms, which were given in eq. \eqref{eq:basistransforms}. This fact necessitates the introduction of the metric $g_{\mu\nu}(x)$, now playing the role as gauge field, which obeys the transformation rule
\begin{equation}
    g_{\mu\nu}(x) \to g_{\mu'\nu'}(x') = J\indices{_{\mu'}^\mu}J\indices{_{\nu'}^\nu}g_{\mu\nu}(x).
\label{eq:metrictrans}\end{equation}
To ensure that scalars formed from basis vectors are also invariant, it is also necessary to introduce the matrix inverse of the metric $(g^{-1})^{\mu\nu} = g^{\mu\nu}$. Likewise, as was demonstrated in eq. \eqref{eq:measure-transform}, the measure $\dd^dx$ acquires a factor of the Jacobian determinant under a diffeomorphism, requiring that it be replaced by the invariant
\begin{equation}
    \metdens\dd^dx.
\end{equation}
In the case of scalar quantum electrodynamics, the introduction of a covariant derivative $D_\mu$ was necessary to correct for the surplus $\ii\theta$ for each derivative of $\Phi$. In the present case, it is the indices that transform. As such, the covariant derivative $\nabla_\mu$ must correct for all all indices, including its own. Considering the action on a vector field $V^\mu$, this requirement amounts to
\begin{equation}
    \nabla_\mu V^\nu \to \nabla_{\mu'}V^{\nu'} = J\indices{_{\mu'}^\mu}J\indices{^{\nu'}_\nu}\nabla_\mu V^\nu.
\end{equation}
This is where the analogy with gauge theories fails to give a unique answer, in that there exists many choices of covariant derivatives if one is willing to have torsion in ones theory. The argument given by Feynman in \cite{Feynman:2003kb} circumvents this (see also \cite{Schwartz:2014sze}). Choosing the Levi--Civita connection, however, gives rise to the covariant derivative
\begin{equation}
    \nabla_\mu V^\nu = \partial_\mu V^\nu + \Gamma\indices{^\nu_{\lambda\mu}}V^\lambda,
\end{equation}
where the Christoffel symbol was defined in eq. \eqref{eq:christoffel}. The invariant matter action can thus be seen to be
\begin{equation}
    S_\text{matter}[\varphi,g] = \int\dx\metdens\bigg[\frac12g^{\mu\nu}\nabla_\mu\varphi\nabla_\nu\varphi - \frac12M^2\varphi\bigg].
\label{eq:invariantrealscalar}\end{equation}
The analogy with the abelian case continues with the construction of the kinetic term. The curvature is defined through the commutator of the covariant derivative,
\begin{equation}
    [\nabla_\mu,\nabla_\nu]V^\rho = R\indices{^\rho_{\lambda\mu\nu}}V^\lambda,
\end{equation}
where the Riemann tensor is now interpreted as the gravitational field strength tensor. It is not possible to construct from the Riemann tensor a scalar that is only bilinear in derivatives of the metric. The best one can do is the Ricci scalar, which lands one directly on the Einstein--Hilbert action from eq. \eqref{eq:ehaction}.\footnote{It is interesting to note, however, that it is possible to get rid of the second derivatives of the metric in the Einstein--Hilbert action by integration-by-parts. See §26 of \cite{Dirac:1996gtr}.} This illustrates the dramatic similarities between general relativity and Yang-Mills theory.

\section{Path integral quantization}\label{sec:pathintegralquant}
I will first provide a generic discussion of path integral quantization, after which I specialize to the gravitational case. Given a quantum field theory defined by an action $I[\phi]$, the central objects of study in a quantum field theory are the Green's functions
\begin{equation}
    G^{i_1\cdots i_m} = \langle\Omega\vert T\phi^{i_1}\cdots\phi^{i_m}\vert\Omega\rangle,
\end{equation}
where $\vert\Omega\rangle$ is the vacuum of the theory and $T$ is the time-ordering operator. The collective index $i$ contains all discrete and continuous labels needed to characterize the field $\phi^i$,
\begin{equation}
    i = \{\text{particle type, spin, Lorentz indices, position, \ldots}\},
\end{equation}
I will assume the fields to be bosonic. The knowledge of all the Green's functions can be encoded in the partition function
\begin{equation}
    Z[J] = Z[0]\sum_{m=0}^\infty\frac{\ii^m}{m!}G^{i_1\cdots i_m}J_{i_1}\cdots J_{i_m},
\label{eq:partitionfunction}\end{equation}
where the $n=0$ term in the sum is one. This quantity is the expectation value of the vacuum in the presence of source, and $Z[0]$ is the expectation value of the vacuum with no sources present. The Einstein convention is used such that an index appearing in upper and lower position is summed over all its discrete labels and integrated over all its continuous ones.\footnote{This notation is inspired by \cite{Cvitanovic:1983eb}.} For instance, the field content later in the thesis will be given by a graviton $h_{\mu\nu}(x)$ and a deflection $z^\mu(\tau)$, in which case
\begin{equation}
    \phi^iJ_i = \int\dx h_{\mu\nu}(x)J^{\mu\nu}(x) + \int\dd\tau\,z^\mu(\tau)f_{\mu}(\tau),
\end{equation}
where $J^{\mu\nu}$ and $f_\mu$ are external sources for gravitons and deflections. Any Green's function can be retrieved from $Z[J]$ by functional differentiation,
\begin{equation}
    G^{i_1\cdots i_m} = \frac{(-\ii)^m}{Z[0]}\frac{\delta^mZ[J]}{\delta J_{i_1}\cdots\delta J_{i_m}}\bigg\vert_{J=0}
\label{eq:greensretrieve}\end{equation}
The partition function can be shown to obey a functional differential equation known as the Dyson--Schwinger equation, which has the formal solution
\begin{equation}
    Z[J] = \int\mathscr{D}\phi\,\ee^{\ii I[\phi] + \ii\phi^iJ_i},
\end{equation}
where the functional measure on the space of field configurations $\mathscr{D}\phi$ is given by
\begin{equation}
    \mathscr{D}\phi = \prod_i\dd\phi^i,
\end{equation}
In a free theory, the action is given by
\begin{equation}
    I_0[\phi] = -\frac12\phi^iI^{(2)}_{ij}\phi^j,
\label{eq:free-action}\end{equation}
where $I^{(2)}_{ij}$ is an invertible, quadratic, differential operator. With this action, the functional integral in the partition function is Gaussian, whence
\begin{align}
    Z_0[J] &= \int\mathscr{D}\phi\,\ee^{-\frac{\ii}{2}\phi^iI^{(2)}_{ij}\phi^j + \ii\phi^iJ_i} \notag\\
    &= Z_0[0]\ee^{\frac{\ii}2J_i\Delta_\text{F}^{ij}J_j}.
\end{align}
The integral is done by Wick rotating to Euclidean space where the integrand is exponentially suppressed, performing the integral, and then Wick rotating back. On the way back to Lorentzian signature, one must ensure that the poles that the inverse of $I^{(2)}_{ij}$ experiences in the complex plane are avoided in a way that is consistent with the vacuum-to-vacuum boundary conditions. This can be done by choosing the Feynman inverse,\footnote{This equality should be understood in terms of a Fourier transform.}
\begin{equation}
    \Delta_\text{F}^{ij} = \frac{1}{-I^{(2)}_{ij}+\ii0}, \qqquad \Delta_\text{F}^{ij}I^{(2)}_{jk} = -\delta^i_k,
\label{eq:genericfeyninverse}\end{equation}
where $\ii0$ is understood as an infinitesimal, positive, imaginary quantity. In an interacting theory the action is given by
\begin{equation}
    I[\phi] = I_0[\phi] + I_\text{int}[\phi],
\end{equation}
where the interaction action admits the series expansion
\begin{equation}
    I_\text{int}[\phi] = \sum_{m=1}^\infty\frac{1}{m!}\frac{\delta I[\phi]}{\delta \phi^{i_1}\cdots\delta\phi^{i_m}}\bigg\vert_{\phi=0}\phi^{i_1}\cdots\phi^{i_m},
\label{eq:actionseries}\end{equation}
and I have excluded constant terms, since they are not dynamical. The partition function is in this case
\begin{align}
    Z[J] &= \int\mathscr{D}\phi\,\ee^{\ii I_\text{int}[\phi]}\ee^{\ii I_0[\phi] + \ii\phi^iJ_i} \\
    &= \ee^{\ii I_\text{int}[-\ii\delta/\delta J]}\int\mathscr{D}\phi\,\ee^{\ii I_0[\phi] + \ii\phi^iJ_i} \\
    &= Z_0[0]\ee^{\ii I_\text{int}[-\ii\delta/\delta J]}\ee^{\frac{\ii}{2}J_i\Delta_\text{F}^{ij}J_j}.
\label{eq:partitionfunctionrewritten}\end{align}
Usually, one is interested not in the Green's functions $G^{i_1\cdots i_m}$, which contain pieces that factorize into disconnected contributions, meaning they do not describe correlations caused by interactions, but in the connected Green's functions $G_\text{(c)}^{i_1\cdots i_m}$. It can be shown that the logarithm of the partition function,
\begin{equation}
    \ii W[J] = \log Z[J],
\end{equation}
is the generating functional for these connected Green's functions in that it admits the series expansion
\begin{equation}
    \ii W[J] = \sum_{m=1}^\infty\frac{\ii^m}{m!}G_\text{(c)}^{i_1\cdots i_m}J_{i_1}\cdots J_{i_m},
\end{equation}
meaning any connected Green's function is given by
\begin{equation}
    G_\text{(c)}^{i_1\cdots i_m} = (-\ii)^m\frac{\delta^m\ii W[J]}{\delta J_{i_1}\cdots\delta J_{i_m}}\bigg\vert_{J=0}.
\end{equation}
In a scattering experiment, the fields must be free at asymptotic past and future infinity,
\begin{subequations}
\begin{alignat}{2}
    \phi^i &\sim \phi^i_\text{in} \qquad&&\text{for}\quad t\to-\infty, \\
    \phi^i &\sim \phi^i_\text{out} \qquad&&\text{for}\quad t\to+\infty,
\end{alignat}\label{eq:asympconds}
\end{subequations}
where the in- and out-fields satisfy the free equation of motion,
\begin{equation}
    -\frac{\delta I_0}{\delta \phi^k}\Bigg\vert_{\phi\to\phi_\text{in/out}} = I^{(2)}_{kj}\phi^j_\text{in/out} = 0.
\end{equation}
Using this, it is possible to write a generating functional for scattering amplitudes as a path integral,
\begin{equation}
    S[\phi_\text{in},\phi_\text{out}] = \int_\eqref{eq:asympconds}\mathscr{D}\phi\,\ee^{\ii I[\phi]},
\end{equation}
where the domain of integration is restricted to those field configurations satisfying eqs. \eqref{eq:asympconds}. According to the theorem of Lehmann--Symanzik--Zimmermann (LSZ), this path integral is related to the partition function by putting external legs on-shell. By defining the amputated Green's function,
\begin{equation}
    G^{i_1\cdots i_n} = G_{j_1\cdots j_n}^\text{amp}\ii\Delta_\text{F}^{j_1i_1}\cdots\ii\Delta_\text{F}^{j_ni_n},
\end{equation}
the partition function from eq. \eqref{eq:partitionfunction} may be written
\begin{equation}
    Z[J] = Z[0]\sum_{m=0}^\infty\frac{\ii^m}{m!}G_{j_1\cdots j_n}^\text{amp}\ii\Delta_\text{F}^{j_1i_1}\cdots\ii\Delta_\text{F}^{j_ni_n}J_{i_1}\cdots J_{i_m}.
\end{equation}
The LSZ theorem can then be stated as
\begin{equation}
    S[\phi_\text{in},\phi_\text{out}] = Z[J]\vert_{\ii\Delta_\text{F}^{ij}J_j\to\phi^i_\text{in/out}}.
\label{eq:smatrixgenerating}\end{equation}
Usually, the free fields admit a mode expansion in eigenstates of the free Hamiltonian, which allows one to retrieve an $S$-matrix element for a process with a given set of in- and out-fields by taking functional derivatives of eq. \eqref{eq:smatrixgenerating} with respect to the creation and annihilation operators, respectively, dividing with any factors remaining from the mode expansion. One must also divide by the vacuum-to-vacuum amplitude. For more details on the functional approach to the $S$-matrix, see \cite{Kim:2023qbl, Bailin:1993smu,Faddeev:1975fct}.

The $S$-matrix elements obtained from the definition in eq. \eqref{eq:smatrixgenerating} contain every piece contributing to a process. Some pieces are not of interest in a scattering problem as they only have support when the momenta in the in-state and out-state are unchanged. The standard way to get rid of these is to define the $T$-matrix as
\begin{equation}
    S = \mathbb{1} + \ii T,
\end{equation}
where the $\mathbb{1}$ now contains the trivial processes. Still, the $T$-matrix contains disconnected pieces, which are unnecessary to compute due the cluster decomposition principle \cite{Weinberg:1995mt}. The disconnected pieces can be discarded simply by replacing $Z[J]$ with $W[J]$ in eq. \eqref{eq:smatrixgenerating}. In practice, $T$-matrix elements can be computed by expanding eq. \eqref{eq:greensretrieve}, with $Z[J]$ given by the expansion of eq. \eqref{eq:partitionfunctionrewritten}, in a series of Feynman diagrams, with edges given by the propagator $\ii\Delta_{\text{F}ij}$ and vertices given by functional derivatives of the interaction action,
\begin{equation}
    \frac{\delta\ii I[\phi]}{\delta \phi^{i_1}\cdots\delta\phi^{i_m}}\Bigg\vert_{\phi=0}.
\end{equation}
In the series, one simply discards disconnected, vacuum and trivial diagrams as one encounters them. The combinatorial factors arising from the different Taylor expansions combine for each diagram into a symmetry factor $S$ that one must divide the diagram by. It is easier to work with the Fourier transformed propagators $\widetilde\Delta_{\text{F}ij}$ and vertices. Schematically then, the Fourier transformed vertices are obtained simply by
\begin{equation}
    \frac{\delta\ii I[\phi]}{\delta \widetilde\phi^{i_1}\cdots\delta\widetilde\phi^{i_m}}\Bigg\vert_{\phi=0},
\label{eq:vertexrule}\end{equation}
with the convention that
\begin{equation}
    \frac{\delta\widetilde\phi^i}{\delta\widetilde\phi^j} = (2\pi)^d\delta_i^j,
\end{equation}
where $d$ is the dimensionality of the space the field is defined on. An easy way to see the usefulness of this convention is the following. Let $F[f]$ be a functional where $f(x)$ is some function of position. First observe that by the chain rule for functional derivatives,
\begin{align}
    \frac{\delta F[f]}{\delta f(-k)} = \int\dd^dx\frac{\delta F[f]}{\delta f(x)}\frac{\delta f(x)}{f(-k)}.
\end{align}
The Jacobian between position and momentum space using the above convention is
\begin{align}
    \frac{\delta f(x)}{\delta f(-k)} &= \int\frac{\dd^dk'}{(2\pi)^d}\ee^{-\ii k'\cdot x}\frac{\delta f(k')}{\delta f(-k)} \\
    &= \ee^{\ii k\cdot x}.
\end{align}
Thus,
\begin{align}
    \frac{\delta F[f]}{\delta f(-k)} &= \int\dd^dx\,\ee^{\ii k\cdot x}\frac{\delta F[f]}{\delta f(x)},
\end{align}
which is simply the Fourier transform. Were it not for the factors of $2\pi$, one would have to multiply an answer obtained in momentum space by a correction factor involving factors of $2\pi$ to obtain the correct Fourier transform of the position-space expression.

\section{Quantization in a general background}\label{sec:quantgeneralback}
The obvious guess for the partition function for general relativity based on the above discussion is
\begin{equation}
    Z[J] = \int\mathscr{D} g\,\ee^{\ii S_\text{EH}[g] + \ii\int\dx g_{\mu\nu}J^{\mu\nu}},
\label{eq:dangerouspathintegral}\end{equation}
where $J^{\mu\nu}(x)$ would be an external source for the metric. This path integral is an active area of research and understanding it lies at the heart of the non-perturbative effects of quantum gravity \cite{Hamber:2009qgr}. Fortunately, it is not necessary to make sense out of eq. \eqref{eq:dangerouspathintegral} if one is satisfied with a perturbative framework, which is the framework relevant to this thesis. A more tractable partition function can be defined by letting the dynamical field variable be a small perturbation to the metric,
\begin{equation}
    g_{\mu\nu} = \bar g_{\mu\nu} + \kappa h_{\mu\nu},
\label{eq:metricdecomp}\end{equation}
and choosing $\bar g_{\mu\nu}$ to satisfy Einstein's equation.\footnote{I remind the reader that $\kappa^2 = 32\pi G_d$ with $G_d$ the $d$-dimensional Newton's constant.} The graviton $h_{\mu\nu}$ is now the fluctuating degree of freedom that will be quantized in the path integral.

Different choices exist for the background field. The popular weak-field expansion is defined by choosing
\begin{equation}
    \bar g_{\mu\nu} = \eta_{\mu\nu}.
\label{eq:choiceflat}\end{equation}
There exists further non-linear extensions of eq. \eqref{eq:choiceflat} which in some cases simplify calculations \cite{Cheung:2020gyp}. I shall consider in addition to this the choice of the Schwarzschild--Tangherlini metric
\begin{equation}
    \bar g_{\mu\nu} = \frac{\Big(1-\frac{\Lambda_d}{4\abs{x_\perp}^{d-3}}\Big)^2}{\Big(1+\frac{\Lambda_d}{4\abs{x_\perp}^{d-3}}\Big)^2}\eta_{\parallel\mu\nu} + \bigg(1 + \frac{\Lambda_d}{4\abs{x_\perp}^{d-3}}\bigg)^\frac{4}{d-3}\eta_{\perp\mu\nu},
\label{eq:choicecurved}\end{equation}
which was derived in chapter \ref{ch:general relativity}. The expansion about this background is readily interpreted as a graviton propagating in a Schwarzschild--Tangherlini background. The weak-field expansion is especially well-suited to calculate amplitudes and thereby observables in the post-Minkowskian expansion. In this thesis, the choice in eq. \eqref{eq:choicecurved} will also be used to calculate amplitudes in the post-Minkowskian expansion. Nevertheless, it has features that might induce hope that going beyond the post-Minkowskian expansion is possible in some cases. 

The decomposition of the metric into a background piece and a perturbation piece in eq. \eqref{eq:metricdecomp} is not unique, which implies there is class of gauge transformations of $h_{\mu\nu}$ under which the expanded theory is invariant. One way to find their form is to simply linearize the transformation law for the metric in eq. \eqref{eq:metrictrans}.\footnote{See \cite{Jakobsen:2020diz} for a complete treatment of this approach.} However, I will utilize an alternative approach \cite{Carroll:2004st}. The idea is to examine carefully how one transitions from the full geometry to the background geometry. Let $M_\text{f}$ and $M_\text{b}$ be manifolds with geometries described by $g_{\mu\nu}$ and $\bar g_{\mu\nu}$, respectively, and let there be a diffeomorphism $f : M_\text{b}\to M_\text{f}$. The decomposition in eq. \eqref{eq:metricdecomp} can then be done properly on $M_\text{b}$ by pulling $g_{\mu\nu}$ back from $M_\text{f}$,
\begin{equation}
    (f^*g)_{\mu\nu} = \bar g_{\mu\nu} + \kappa h_{\mu\nu}.
\label{eq:pullbackmetricdef}\end{equation}
Now, $f$ must be such that the condition
\begin{equation}
    \kappa\abs{h_{\mu\nu}} \ll 1
\label{eq:smallhcond}\end{equation}
is satisfied; there will still be freedom in the choice of $f$ after restricting to these cases. To examine how it looks, I can let $\psi_\epsilon : M_\text{b}\to M_\text{b}$ be a one-parameter family of diffeomorphisms generated by the vector $\xi^\mu$. This induces a family of perturbations defined through the pullback by $\psi_\epsilon$ of $(f^*g)_{\mu\nu}$,
\begin{equation}
    h^{(\epsilon)}_{\mu\nu} = \kappa^{-1}(\psi_\epsilon^*(f^*g))_{\mu\nu} - \kappa^{-1}\bar g_{\mu\nu},
\end{equation}
As $\psi_0$ is the identity map, choosing $\epsilon \ll 1$ ensures that eq. \eqref{eq:smallhcond} holds. Now, eq. \eqref{eq:pullbackmetricdef} and the linearity of the pullback on tensors gives
\begin{equation}
    h^{(\epsilon)}_{\mu\nu} = (\psi_\epsilon^*h)_{\mu\nu} + \kappa^{-1}(\psi_\epsilon^*\bar g)_{\mu\nu} - \kappa^{-1}\bar g_{\mu\nu}.
\end{equation}
By adding and subtracting $h_{\mu\nu}$, the graviton before pulling back, the above expression becomes
\begin{equation}
    h^{(\epsilon)}_{\mu\nu} = h_{\mu\nu} + (\psi_\epsilon^*h)_{\mu\nu} - h_{\mu\nu} + \kappa^{-1}(\psi_\epsilon^*\bar g)_{\mu\nu} - \kappa^{-1}\bar g_{\mu\nu}.
\end{equation}
As the Lie derivative along $\xi$ of a two-form $\omega_{\mu\nu}$ is defined as
\begin{equation}
    \pounds_\xi \omega_{\mu\nu} = \lim_{\epsilon\to0}\frac{(\psi_\epsilon^*\omega)_{\mu\nu} - \omega_{\mu\nu}}{\epsilon},
\end{equation}
the assumption that $\epsilon$ was small implies that
\begin{equation}
    h^{(\epsilon)}_{\mu\nu} = h_{\mu\nu} + \epsilon\pounds_\xi h_{\mu\nu} + \epsilon\kappa^{-1}\pounds_\xi\bar g_{\mu\nu}.
\end{equation}
The Lie derivative of the background metric is $\pounds_\xi\bar g_{\mu\nu} = 2\bar\nabla_{(\mu}\xi_{\nu)}$, so in the end I obtain
\begin{align}
    h^{(\epsilon)}_{\mu\nu} &= h_{\mu\nu} + 2\epsilon\kappa^{-1}\bar\nabla_{(\mu}\xi_{\nu)} + \epsilon\pounds_\xi h_{\mu\nu},
\label{eq:hgaugetrans}\end{align}
where $\bar\nabla_{\mu}$ is the background-covariant derivative. The second term is the gauge part analogous to eq. \eqref{eq:photongaugetrans} while the third comes from the transformation properties of tensors under infinitesimal diffeomorphisms. Given that both $\kappa\abs{h_{\mu\nu}}$ and $\epsilon$ were assumed to be much less than unity, invariance under eq. \eqref{eq:hgaugetrans} to leading order requires invariance under the replacement
\begin{equation}
    h_{\mu\nu} \to h_{\mu\nu} + 2\epsilon\kappa^{-1}\bar\nabla_{(\mu}\xi_{\nu)}.
\end{equation}
If the background metric is taken to be flat, then $\bar\nabla_{\mu} = \partial_\mu$, and the transformation is known as a linear diffeomorphism. The above considerations show in particular that weak-field gravity is invariant under linear diffeomorphisms. This characterizes the gauge freedom of $h_{\mu\nu}$ and displays a conspicuous similarity to the transformation properties of the gauge field $A_\mu$ in eq. \eqref{eq:photongaugetrans}. In momentum space, linear diffeomorphisms become
\begin{equation}
    \varepsilon_{\mu\nu} \to \varepsilon_{\mu\nu} + 2k_{(\mu}\zeta_{\nu)},
\end{equation}
where $k_\mu$ and $\varepsilon_{\mu\nu}$ is the momentum and polarization tensor of the graviton, respectively, and $\zeta_\mu$ is an arbitrary function of $k_\mu$.

\subsection{Einstein--Hilbert action to quadratic order}
To proceed with the quantization, it is necessary to know the gravitational version of eq. \eqref{eq:free-action}, i.e., the expansion of the Einstein--Hilbert action to second order in the graviton. In section \ref{sec:actionformulation}, I computed the first order variation
\begin{equation}
    \delta S_\text{EH} = \frac{2}{\kappa^2}\int\dx\metdens\big(G^{\mu\nu} + \nabla^2\delta g - \nabla_\mu\nabla_\nu g^{\mu\nu}\big).
\end{equation}
The Einstein--Hilbert action admits a series expansion like in eq. \eqref{eq:actionseries}, so the piece quadratic in the graviton is
\begin{equation}
    S_\text{EH}[h]\vert_{h^2} = \frac12\int\dd^dx_1\dd^dx_2\,\frac{\delta^2S_\text{EH}}{\delta h_{\mu_1\nu_1}(x_1)\delta h_{\mu_2\nu_2}(x_2)}\Bigg\vert_{h=0}h_{\mu_1\nu_1}(x_1)h_{\mu_2\nu_2}(x_2).
\end{equation}
I find the simplest way to obtain this is by computing
\begin{align}
    S_\text{EH}[h]\vert_{h^2} &= \frac12\delta^2S_\text{EH}\bigg\vert_{{}^{\delta g\to\kappa h}_{g\to\bar g}} \notag \\
    &= -\frac{1}{\kappa^2}\int\dx\big(\delta^2\metdens R + 2\delta\metdens\delta R + \metdens\delta^2R\big)\bigg\vert_{{}^{\delta g\to\kappa h}_{g\to\bar g}}.
\label{eq:ehsecondorder}\end{align}
The second variation of the Ricci tensor is obtained by varying the first variation from eq. \eqref{eq:ricci-variation},
\begin{equation}
    \delta^2R = -\delta(\delta g^{\mu\nu}R_{\mu\nu}) + \delta(\nabla_\mu\nabla_\nu \delta g^{\mu\nu}) - \delta(\nabla^2\delta g).
\end{equation}
The first term is quite benign. The only additional piece needed is that
\begin{align}
    \delta(\delta g^{\mu\nu}) &= \big(\delta (g^{\mu\rho})g^{\nu\sigma} + g^{\mu\rho}\delta (g^{\nu\sigma})\big)\delta g_{\rho\sigma} \notag\\
    &= -2\delta g^{\mu\lambda}\delta g\indices{_\lambda^\nu}.
\end{align}
Using this, the term evaluates to
\begin{align}
    -\delta(\delta g^{\mu\nu}R_{\mu\nu}) &= 2\delta g^{\mu\sigma}\delta g\indices{_\sigma^\nu}R_{\mu\nu} - \delta g^{\mu\nu}\delta R_{\mu\nu} \notag\\
    &= 2\delta g^{\mu\sigma}\delta g\indices{_\sigma^\nu}R_{\mu\nu} - \delta g^{\mu\nu}\nabla_\sigma\nabla_\mu \delta g\indices{_{\nu}^\sigma} + \frac12\delta g^{\mu\nu}\nabla_\mu\nabla_\nu \delta g + \frac12\delta g^{\mu\nu}\nabla^2\delta g_{\mu\nu}.
\end{align}
The symmetrization on the second term has been dropped because of the contraction with $\delta g^{\mu\nu}$. A measure I shall take before computing the next two terms is determining the variation of the covariant divergence on an arbitrary vector field $V^\mu$. For this purpose, it is useful to recall the identity
\begin{equation}
    \nabla_\mu V^\mu = \frac{1}{\metdens}\partial_\mu\big(\metdens V^\mu\big).
\end{equation}
This can be straightforwardly varied to give
\begin{align}
    \delta(\nabla_\mu V^\mu) &= \delta\bigg(\frac{1}{\metdens}\bigg)\partial_\mu\big(\metdens V^\mu\big) + \frac{1}{\metdens}\partial_\mu\big(\delta(\metdens)V^\mu\big) + \frac{1}{\metdens}\partial_\mu\big(\metdens\delta V^\mu\big) \notag\\
    &= -\frac{1}{2\metdens}g^{\kappa\lambda}\delta g_{\kappa\lambda}\partial_\mu\big(\metdens V^\mu\big) + \frac{1}{2\metdens}\partial_\mu\big(g^{\kappa\lambda}\delta g_{\kappa\lambda}\metdens V^\mu\big) + \nabla_\mu\delta V^\mu \notag\\
    &= \frac12V^\mu\nabla_\mu \delta g + \nabla_\mu\delta V^\mu.
\label{eq:variation-of-vector-field}\end{align}
The last term in this identity is a total covariant derivative, meaning it contributes only a boundary term to the action which can be made to vanish under appropriate boundary conditions on the graviton field. Additionally, I am only interested in the expansion to second order, so the functional form of $\delta V^\mu$ is irrelevant.\footnote{To go to higher order, such terms would need to be kept for the calculation of the next variation, since the variation operator and the covariant total derivative as showcased do not commute.} The third term of the second variation of the Ricci scalar corresponds to the case where $V^\mu = - g^{\mu\nu}\nabla_\nu \delta g$. Inserting this in eq. \eqref{eq:variation-of-vector-field} results in
\begin{equation}
    -\delta(\nabla^2\delta g) = -\frac12\nabla^\mu \delta g\nabla_\mu \delta g - \nabla_\mu\delta(\nabla^\mu \delta g).
\end{equation}
The second term corresponds to the case where $V^\mu = \nabla_\nu \delta g^{\nu\mu}$, and usage of the identity gives
\begin{align}
    \delta(\nabla_\mu\nabla_\nu \delta g^{\mu\nu}) = \frac12\nabla_\nu\delta g^{\mu\nu}\nabla_\mu\delta g + \nabla_\mu\delta(\nabla_\nu\delta g^{\mu\nu}).
\end{align}
The second variation of the Ricci scalar thus becomes
\begin{equation}
\begin{aligned}
    \delta^2R &= 2\delta g^{\mu\sigma}\delta g\indices{_\sigma^\nu}R_{\mu\nu} - \delta g^{\mu\nu}\nabla_\sigma\nabla_\mu\delta g\indices{_{\nu}^\sigma} \\
    &\hspace{.5cm}+ \frac12\delta g^{\mu\nu}\nabla_\mu\nabla_\nu\delta g + \frac12\nabla_\mu\delta g^{\mu\nu}\nabla_\nu\delta g \\
    &\hspace{.5cm}+ \frac12\delta g^{\mu\nu}\nabla^2\delta g_{\mu\nu} -\frac12\nabla^\mu \delta g\nabla_\mu \delta g \\
    &\hspace{.5cm}+ \nabla_\mu\delta(\nabla_\nu\delta g^{\mu\nu} - \nabla^\mu \delta g),
\end{aligned}
\end{equation}
where it is clear that an integration by parts will simplify the second line. As the last line is a total derivative, I will drop it from here on out.

The second variation of the metric density can be obtained by varying eq. \eqref{eq:metdens-variation}, which results in
\begin{align}
    \delta^2\metdens &= \frac12\delta\sqrt{\abs{g}}g^{\mu\nu}\delta g_{\mu\nu} + \frac12\sqrt{\abs{g}}\delta (g^{\mu\nu})h_{\mu\nu} \notag\\
    &= \frac14\sqrt{\abs{g}}(\delta g^2 - 2\delta g^{\mu\nu}\delta g_{\mu\nu}).
\end{align}
The middle term of eq. \eqref{eq:ehsecondorder} is
\begin{equation}
    2\delta\metdens\delta R = -\metdens\big(\delta g\delta g^{\mu\nu}R_{\mu\nu} - \delta g\nabla_\mu\nabla_\nu\delta g^{\mu\nu} + \delta g\nabla^2\delta g\big)
\end{equation}
Now, the pieces can be assembled by taking $\delta g_{\mu\nu} = \kappa h_{\mu\nu}$ and evaluating the background quantities on $\bar g_{\mu\nu}$. There is a Ricci scalar term,
\begin{equation}
    S_\text{EH}[h]\vert_{h^2} \supset \frac14\int\dx\bmetdens(2h^{\mu\nu}h_{\mu\nu} - h^2)\bar R,
\end{equation}
and a Ricci tensor term,
\begin{equation}
    S_\text{EH}[h]\vert_{h^2} \supset \int\dx\bmetdens(hh^{\mu\nu} - 2h^{\mu\sigma}h\indices{_\sigma^\nu})\bar R_{\mu\nu}.
\end{equation}
Then, there is a large collection of covariant derivative terms,
\begin{equation}
\begin{aligned}
    S_\text{EH}[h]\vert_{h^2} \supset \int\dx\bmetdens\Big[&h^{\mu\nu}\bar\nabla_\sigma\bar\nabla_\mu h\indices{_{\nu}^\sigma} - \frac12h^{\mu\nu}\bar\nabla^2h_{\mu\nu} \\
    &+\frac12\bar\nabla^\mu h\bar\nabla_\mu h + h\bar\nabla^2h \\
    &- \frac12h^{\mu\nu}\bar\nabla_\mu\bar\nabla_\nu h - \frac12\bar\nabla_\mu h^{\mu\nu}\bar\nabla_\nu h - h\bar\nabla_\mu\bar\nabla_\nu h^{\mu\nu}\Big].
\end{aligned}
\end{equation}
The second and third line simplify to a total of two terms under an integration by parts. Carrying this out and collecting everything results in
\begin{equation}
\begin{aligned}
    S_\text{EH}[h]\vert_{h^2} &= \int\dx\bmetdens \\
    &\hspace{.8cm}\times\Big[\frac14(2h^{\mu\nu}h_{\mu\nu} - h^2)\bar R + (hh^{\mu\nu} - 2h^{\mu\sigma}h\indices{_\sigma^\nu})\bar R_{\mu\nu} -\bar\nabla_\sigma h^{\mu\nu}\bar\nabla_\mu h\indices{_\nu^\sigma} \\
    &\hspace{1.4cm} + \frac12\bar\nabla_\sigma h^{\mu\nu}\bar\nabla^\sigma h_{\mu\nu} - \frac12\bar\nabla^\mu h\bar\nabla_\mu h + \bar\nabla_\mu h \bar\nabla_\nu h^{\mu\nu}\Big].
\end{aligned}
\label{eq:ehquadraticnotgf}\end{equation}

\subsection{Gauge-fixed partition function}
At this point, it may seem that the theory of a graviton $h_{\mu\nu}$ living on background $\bar g_{\mu\nu}$ could readily be quantized---however, as discussed above the expanded theory inherits from the diffeomorphism invariance of the full theory an invariance under the transformation in eq. \eqref{eq:hgaugetrans}. This presents a problem as gauge theory path integrals are generically divergent. The intuitive reason for this is that the path integral sums over all field configurations, even those describing the same physical situation but being related by a gauge transformation. This means the classical equations of motion for a given source are degenerate such that the Green's function for the field is not well-defined. It is thus not possible to perform the Gaussian path integral of the free theory. To remedy this, one must divide by the number of times one overcounts in the integral, which is given by the (infinite) volume of the gauge group encompassing the transformations in eq. \eqref{eq:hgaugetrans}. Schematically then, the proper definition of the partition function should include a division by this factor
\begin{equation}
    Z[J] = \int\frac{\mathscr{D} h}{\operatorname{vol}\text{gauge}}\ee^{\ii S_\text{EH}[h_{\mu\nu}] + \ii\int\dd^dx\,h_{\mu\nu}J^{\mu\nu}}.
\label{eq:partitionfunctionbeforefp}\end{equation}
A practical way to proceed from here is given by the Faddeev--Popov procedure, which provides a clever way of inserting unity into eq. \eqref{eq:partitionfunctionbeforefp} such that one may factor out of the path integral an integral over gauge transformations. This cancels the volume in the denominator, leaving a finite result which will depend upon a Faddeev-Popov determinant. Effectively, one just ends up with a path integral restricted to a single gauge orbit \cite{Faddeev:1967fc}. This is known as fixing the gauge. The most popular and convenient choice is de Donder gauge, specified by the condition
\begin{equation}
    G_\mu = \bar\nabla_\nu h\indices{_\mu^\nu} - \frac12\bar\nabla_\mu h = 0,
\label{eq:dedondergauge}\end{equation}
which is the harmonic gauge condition,
\begin{equation}
    g^{\rho\sigma}\Gamma_{\mu\rho\sigma} = 0
\label{eq:harmonicgauge}\end{equation}
expanded around the background metric. Harmonic gauge is widely used in the treatment of gravitational waves in classical general relativity. In a coordinate system where eq. \eqref{eq:harmonicgauge} is satisfied, one can show that all the coordinate functions satisfy d'Alembert's equation. One can interpret this as the coordinate system having ``wavy'' Cartesian coordinate axes.

To apply the Faddeev--Popov procedure as discussed in \cite{Bailin:1993smu}, some definitions must be made. The gauge transformations for the graviton I derived in eq. \eqref{eq:hgaugetrans} may by absorbing $\epsilon$ in $\xi$ be written
\begin{equation}
    h^{(\xi)}_{\mu\nu} = h_{\mu\nu} + 2\kappa^{-1}\bar\nabla_{(\mu}\xi_{\nu)} + \pounds_\xi h_{\mu\nu}.
\end{equation}
Using this I define
\begin{equation}
    G_\mu^{(\xi)} = \bar\nabla_\nu h_\mu^{(\xi)\nu} - \frac12\bar\nabla_\mu h^{(\xi)}
\end{equation}
Then, by the Faddeev-Popov method, the gauge-fixed partition function can be written
\begin{equation}
    Z[J] = \int\mathscr{D} h\,\Delta[h_{\mu\nu}]\delta[G_\mu - f_\mu]\ee^{\ii S_\text{EH}[h] + \ii\int\dx h_{\mu\nu}J^{\mu\nu}},
\label{eq:partitionfunctionafterfp}\end{equation}
where
\begin{equation}
    \Delta[h_{\mu\nu}] = \det\bigg[\kappa\frac{\delta G_\mu^{(\xi)}}{\delta\xi^\nu}\bigg]
\end{equation}
is the Faddeev--Popov determinant, and $\delta[G_\mu - f_\mu]$ is a functional delta function. Because the step from eq. \eqref{eq:partitionfunctionbeforefp} to eq. \eqref{eq:partitionfunctionafterfp} involved inserting unity, the partition function is independent of the choice of $f_\mu$. Specifically, the partition function can be integrated over $f_\mu$ with a Gaussian kernel. This will change the normalization of the partition function but will not impact correlation functions. Doing this, the functional delta function becomes
\begin{equation}
    \int\mathscr{D} h\,\ee^{\ii\int\dx\bmetdens f_\mu f^\mu}\delta[G_\mu - f_\mu] = \ee^{\ii S_\text{gf}[h]}.
\end{equation}
where
\begin{equation}
    S_\text{gf}[h] = \int\dx\bmetdens G_\mu G^\mu.
\end{equation}
The determinant may be represented as a path integral over complex Grassmann-valued vector ghost fields $\bar b^\mu$ and $b^\mu$,
\begin{equation}
    \Delta[h_{\mu\nu}] = \int\mathscr{D} b\mathscr{D}\bar b\,\ee^{\ii S_\text{gh}[\bar b, b, h]},
\end{equation}
where $S_\text{gh}[\bar b, b, h]$ is the ghost action,
\begin{equation}
    S_\text{gh}[\bar b, b, h] = -\int\dx\bmetdens\kappa\bar b^\mu\frac{\delta G_\mu^{(\xi)}}{\delta\xi^\nu}b^\nu.
\end{equation}
The differential operator comes out to be
\begin{align}
    \kappa\frac{\delta G_\mu^{(\xi)}}{\delta\xi^\nu} &= (\bar g_{\mu\nu}\bar\nabla^2 + \bar R_{\mu\nu}) + \kappa\frac{\delta}{\delta\xi^\nu}\Big[\bar\nabla^{\rho}\pounds_\xi h_{\mu\rho} - \frac12\bar\nabla_\mu\bar g^{\rho\sigma}\pounds_\xi h_{\rho\sigma}\Big].
\label{eq:ghostlagrangian}\end{align}
The first term gives rise to the background-covariant ghost propagator, while the second gives rise to a background-covariant cubic interaction between ghost, anti-ghost and graviton.

After the dust has settled, I arrive at the gauge-fixed partition function
\begin{equation}
    Z[J] = \int\mathscr{D} h\,\ee^{\ii S_\text{G}[h] + \ii S_\text{gh}[\bar b, b, h]},
\end{equation}
where
\begin{equation}
    S_\text{G}[h] = S_\text{EH}[h] + S_\text{gf}[h]
\label{eq:gravitationalaction}\end{equation}
is the gauge-fixed Einstein--Hilbert action. As is evident from their Lagrangian, the ghosts will only every appear in graviton loops, which, as will be discussed below, do not contribute to the classical limit. As a matter of fact, ghosts play no role in the modern unitarity-based approach to quantum field theory \cite{Badger:2023eqz}. In light of this, it may seem superfluous to include such a detailed discussion as the one given above. However, as I was unable to find a clear and consistent treatment of these aspects in the existing literature, I have chosen to include it here.

\section{Feynman rules in the weak-field expansion}\label{sec:feynman rules in the weak-field expansion}
I will now derive the Feynman rules for weak-field gravity, where $\bar g_{\mu\nu} = \eta_{\mu\nu}$. In this limit, the Ricci scalar and Ricci tensor vanish and covariant derivatives become partial derivatives, so eq. \eqref{eq:ehquadraticnotgf} becomes
\begin{equation}
    S_\text{EH}[h]\vert_{h^2} = \int\dx\Big[\frac12\partial_\sigma h^{\mu\nu}\partial^\sigma h_{\mu\nu} - \frac12\partial_\mu h\partial^\mu h + \partial_\mu h\partial_\nu h^{\mu\nu} - \partial_\sigma h^{\mu\nu}\partial_\mu h\indices{_\nu^\sigma}\Big].
\label{eq:fierzpauli}\end{equation}
This is known as the (massless) Fierz--Pauli action. As mentioned in the introduction of this chapter, Feynman proved that general relativity is the unique self-consistent non-linear, extension of this action. Indices are now raised and lowered with the Minkowski metric. Once the gauge-fixing action is added to eq. \eqref{eq:fierzpauli}, one obtains
\begin{equation}
    S_\text{G}\vert_{h^2} = -\frac12\int\dx h_{\mu\nu}\Big[I^{\mu\nu\,\rho\sigma} - \frac12\eta^{\mu\nu}\eta^{\rho\sigma}\Big]\partial^2h_{\rho\sigma}
\label{eq:quadraticactionflatgf}\end{equation}
after integrating by parts. I introduced the identity tensor in the space of symmetric pairs of indices
\begin{equation}
    I^{\mu\nu\,\rho\sigma} = \eta^{\mu(\rho}\eta^{\sigma)\nu}, \qqquad I\indices{^{\mu\nu}_{\rho\sigma}}T^{\rho\sigma} = T^{\mu\nu},
\end{equation}
where $T^{\mu\nu}$ is any symmetric tensor. The action is now in the form of eq. \eqref{eq:free-action}, so the propagator is found by solving eq. \eqref{eq:genericfeyninverse}. 
By introducing the Fourier transformed graviton
\begin{equation}
    h_{\mu\nu}(x) = \int_k\ee^{-\ii k\cdot x}h_{\mu\nu}(k), \qqquad \int_k = \int\frac{\dd^dk}{(2\pi)^d},
\label{eq:gravitonfourier}\end{equation}
the equation to be solved can be written
\begin{equation}
    -k^2\Big[I^{\mu\nu\,\kappa\lambda} - \frac12\eta^{\mu\nu}\eta^{\kappa\lambda}\Big]\Delta_{\text{F}\kappa\lambda\,\rho\sigma}(k) = -I\indices{^{\mu\nu}_{\rho\sigma}}
\end{equation}
It is not hard to see that
\begin{equation}
    \Delta_{\text{F}\mu\nu\,\rho\sigma} = \frac{P_{\mu\nu\,\rho\sigma}}{k^2+\ii0}, \qqquad P_{\mu\nu\,\rho\sigma} = I_{\mu\nu\,\rho\sigma} - \frac{1}{d-2}\eta_{\mu\nu}\eta_{\rho\sigma}
\end{equation}
solves this equation. This propagator is known as the de Donder propagator and $P_{\mu\nu\,\rho\sigma}$ as the de Donder projector. In diagrams, it acquires a factor of $\ii$,
\begin{equation}
    \begin{tikzpicture}[baseline={(current bounding box.center)}]
        \coordinate (x) at (-.5,0);
        \coordinate (y) at (1.5,0);
        \draw [photon] (x) -- (y) node [midway, below] {$k$};
        \draw [fill] (x) node [above] {$h_{\mu\nu}(k)$};
        \draw [fill] (y) node [above] {$h_{\rho\sigma}(k)$};
    \end{tikzpicture} = \frac{\ii P_{\mu\nu\,\rho\sigma}}{k^2 + \ii0}.
\label{eq:gravitonpropagator}\end{equation}
The complete, expanded Einstein--Hilbert action takes the form of an infinite series
\begin{equation}
    S_\text{EH} = S_\text{EH}[h]\vert_{h^2} + \sum_{i=1}^\infty\big[\kappa^m\partial^2h^{m+2}\big].
\label{eq:weakfieldschematiceh}\end{equation}
Following eq. \eqref{eq:vertexrule}, a momentum space $m$-point graviton vertex is defined as
\begin{equation}
    \ii\kappa^{m-2}\hat\delta^{(d)}(k_{1\cdots m})V^{\mu_1\nu_1\cdots\mu_m\nu_m}(k_1,\ldots,k_m) = \frac{\delta^m\ii S_\text{G}}{\delta h_{\mu_1\nu_1}(-k_1)\cdots\delta h_{\mu_m\nu_m}(-k_m)}\Bigg\vert_{h=0},
\label{eq:gravitonvertices}\end{equation}
where I defined the sum of vertex momenta
\begin{equation}
    k_{1\cdots m} = \sum_{i=1}^mk_i,
\end{equation}
and the $d$-dimensional hatted delta function
\begin{equation}
    \hat\delta^{(d)}(k) = (2\pi)^d\delta^{(d)}(k).
\end{equation}
It is useful to factor out the imaginary unit, the gravitational coupling constant, and the momen\-tum-conserving delta function, such that one may more easily keep track of these in computations. The vertices derived from eq. \eqref{eq:gravitonvertices} have \emph{outgoing} momenta being positive. To see this, recall that in the mode expansion of a quantum field, the annihilation operator is associated with $\ee^{-\ii k\cdot x}$ while the creation operator is associated with the conjugate $\ee^{\ii k\cdot x}$. The Fourier convention in eq. \eqref{eq:gravitonfourier} in combination with the sign of the momenta on the right-hand side of eq. \eqref{eq:gravitonvertices} then implies the statement. The vertices are symmetric under interchanges of pairs of indices $\mu_i\nu_i \leftrightarrow \mu_j\nu_j$ due to the Bose statistics of the graviton field, as well as under interchanges within index pairs $\mu_i \leftrightarrow \nu_i$ due to the symmetry of the graviton.

The cubic vertex is represented diagrammatically by
\begin{equation}
    \begin{tikzpicture}[baseline={(current bounding box.center)}]
        \coordinate (out1) at (1,0);
        \coordinate (out2) at (-.5,.866);
        \coordinate (out3) at (-.5,-.866);
        \coordinate (x) at (0,0);
        \draw [photon] (x) -- (out1) node [right] {$h_{\mu_3\nu_3}(k_3)$};
        \draw [photon] (x) -- (out2) node [above left] {$h_{\mu_2\nu_2}(k_2)$};
        \draw [photon] (x) -- (out3) node [below left] {$h_{\mu_1\nu_1}(k_1)$};
        \draw [fill] (x) circle (.04);
    \end{tikzpicture} = \ii\kappa\hat\delta^{(d)}(k_{123})V^{\mu_1\nu_1\,\mu_2\nu_2\,\mu_3\nu_3}(k_1, k_2, k_3).
\end{equation}
It is extremely noteworthy that this vertex, when contracted with physical polarization tensors, reduces to the square of the cubic vertex from quantum chromodynamics. This is a manifestation of the by now famous statement that gravity in some sense is the square of gauge theory \cite{Bern:2019prr}. This has powerful theoretical and computational ramifications, which shall not be explored in this thesis. I shall also need the quartic vertex, which diagrammatically is
\begin{equation}
    \begin{tikzpicture}[baseline={(current bounding box.center)}]
        \coordinate (out1) at (-.71,-.71);
        \coordinate (out2) at (-.71,.71);
        \coordinate (out3) at (.71,.71);
        \coordinate (out4) at (.71,-.71);
        \coordinate (x) at (0,0);
        \draw [photon] (x) -- (out1) node [below left] {$h_{\mu_1\nu_1}(k_1)$};
        \draw [photon] (x) -- (out2) node [above left] {$h_{\mu_2\nu_2}(k_2)$};
        \draw [photon] (x) -- (out3) node [above right] {$h_{\mu_3\nu_3}(k_3)$};
        \draw [photon] (x) -- (out4) node [below right] {$h_{\mu_4\nu_4}(k_4)$};
        \draw [fill] (x) circle (.04);
    \end{tikzpicture} = \ii\kappa^2\hat\delta^{(d)}(k_{1234})V^{\mu_1\nu_1\,\mu_2\nu_2\,\mu_3\nu_3\,\mu_4\nu_4}(k_1, k_2, k_3, k_4).
\end{equation}
The fully expanded and maximally simplified form of the cubic vertex consists of a minimum of 171 terms, due the symmetries the vertex needs to respect. For the quartic vertex, this number is 2850 \cite{DeWitt:1967uc}. The actual number encountered when deriving them may be higher due to unrealized simplifications from momentum conservation. In the past, these vertices were painstakingly computed by expanding the Einstein--Hilbert action to the required order by hand.\footnote{Or by other more refined, but still laborious, methods. See \cite{DeWitt:1967ub}.} These heroic efforts by early workers required a significant amount of algebraic labor and attention to detail, but was of immense importance to the understanding of perturbative quantum gravity. The cubic and quartic vertices were first computed by DeWitt \cite{DeWitt:1967uc} with a correction being offered subsequently by Berends and Gastmans \cite{Berends:1975hg}. Further corrections to the quartic vertex were proposed by Sannan \cite{Sannan:1986gr}. It is possible to work with non-linear gauges where the vertices simplify. See for instance \cite{Kalin:2020mvi, Driesse:2024feo}.

With the advent of modern computational tools, it is fortunately no longer necessary to calculate these vertices by hand. I have made use of the \texttt{Mathematica} packages \texttt{xTensor} and \texttt{xPert} to carry out the expansions and to manipulate expressions. They are part of the wonderful computer tensor algebra bundle \texttt{xAct} \cite{xAct, Brizuela:2008ra}.

From eq. \eqref{eq:weakfieldschematiceh}, it is observed that the non-linearity of the Einstein--Hilbert action leads to an infinite set of interaction vertices, quadratic in the graviton momentum. This poses no serious problem for the perturbation theory besides making it cumbersome and unwieldy. They take the schematic form
\begin{equation}
    \begin{tikzpicture}[baseline={(current bounding box.center)}]
        \coordinate (out1) at (1,0);
        \coordinate (out2) at (.309,.951);
        \coordinate (out3) at (-.809,.588);
        \coordinate (out4) at (-.809,-.588);
        \coordinate (out5) at (.309,-.951);
        \coordinate (x) at (0,0);
        \draw [photon] (x) -- (out1);
        \draw [photon] (x) -- (out2);
        \draw [photon] (x) -- (out3);
        \draw [photon] (x) -- (out4);
        \draw [photon] (x) -- (out5);
        \draw [fill] (x) circle (.04);
    \end{tikzpicture} \sim \ii\kappa^3k^2, \qquad \begin{tikzpicture}[baseline={(current bounding box.center)}]
        \coordinate (out1) at (1,0);
        \coordinate (out2) at (.5,.866);
        \coordinate (out3) at (-.5,.866);
        \coordinate (out4) at (-1,0);
        \coordinate (out5) at (-.5,-.866);
        \coordinate (out6) at (.5,-.866);
        \coordinate (x) at (0,0);
        \draw [photon] (x) -- (out1);
        \draw [photon] (x) -- (out2);
        \draw [photon] (x) -- (out3);
        \draw [photon] (x) -- (out4);
        \draw [photon] (x) -- (out5);
        \draw [photon] (x) -- (out6);
        \draw [fill] (x) circle (.04);
    \end{tikzpicture} \sim \ii\kappa^4k^2, \qquad \cdot\,\cdot\,\cdot\,.
\end{equation}

\section{Feynman rules in an arbitrary background}\label{sec:feynmanrulesgeneralback}
Having dealt with the weak-field limit, I can now tackle the arbitrary background case. When the background is kept arbitrary, one must deal directly with the action in eq. \eqref{eq:ehquadraticnotgf}. This means the gauge-fixing condition in eq. \eqref{eq:dedondergauge} must be used, implying one must take as the gauge-fixing action
\begin{equation}
    S_\text{gf}[h] = \int\dx\bmetdens\Big[\bar\nabla_\sigma h\indices{_\nu^\sigma}\bar\nabla_\mu h^{\mu\nu} - \bar\nabla_\mu h\bar\nabla_\nu h^{\mu\nu} + \frac14\bar\nabla_\mu h\bar\nabla^\mu h\Big].
\label{eq:covariantgaugeaction}\end{equation}
Indices are now raised and lowered with the background metric $\bar g_{\mu\nu}$. When added to eq. \eqref{eq:ehquadraticnotgf}, some covariant derivative terms combine nicely. The first term of eq. \eqref{eq:covariantgaugeaction} and the third term of the second line of eq. \eqref{eq:ehquadraticnotgf} can be integrated by parts to give
\begin{align}
     h^{\mu\nu}\bar\nabla_\sigma\bar\nabla_\mu h\indices{_\nu^\sigma} - h^{\mu\nu}\bar\nabla_\mu\bar\nabla_\sigma h\indices{_\nu^\sigma} &= h^{\mu\nu}[\bar\nabla_\sigma,\bar\nabla_\mu]h\indices{_\nu^\sigma} \notag\\
     &= \bar R_{\mu\nu}h^{\nu\sigma}h\indices{_\sigma^\mu} - \bar R_{\mu\rho\nu\sigma}h^{\mu\nu}h^{\rho\sigma}.
\end{align}
In addition, the second term of eq. \eqref{eq:covariantgaugeaction} and the last term of eq. \eqref{eq:ehquadraticnotgf}, leaving the last term of eq. \eqref{eq:covariantgaugeaction} to combine with the penultimate term in eq. \eqref{eq:ehquadraticnotgf}. This yields the action
\begin{equation}
\begin{aligned}
    S_\text{G}\big\vert_{h^2} = \int\dx\bmetdens\Big[&\frac14(2h^{\mu\nu}h_{\mu\nu} - h^2)\bar R  - \bar R_{\mu\rho\nu\sigma}h^{\mu\nu}h^{\rho\sigma} \\
    &+(hh^{\mu\nu} - h^{\mu\sigma}h\indices{_\sigma^\nu})\bar R_{\mu\nu} + \frac12\bar\nabla_\sigma h_{\mu\nu}\bar\nabla^\sigma h^{\mu\nu} - \frac14\bar\nabla_\mu h\bar\nabla^\mu h\Big].
\end{aligned}
\label{eq:ehactionquadraticgaugefixed}\end{equation}
This result can also be found in \cite{Donoghue:2017pgk,Kosmopoulos:2023bwc,Cheung:2024byb}. One approach from here would be to integrate eq. \eqref{eq:ehactionquadraticgaugefixed} by parts and try to invert the differential operator appearing. This is a, if not impossible, then very difficult problem, even when the background metric is chosen to be the Schwarzschild--Tangherlini one. Another approach is to separate it into a free piece and an interacting piece by adding and subtracting eq. \eqref{eq:quadraticactionflatgf}. This yields the background-independent free action
\begin{equation}
    S_\text{G}^\text{free}[h] = -\frac12\int\dx h_{\mu\nu}\Big[I^{\mu\nu\,\rho\sigma} - \frac12\eta^{\mu\nu}\eta^{\rho\sigma}\Big]\partial^2h_{\rho\sigma},
\label{eq:freeaction}\end{equation}
which gives rise to the same de Donder propagator from eq. \eqref{eq:gravitonpropagator}, and the background-dependent interaction action
\begin{equation}
\begin{aligned}
    S_\text{G}^\text{int}[h] = \int\dx\bmetdens\Big[&\frac14(2h^{\mu\nu}h_{\mu\nu} - h^2)\bar R  - \bar R_{\mu\rho\nu\sigma}h^{\mu\nu}h^{\rho\sigma} \\
    &+(hh^{\mu\nu} - h^{\mu\sigma}h\indices{_\sigma^\nu})\bar R_{\mu\nu} + \frac12\bar\nabla_\sigma h_{\mu\nu}\bar\nabla^\sigma h^{\mu\nu} - \frac14\bar\nabla_\mu h\bar\nabla^\mu h\Big]\textbf{} \\
    &+\frac14\int\dx(2\eta^{\kappa\lambda}I^{\mu\nu\,\rho\sigma} - \eta^{\kappa\lambda}\eta^{\mu\nu}\eta^{\rho\sigma})\partial_\kappa h_{\mu\nu}\partial_\lambda h_{\rho\sigma}.
\end{aligned}
\label{eq:interactionaction}\end{equation}
I integrated the free action by parts in the above and rewrote it slightly. I also exposed the metric factors in the free action for clarity. Notice the similarity between the covariant derivative terms and the free terms. This manifests the fact that the role of the free action in the above is to ensure that $S_\text{G}^\text{int}[h]\vert_{\kappa^0} = 0$.

To develop the perturbation theory, it is nice to write the interaction action in a form which exhibits explicitly what terms contain zero, one, and two partial derivatives of the graviton. This is achieved by writing
\begin{align}
    S_\text{G}^\text{int}[h] = \int\mathrm{d}^dx\,\Big[\bar{\mathfrak{C}}_{[\partial^0]}^{\mu_1\nu_1\,\mu_2\nu_2\,\gamma\delta}\partial_\gamma h_{\mu_1\nu_1}\partial_\delta h_{\mu_2\nu_2} + \bar{\mathfrak{C}}_{[\partial^1]}^{\mu_1\nu_1\,\mu_2\nu_2\,\delta}h_{\mu_1\nu_1}\partial_\delta h_{\mu_2\nu_2} + \bar{\mathfrak{C}}_{[\partial^2]}^{\mu_1\nu_1\,\mu_2\nu_2}h_{\mu_1\nu_1}h_{\mu_2\nu_2}\Big].
\label{eq:quad-act-int-short}\end{align}
By introducing the tensor
\begin{equation}
    T\indices*{^{\mu_1\nu_1\,\mu_2\nu_2}_{\rho_1\sigma_1\,\rho_2\sigma_2}} = I\indices*{^{\mu_1(\mu_2}_{\rho_1\sigma_1}}I\indices*{^{\nu_2)\nu_1}_{\rho_2\sigma_2}} - \frac12I\indices*{^{\mu_1\nu_1}_{\rho_1\sigma_1}}I\indices*{^{\mu_2\nu_2}_{\rho_2\sigma_2}},
\end{equation}
one can by a tedious computation show that
\begin{subequations}
\begin{align}
    \bar{\mathfrak{C}}_{[\partial^0]}^{\mu_1\nu_1\,\mu_2\nu_2\,\gamma\delta} &= \frac18(\bmetdens\bar g^{\gamma\delta}\bar g^{\rho_1\sigma_1}\bar g^{\rho_2\sigma_2} - \eta^{\gamma\delta}\eta^{\rho_1\sigma_1}\eta^{\rho_2\sigma_2})T\indices*{^{\mu_1\nu_1\,\mu_2\nu_2}_{\rho_1\sigma_1\,\rho_2\sigma_2}}, \\
    \bar{\mathfrak{C}}_{[\partial^1]}^{\mu_1\nu_1\,\mu_2\nu_2\,\delta} &= - \frac12\bmetdens\bar g^{\rho_2\sigma_2}T\indices*{^{\mu_2\nu_2\,\delta(\nu_1}_{\rho_1\sigma_1\,\rho_2\sigma_2}}\bar\Gamma^{\mu_1)\rho_1\sigma_1}, \\
    \bar{\mathfrak{C}}_{[\partial^2]}^{\mu_1\nu_1\,\mu_2\nu_2} &= \operatorname{sym} P_2\frac18\bmetdens\Big(4\bar g^{\rho_1\sigma_1}\bar g^{\rho_2\sigma_2}\bar\Gamma\indices{^{\mu_1}_{\gamma\lambda}}\bar\Gamma\indices{^{\mu_2}_{\delta}^{\lambda}}T\indices*{^{\nu_1\gamma\,\nu_2\delta}_{\rho_1\sigma_1\,\rho_2\sigma_2}} + \bar R\bar g^{\rho_1\sigma_1}\bar g^{\rho_2\sigma_2}T\indices*{^{\mu_1\nu_1\,\mu_2\nu_2}_{\rho_1\sigma_1\,\rho_2\sigma_2}} \notag\\
    &\hspace{4.5em}+ 16\bar g^{\mu_1[\nu_1}\bar R^{\mu_2]\nu_2} - 8\bar R^{\mu_1\mu_2\nu_1\nu_2}\Big),
\end{align}
\end{subequations}
where $P_2$ denotes symmetrization in the index pairs $\mu_1\nu_1$ and $\mu_2\nu_2$ and sym denotes symmetrization between $\mu_i$ and $\nu_i$.

Fourier transforming the gravitons in eq. \eqref{eq:quad-act-int-short}, I obtain
\begin{equation}
\begin{aligned}
    S_\text{G}^\text{int}[h] &= \int_{k_1,k_2}h_{\mu_1\nu_1}(-k_1)h_{\mu_2\nu_2}(-k_2)\int\mathrm{d}^dx\,\ee^{\ii(k_1+k_2)\cdot x} \\
    &\times\bigg(-\bar{\mathfrak{C}}_{[\partial^0]}^{\mu_1\nu_1\,\mu_2\nu_2\,\gamma\delta}k_{1\gamma}k_{2\delta} + \frac{\ii}{2}\Big[\bar{\mathfrak{C}}_{[\partial^1]}^{\mu_1\nu_1\,\mu_2\nu_2\,\delta}k_{1\delta} + \bar{\mathfrak{C}}_{[\partial^1]}^{\mu_2\nu_2\,\mu_1\nu_1\,\delta}k_{2\delta}\Big] + \bar{\mathfrak{C}}_{[\partial^2]}^{\mu_1\nu_1\,\mu_2\nu_2}\bigg).
\end{aligned}
\end{equation}
This implies that the two-point interaction vertex in a generic background is given by
\begin{equation}
\begin{aligned}
    &\bar V^{\mu_1\nu_1\,\mu_2\nu_2}(k_1, k_2) = \int\mathrm{d}^dx\,\ee^{\ii q\cdot x} \\
    &\hspace{2em}\times\big(-2\ii\bar{\mathfrak{C}}_{[\partial^0]}^{\mu_1\nu_1\,\mu_2\nu_2\,\gamma\delta}k_{1\gamma}k_{2\delta} + \bar{\mathfrak{C}}_{[\partial^1]}^{\mu_1\nu_1\,\mu_2\nu_2\,\delta}k_{1\delta} + \bar{\mathfrak{C}}_{[\partial^1]}^{\mu_2\nu_2\,\mu_1\nu_1\,\delta}k_{2\delta} + 2\ii\bar{\mathfrak{C}}_{[\partial^2]}^{\mu_1\nu_1\,\mu_2\nu_2}\big).
\end{aligned}
\label{eq:curvedgraviton2ptvertex}\end{equation}

\subsection{Specifying to the Schwarzschild--Tangherlini solution}
The stage is now set for the development of the Feynman rules when the background metric is the Schwarzschild--Tangherlini one
\begin{equation}
    \bar g_{\mu\nu} = \frac{\Big(1-\frac{\Lambda_d}{4\abs{x_\perp}^{d-3}}\Big)^2}{\Big(1+\frac{\Lambda_d}{4\abs{x_\perp}^{d-3}}\Big)^2}\eta_{\parallel\mu\nu} + \bigg(1 + \frac{\Lambda_d}{4\abs{x_\perp}^{d-3}}\bigg)^\frac{4}{d-3}\eta_{\perp\mu\nu}.
\label{eq:choicecurved}\end{equation}
As demonstrated in section \ref{sec:schwarzschild--tangherlini}, the Schwarzschild--Tangherlini metric is sourced by a point particle. It is thus quite natural to diagrammatically depict the vertex as,
\begin{equation}
    \begin{tikzpicture}[baseline={(current bounding box.center)}]
        \coordinate (in) at (-1,0);
        \coordinate (out) at (1,0);
        \coordinate (x) at (0,0);
        \coordinate (gin) at (-1,-1.2);
        \coordinate (gout) at (1,-1.2);
        \coordinate (v) at (0,-1.2);
        \draw [dotted, thick] (in) -- (x);
        \draw [dotted, thick] (x) -- (out);
        \draw [photon2] (x) -- (v) node [midway, left] {$q\!\downarrow$};
        \draw [photon] (v) -- (gin) node [left, below=.5em] {$h_{\mu_1\nu_1}(k_1)$};
        \draw [photon] (v) -- (gout) node [right, below=.5em] {$h_{\mu_2\nu_2}(k_2)$};
        \draw [fill] (v) circle (.08);
        \draw [fill] (x) circle (.08);
    \end{tikzpicture} = \bar V^{\mu_1\nu_1\,\mu_2\nu_2}(k_1, k_2),
\label{eq:exacttwopointvertex}\end{equation}
where the faint dotted line represents the worldline of the point particle. I emphasize that the vertex in eq. \eqref{eq:exacttwopointvertex} is a two-point vertex from the point of view of Feynman rules, and the fat graviton line and the worldline is drawn only to indicate the origin of the interaction from the position-dependent background metric.

Recall that the goal is to develop these rules in a post-Minkowskian expansion, such that it may be verified that the Compton amplitude obtained with them is the same as the one obtained with the flat space rules discussed in section \ref{sec:feynman rules in the weak-field expansion}. This is a significant simplification, as it turns the difficult Fourier transform into an easily doable one. It also allows effectively working in flat space, meaning that indices at the expanded level can be raised and lowered with the Minkowski metric. The direct evaluation of the vertex in eq. \eqref{eq:curvedgraviton2ptvertex} is achieved by inserting the expanded metric
\begin{equation}
    \bar g_{\mu\nu}(x) = \eta_{\mu\nu} - \frac{\Lambda_d}{\abs{x_\perp}^{d-3}}\bigg(\eta_{\parallel\mu\nu} - \frac{\eta_{\perp\mu\nu}}{d-3}\bigg) + \frac{\Lambda_d^2}{\abs{x_\perp}^{2(d-3)}}\bigg(\frac{\eta_{\parallel\mu\nu}}{2} - \frac{(d-7)\eta_{\perp\mu\nu}}{8(d-3)^2}\bigg) + \cdots,
\end{equation}
in the vertex. Then, the Fourier transform is performed order by order in $\Lambda_d$, which is equivalent to an expansion in $\kappa^2$ or Newton's constant. I will demonstrate in the following that the $\Lambda_d^n$ order vertex effectively is an $(n-1)$-loop integral. First, however, it is useful to get an idea of the structure of the expanded vertex at, say, order $\Lambda_d^n$, by investigating the types of terms that appear. These can be categorized by which gothic coefficient they originate from. From $\bar{\mathfrak{C}}_{[\partial^0]}$, which contains no derivatives on the metric, the general structure one observes is just
\begin{equation}
    \frac{\Lambda_d^n}{\abs{x_\perp}^{n(d-3)}}.
\label{eq:structures0}\end{equation}
From $\bar{\mathfrak{C}}_{[\partial^1]}$, which contains a single partial derivative, one has structures of the form
\begin{equation}
    \partial_\mu\frac{\Lambda_d^n}{\abs{x_\perp}^{n(d-3)}}, ~~~ \frac{\Lambda_d}{\abs{x_\perp}^{d-3}}\partial_\mu\frac{\Lambda_d^{n-1}}{\abs{x_\perp}^{(n-1)(d-3)}}, ~~~ \frac{\Lambda_d^2}{\abs{x_\perp}^{2(d-3)}}\partial_\mu\frac{\Lambda_d^{n-2}}{\abs{x_\perp}^{(n-2)(d-3)}}, ~\cdots.
\label{eq:structures1}\end{equation}
These can always trivially be reduced to the form
\begin{equation}
    \frac{\Lambda_d^{n-1}}{\abs{x_\perp}^{(n-1)(d-3)}}\partial_\mu\frac{\Lambda_d}{\abs{x_\perp}^{d-3}}
\end{equation}
by the Leibniz rule. In $\bar{\mathfrak{C}}_{[\partial^2]}$ there are two derivatives, so one encounters terms of the type
\begin{equation}
\begin{alignedat}{3}
    &\partial_\mu\partial_\nu\frac{\Lambda_d^n}{\abs{x_\perp}^{n(d-3)}}, ~~~ &&\frac{\Lambda_d}{\abs{x_\perp}^{d-3}}\partial_\mu\partial_\nu\frac{\Lambda_d^{n-1}}{\abs{x_\perp}^{(n-1)(d-3)}}, ~~~ &&\frac{\Lambda_d^2}{\abs{x_\perp}^{2(d-3)}}\partial_\mu\partial_\nu\frac{\Lambda_d^{n-2}}{\abs{x_\perp}^{(n-2)(d-3)}}, ~\cdots, \\
    &\partial_\mu\partial_\nu\frac{\Lambda_d^n}{\abs{x_\perp}^{n(d-3)}}, ~~~ &&\partial_\mu\frac{\Lambda_d}{\abs{x_\perp}^{d-3}}\partial_\nu\frac{\Lambda_d^{n-1}}{\abs{x_\perp}^{(n-1)(d-3)}}, ~~~ &&\partial_\mu\frac{\Lambda_d^2}{\abs{x_\perp}^{2(d-3)}}\partial_\nu\frac{\Lambda_d^{n-2}}{\abs{x_\perp}^{(n-2)(d-3)}}, ~\cdots,
\end{alignedat}
\label{eq:structures2}\end{equation}
where the derivative acts only on the object immediately following it. The above sample should be enough to illustrate the pattern: one encounters generically all possible distributions of the two derivatives among the $n$ factors of $\abs{x_\perp}^{-(d-3)}$. Again, by the Leibniz rule these can always be reduced to either of the two forms
\begin{equation}
    \frac{\Lambda_d^{n-1}}{\abs{x_\perp}^{(n-1)(d-3)}}\partial_\mu\partial_\nu\frac{\Lambda_d}{\abs{x_\perp}^{d-3}}, \qqquad \frac{\Lambda_d^{n-2}}{\abs{x_\perp}^{(n-2)(d-3)}}\partial_\mu\frac{\Lambda_d}{\abs{x_\perp}^{d-3}}\partial_\nu\frac{\Lambda_d}{\abs{x_\perp}^{d-3}}.
\end{equation}
To take the Fourier transform of these structures, it is first necessary to compute the integral
\begin{equation}
    \int\mathrm{d}^dx\,\ee^{\ii q\cdot x}\frac{\Lambda_d}{\abs{x_\perp}^n},
\end{equation}
which can be computed as follows. First, make the decomposition
\begin{equation}
    x^\mu = x_\parallel^\mu + x_\perp^\mu, \qqquad \dd^dx = \dd^{d-1}x_\perp\dd x_\parallel.
\label{eq:decomppos}\end{equation}
Then, the integration along the velocity $v^\mu$, trivially evaluates to a delta function,
\begin{align}
    \int\mathrm{d}^dx\,\ee^{\ii q\cdot x}\frac{\Lambda_d}{\abs{x_\perp}^n} &= \int\mathrm{d}^{d-1}x_\perp\mathrm{d}x_\parallel\,\ee^{\ii q\cdot x}\frac{\Lambda_d}{\abs{x_\perp}^n} \notag\\
    &= \hat\delta(q\cdot v)\int\mathrm{d}^{d-1}x_\perp\,\ee^{\ii q\cdot x_\perp}\frac{\Lambda_d}{\abs{x_\perp}^n}.
\end{align}
To do the remaining integral, it is useful to exponentiate the denominator with Schwinger's trick,
\begin{equation}
    \frac{1}{D^m} = \frac{1}{\Gamma(m)}\int_0^\infty\dd\alpha\,\alpha^{m-1}\ee^{-\alpha D},
\end{equation}
which is valid when $D > 0$ (to ensure the convergence of the right-hand side). Using the tautological $\abs{x_\perp}^n = (\abs{x_\perp}^2)^\frac{n}{2}$ and introducing a Schwinger parameter, I get
\begin{align}
    \int\mathrm{d}^{d-1}x_\perp\,\ee^{\ii q\cdot x_\perp}\frac{\Lambda_d}{\abs{x_\perp}^n} &= \frac{\Lambda_d}{\Gamma\big(\frac{n}{2}\big)}\int_0^\infty\mathrm{d}\alpha\,\alpha^{\frac{n}{2}-1}\int\mathrm{d}^{d-1}x_\perp\,\ee^{-\alpha\abs{x_\perp}^2+\ii q\cdot x_\perp} \notag\\
    &= \frac{\pi^\frac{d-1}{2}\Lambda_d}{\Gamma\big(\frac{n}{2}\big)}\int_0^\infty\mathrm{d}\alpha\,\alpha^{\frac{n-d-1}{2}}\exp\frac{q^2}{4\alpha} \notag\\
    &= \frac{(4\pi)^\frac{d-1}{2}\Gamma\big(\frac{d-1-n}{2}\big)}{2^n\Gamma\big(\frac{n}{2}\big)}\frac{\Lambda_d}{(-q^2)^\frac{d-1-n}{2}}.
\end{align}
Notice that $q^2<0$ due to the delta function, so the integral in the second line is convergent. The desired result is then obtained by substituting $n = d - 3$. To wit,
\begin{align}
    \int\mathrm{d}^dx\,\ee^{\ii q\cdot x}\frac{\Lambda_d}{\abs{x_\perp}^{d-3}} &= \frac{4\pi^\frac{d-1}{2}\Lambda_d}{\Gamma\big(\frac{d-3}{2}\big)}\frac{\hat\delta(q\cdot v)}{-q^2} \notag\\
    &= \hat\delta(q\cdot v)\frac{(d-3)\Omega_{d-2}\Lambda_d}{-q^2},
\end{align}
where I used that $\Gamma(n+1) = n\Gamma(n)$ and the surface area of the $(d-2)$-sphere $\Omega_{d-2}$ was defined in eq. \eqref{eq:sphere-surface-area}. By using the definition of $\Lambda_d$ in eq. \eqref{eq:lambdaddef}, I find that the constants combine elegantly into
\begin{equation}
    (d-3)\Omega_{d-2}\Lambda_d = \frac{\kappa^2M}{2}\frac{d-3}{d-2}.
\end{equation}
It is noteworthy that the factor of $(d-2)^{-1}$ normally associated with the de Donder propagator appears here. Also significant is the fact that the injected momentum enters with a power of $-2$. The effect of any number of derivatives in the Fourier transform is simply to pull down factors of momentum from the exponential, such that
\begin{equation}
    \int\mathrm{d}^dx\,\ee^{\ii q\cdot x}\partial_{\mu_1}\cdots\partial_{\mu_n}\frac{\Lambda_d}{\abs{x_\perp}^{d-3}} = \frac{\kappa^2M}{2}\frac{d-3}{d-2}\frac{(-\ii)^nq_{\mu_1}\cdots q_{\mu_n}}{-q^2}.
\label{eq:thefourierineed}\end{equation}
This is easily seen by either Fourier transforming $\abs{x_\perp}^{-(d-3)}$ before differentiating or integrating eq. \eqref{eq:thefourierineed} repeatedly by parts.

The last thing that is needed to determine the Fourier transforms of the structures in eqs. \eqref{eq:structures0}, \eqref{eq:structures1}, \eqref{eq:structures2} is the convolution theorem: a product in position space is a convolution in momentum space. Knowing this, the Fourier transform of eq. \eqref{eq:structures0} can be computed to be
\begin{align}
    \int\mathrm{d}^dx\,\ee^{\ii q\cdot x}\frac{\Lambda_d^n}{\abs{x_\perp}^{n(d-3)}} = \hat\delta(q\cdot v)\mathcal{N}_d\int_{\ell_1,\ldots,\ell_{n-1}}\frac{(-1)^n\hat\delta(\ell_1\cdot v)\cdots\hat\delta(\ell_{n-1}\cdot v)}{\ell_1^2\cdots\ell_{n-1}^2(q-\ell_{1\cdots (n-1)})^2},
\label{eq:nth-order-fan-integral}\end{align}
where
\begin{equation}
    \ell_{1\cdots (n-1)} = \sum_{i=1}^{n-1}\ell_i, \qqquad \mathcal{N}_d = \frac{(\kappa^2M)^n}{2^n}\frac{(d-3)^n}{(d-2)^n}.
\end{equation}
The type of integral occurring in eq. \eqref{eq:nth-order-fan-integral} is called a fan integral due to the diagram topology it represents. It can be evaluated exactly for any $n$ \cite{Cheung:2024byb}; I will keep it unevaluated as this allows a unified treatment of the integrand of the Compton amplitude. This is both procedurally convenient and important for checking the Ward identity at the integrand level.

Using eq. \eqref{eq:thefourierineed}, the Fourier transform of the structures in eqs. \eqref{eq:structures1} and \eqref{eq:structures2} are determined to be
\begin{align}
    \int&\mathrm{d}^dx\,\ee^{\ii q\cdot x}\frac{\Lambda_d^{n-1}}{\abs{x_\perp}^{(n-1)(d-3)}}\partial^\mu\frac{\Lambda_d}{\abs{x_\perp}^{d-3}} \notag\\ &= \hat\delta(q\cdot v)\mathcal{N}_d\int_{\ell_1,\ldots,\ell_{n-1}}\frac{(-1)^n\hat\delta(\ell_1\cdot v)\cdots\hat\delta(\ell_{n-1}\cdot v)(-\ii\ell^\mu_{n-1})}{\ell_1^2\cdots\ell_{n-1}^2(q-\ell_{1\cdots (n-1)})^2}, \label{eq:nth-order-fan-integral1der}\\
    \int&\mathrm{d}^dx\,\ee^{\ii q\cdot x}\frac{\Lambda_d^{n-1}}{\abs{x_\perp}^{(n-1)(d-3)}}\partial^\mu\partial^\nu\frac{\Lambda_d}{\abs{x_\perp}^{d-3}} \notag\\ &= \hat\delta(q\cdot v)\mathcal{N}_d\int_{\ell_1,\ldots,\ell_{n-1}}\frac{(-1)^n\hat\delta(\ell_1\cdot v)\cdots\hat\delta(\ell_{n-1}\cdot v)(-\ell^\mu_{n-1}\ell^\nu_{n-1})}{\ell_1^2\cdots\ell_{n-1}^2(q-\ell_{1\cdots (n-1)})^2}, \label{eq:nth-order-fan-integral2der1} \\
    \int&\mathrm{d}^dx\,\ee^{\ii q\cdot x}\frac{\Lambda_d^{n-2}}{\abs{x_\perp}^{(n-2)(d-3)}}\partial^\mu\frac{\Lambda_d}{\abs{x_\perp}^{d-3}}\partial^\nu\frac{\Lambda_d}{\abs{x_\perp}^{d-3}} \notag\\ &= \hat\delta(q\cdot v)\mathcal{N}_d\int_{\ell_1,\ldots,\ell_{n-1}}\frac{(-1)^n\hat\delta(\ell_1\cdot v)\cdots\hat\delta(\ell_{n-1}\cdot v)(-\ell^\mu_{n-1}(q-\ell_{1\cdots (n-1)})^\nu)}{\ell_1^2\cdots\ell_{n-1}^2(q-\ell_{1\cdots (n-1)})^2}. \label{eq:nth-order-fan-integral2der2}
\end{align}
Putting all this together, the expansion of the momentum space vertex can be performed,
\begin{equation}
    \bar V^{\mu_1\nu_1\,\mu_2\nu_2}(k_1, k_2) = \sum_{i=1}^\infty\bar V_{(i)}^{\mu_1\nu_1\,\mu_2\nu_2}(k_1, k_2),
\end{equation}
where it is now clear from the preceding discussion that $V_{(i)}^{\mu_1\nu_1\,\mu_2\nu_2}(k_1, k_2)$ is an $(i-1)$-loop integral, as was claimed. Diagrammatically, the expansion can be depicted as
\begin{equation}
    \begin{tikzpicture}[baseline={(current bounding box.center)}]
        \coordinate (in) at (-1,0);
        \coordinate (out) at (1,0);
        \coordinate (x) at (0,0);
        \coordinate (gin) at (-1,-1.2);
        \coordinate (gout) at (1,-1.2);
        \coordinate (v) at (0,-1.2);
        \draw [dotted, thick] (in) -- (x);
        \draw [dotted, thick] (x) -- (out);
        \draw [photon2] (x) -- (v) node [midway, left, black] {$q\!\downarrow$};
        \draw [photon] (v) -- (gin) node [left, below=.5em] {$h_{\mu_1\nu_1}(k_1)$};
        \draw [photon] (v) -- (gout) node [right, below=.5em] {$h_{\mu_2\nu_2}(k_2)$};
        \draw [fill] (v) circle (.08);
        \draw [fill] (x) circle (.08);
    \end{tikzpicture} = \sum_{i=1}^\infty\begin{tikzpicture}[baseline={(current bounding box.center)}]
        \coordinate (in) at (-1,0);
        \coordinate (out) at (1,0);
        \coordinate (x) at (0,0);
        \coordinate (gin) at (-1,-1.2);
        \coordinate (gout) at (1,-1.2);
        \coordinate (v) at (0,-1.2);
        \draw [dotted, thick] (in) -- (x);
        \draw [dotted, thick] (x) -- (out);
        \draw [photon2] (x) -- (v) node [midway, left, black] {$q\!\downarrow$};
        \draw [photon] (v) -- (gin) node [left, below=.5em] {$h_{\mu_1\nu_1}(k_1)$};
        \draw [photon] (v) -- (gout) node [right, below=.5em] {$h_{\mu_2\nu_2}(k_2)$};
        \draw [fill] (v) circle (.14) node {\color{white}\tiny$i$};
        \draw [fill] (x) circle (.08);
    \end{tikzpicture}.
\label{eq:2-pt-expanded-diagram}\end{equation}
For the computation of the 2PM Compton amplitude, the first and second order vertex are needed. The first can be written
\begin{equation}
    \begin{tikzpicture}[baseline={(current bounding box.center)}]
        \coordinate (in) at (-1,0);
        \coordinate (out) at (1,0);
        \coordinate (x) at (0,0);
        \coordinate (gin) at (-1,-1.2);
        \coordinate (gout) at (1,-1.2);
        \coordinate (v) at (0,-1.2);
        \draw [dotted, thick] (in) -- (x);
        \draw [dotted, thick] (x) -- (out);
        \draw [photon2] (x) -- (v) node [midway, left, black] {$q\!\downarrow$};
        \draw [photon] (v) -- (gin) node [left, below=.5em] {$h_{\mu_1\nu_1}(k_1)$};
        \draw [photon] (v) -- (gout) node [right, below=.5em] {$h_{\mu_2\nu_2}(k_2)$};
        \draw [fill] (v) circle (.14) node {\color{white}\tiny$1$};
        \draw [fill] (x) circle (.08);
    \end{tikzpicture} = \ii\hat\delta(q\cdot v)\frac{\kappa^2M}{q^2}\bar N_{(1)}^{\mu_1\nu_1\,\mu_2\nu_2},
\label{eq:2-pt-kappa2}\end{equation}
where $\bar N_{(1)}^{\mu_1\nu_1\,\mu_2\nu_2}$ is a numerator containing the remaining pieces of the vertex including the tensor structure. For the next-order vertex, there is one integral:
\begin{equation}
    \begin{tikzpicture}[baseline={(current bounding box.center)}]
        \coordinate (in) at (-1,0);
        \coordinate (out) at (1,0);
        \coordinate (x) at (0,0);
        \coordinate (gin) at (-1,-1.2);
        \coordinate (gout) at (1,-1.2);
        \coordinate (v) at (0,-1.2);
        \draw [dotted, thick] (in) -- (x);
        \draw [dotted, thick] (x) -- (out);
        \draw [photon2] (x) -- (v) node [midway, left, black] {$q\!\downarrow$};
        \draw [photon] (v) -- (gin) node [left, below=.5em] {$h_{\mu_1\nu_1}(k_1)$};
        \draw [photon] (v) -- (gout) node [right, below=.5em] {$h_{\mu_2\nu_2}(k_2)$};
        \draw [fill] (v) circle (.14) node {\color{white}\tiny$2$};
        \draw [fill] (x) circle (.08);
    \end{tikzpicture} = \ii\hat\delta(q\cdot v)(\kappa^2M)^2\int_\ell\frac{\hat\delta(\ell\cdot v)\bar N_{(2)}^{\mu_1\nu_1\,\mu_2\nu_2}}{\ell^2(q-\ell)^2}.
\label{eq:2-pt-kappa4}\end{equation}
The numerators are quite complicated; I will not report them here, as one does not gain much from staring at them. I will mention for clarity, however, that they in general contain all combinations of $\eta_\parallel^{\mu\nu}$, $\eta_\perp^{\mu\nu}$, $k_1^\mu$, $k_2^\mu$ and $q^\mu$ that satisfy the symmetries of the vertex. For $N_{(2)}^{\mu_1\nu_1\,\mu_2\nu_2}$, the loop momentum $\ell^\mu$ may also appear.

\section{Gravity as an effective field theory}\label{sec:gravityaseft}
Now that the Feynman rules in the weak-field expansion and in a curved background have been discussed in depth, I find it appropriate to describe the problems that have been discovered with quantum gravity before continuing.

I will first describe by a power-counting argument why one does not expect pure Einstein gravity\footnote{The term Einstein gravity refers to the theory of general relativity as described by the Einstein--Hilbert action.} to be well-behaved in the ultraviolet regime. Recall from the discussion of the Feynman rules that the graviton propagator goes as $k^{-2}$ while the graviton vertices go as $k^2$. As each loop integral contains a factor of $\dd^dk$, the superficial degree of divergence of a diagram with $I$ internal edges, $V$ vertices, and $L$ loops is
\begin{equation}
    \omega = dL + 2V - 2I.
\end{equation}
The familiar Feynman diagram relation
\begin{equation}
    L = 1 + I - V
\end{equation}
implies that
\begin{equation}
    \omega = 2 + (d-2)L,
\end{equation}
which depends only on the number of loops. Thus, for dimensions larger than two, the superficial degree of divergence rises linearly with the number of loops, meaning one expects increasingly worse ultraviolet behavior. In fact, one should expect this based on Newton's constant having negative mass dimension. More loops imply more factors of Newton's constant which, from dimensional analysis, requires a commensurate increase in the number of loop momenta in the numerator.

The divergences of Einstein gravity were first studied using the background field method developed in \cite{DeWitt:1967ub} and reviewed in \cite{Abbott:1981ke}. In this method, one computes an effective action for the background metric by integrating out the graviton from eq. \eqref{eq:ehactionquadraticgaugefixed}. In the integration, one can then discard anything that is not divergent, hereby isolating the divergences. The one-loop divergence was first computed by 't Hooft and Veltman \cite{tHooft:1974toh}, while the two-loop divergence was found by Goroff and Sagnotti \cite{Goroff:1985th}. These divergences require the addition to the Einstein--Hilbert Lagrangian of the counterterm Lagrangians
\begin{subequations}
\begin{align}
    \mathcal{L}_\text{ct}^\text{one-loop} &= \frac{\metdens}{8\pi^2(d-4)}\Big[\frac{1}{120}R^2 + \frac{7}{20}R_{\mu\nu}R^{\mu\nu}\Big], \\
    \mathcal{L}_\text{ct}^\text{two-loop} &= \frac{\metdens}{(16\pi^2)^2(d-4)}\frac{209}{2880}R\indices{_{\mu\nu}^{\rho\sigma}}R\indices{_{\rho\sigma}^{\kappa\lambda}}R\indices{_{\kappa\lambda}^{\mu\nu}}.
\end{align}
\end{subequations}
Coupling gravity to matter necessitates the addition of additional counterterms \cite{Hamber:2009qgr}. 

It is a striking fact that Einstein gravity breaks down in the ultraviolet, as it strongly indicates that new physics lurks beyond the veil of current understanding. Nevertheless, this breakdown does not endanger the predictability of the theory at low energies; it merely suggests the existence of heavy degrees of freedom of whose ignorance one must be earnest when doing calculations. In fact, from the modern point of view of effective field theory, any (pedestrian) quantum field theory---even a renormalizable one---should be equipped with a certain energy regime of validity \cite{Weinberg:1995mt}.

Incorporating the ignorance of unknown high-energy physics is done by including all higher-derivative operators in the Lagrangian that are compatible with the symmetries of the theory. The intuition for why this is so is the following. Consider a four-dimensional theory of a massless field $\phi$ interacting with a heavy field $\Phi$ of mass $M$. The action of this theory could for instance be
\begin{equation}
    S[\phi,\Phi] = \int\dd^4x\bigg(-\frac12\phi\partial^2\phi -\frac12\Phi(\partial^2 + M^2)\Phi - \lambda\Phi\phi^2\bigg).
\end{equation}
One can now perform the integration over the heavy field in the partition function. This is easy in this particular theory, as the heavy field appears quadratically in the action making the integral Gaussian. Neglecting some details, this will yield
\begin{align}
    S_\text{eff}[\phi] &= \int\dd^4x\bigg(-\phi\partial^2\phi + \frac12\phi^2\frac{1}{\partial^2 + M^2}\phi^2\bigg) \notag\\
    &= \int\dd^4x\bigg(-\phi\partial^2\phi + \frac{1}{2M^2}\sum_{n=0}^\infty\phi^2\Big[-\frac{\partial^2}{M^2}\Big]^n\phi^2\bigg).
\label{eq:toymodeleft}\end{align}
There arises a tower of higher-derivative interaction terms suppressed by increasing powers of $M^2$. A standard model example of this is Fermi's contact interaction, which arises when one integrates out the W boson, turning two three-point interactions into a four-point interaction. Exporting this idea to gravity, one should augment the Einstein--Hilbert action by including every possible contraction of the Riemann tensor,
\begin{equation}
    S_\text{EH}[g] = \int\dx\metdens\bigg(-\frac{\Lambda}{M_\text{Pl}^d} - \frac{2}{\kappa^2}R + \frac{c_1}{M^{d-4}_\text{Pl}}R^2 + \frac{c_2}{M^{d-4}_\text{Pl}}R_{\mu\nu}R^{\mu\nu} + \cdots\bigg).
\label{eq:greft}\end{equation}
Here, factors of the Planck mass $M_\text{Pl} \sim 10^{19}\,\text{GeV}$ were included to make the coefficients of each term dimensionless. The growing suppression of higher-derivative operators by $M_\text{Pl}$ is similar to eq. \eqref{eq:toymodeleft}, and the Planck mass is chosen for this purpose exactly because it is believed to be the scale at which quantum gravity effects generically become important. The coefficients $c_i$ parameterize our ignorance of the ultraviolet completion of gravity and must be determined by experiment or by knowledge of a (candidate) ultraviolet completion, such as string theory. They are known as Wilson coefficients, and they encode all information about the heavy degrees of freedom that is relevant to the dynamics of the light degrees of freedom, i.e. the graviton, at the $i^\text{th}$ order in the energy expansion in eq. \eqref{eq:greft}.

In this thesis, it is the interaction of gravity with a non-spinning black hole that is under study. The quantum field-theoretic avatar for this object is a scalar field of macroscopic mass. By the principles of effective field theory, one must include also in the scalar action all possibilities allowed by diffeomorphism invariance
\begin{align}
    S_\phi[\phi,g] = \int\dx\metdens\bigg(&-V(\phi) + \frac12g^{\mu\nu}\partial_\mu\phi\partial_\nu\phi - \frac12m^2\phi^2 - \xi\phi^2R \notag\\
    &+ \frac{d_1}{M_\text{Pl}^2}Rg^{\mu\nu}\partial_\mu\phi\partial_\nu\phi + \frac{d_2}{M_\text{Pl}^2}R^{\mu\nu}\partial_\mu\phi\partial_\nu\phi + \cdots\bigg).
\label{eq:scalareft}\end{align}
The couplings were again made dimensionless by the inclusion of a factor of the Planck mass. These higher-order operators again encode the unknown dynamics of heavy degrees of freedom. For the massive scalar field, these are readily interpreted as coming from the finite-size of the object one is modeling and thus correspond to tidal effects \cite{Haddad:2020que}. The Wilson coefficients of eq. \eqref{eq:scalareft} are thus related to the Love numbers of general relativity \cite{Bern:2020uwk}.

There is a multitude of interesting things to say about the treatment of general relativity as an effective field theory and about effective field theory in general. The most important implication for the purposes of this thesis is for the validity of the low-energy expansion. It implies one can extract results from the theory for a given process in the low-energy regime in a well-defined way by assuming that the characteristic curvatures are small \cite{Bjerrum-Bohr:2022ows}. It has been known that any massless spin-two field theory must reduce to general relativity when the characteristic distances are large \cite{Weinberg:1965rz}. Indeed, the effective field theory approach has been used to great success as detailed in the introduction of this thesis.

\section{Approach to classical physics}\label{sec:classicalphysics}
In this section, I will briefly describe how the classical limit is taken from the point of view of scattering amplitudes. I do this to set the stage for the developments in the following chapter, where the worldline quantum field theory is discussed. Worldline quantum field theory represents an alternative and efficient way to take the classical limit.

The partition function of a quantum field theory is a highly oscillatory object. Any field configuration not satisfying the equations of motion $\delta S = 0$---which is the stationary point of the phase of the integrand---is suppressed due to this oscillatory behavior. To take the classical limit, one usually reintroduces Planck's constant $\hbar$, which in turn reintroduces the difference between the dimensions of energy [M] and length [L], as $[\hbar] = [\text{M}][\text{L}]$. With $\hbar$ turned on, the classical limit can be taken by taking $\hbar \to 0$. For a given scattering amplitude, the pertinent question for taking the limit is thus how its constituent parts scale with $\hbar$, which one can deduce from dimensional analysis. The action divided by Planck's constant must be dimensionless,
\begin{equation}
    \frac{[S]}{[\hbar]} = 1.
\end{equation}
As $[S] = [\dd^dx][\mathcal{L}]$, the Lagrangian must have units of
\begin{equation}
    [\mathcal{L}] = [\text{L}]^{-d}[\hbar].
\end{equation}
From elementary quantum mechanics, it is natural to associate a factor of $\hbar$ to each partial derivative. As the kinetic term for a scalar field $\phi$ is of the form $\hbar^2\partial^2\phi^2$, the field has to have dimensions
\begin{equation}
    [\phi] = [\text{L}]^{1-\frac{d}{2}}[\hbar]^{-\frac12}.
\end{equation}
This implies that mass terms require no extra factors of $\hbar$. The same analysis holds for the graviton, such that
\begin{equation}
    [h_{\mu\nu}] = [\text{L}]^{1-\frac{d}{2}}[\hbar]^{-\frac12}.
\end{equation}
This occurrence of $\hbar$ in the action implies that any momentum appearing in a computation must be multiplied with $\hbar$. Now, the cubic interaction term between graviton and scalar field is of the form $\kappa h_{\mu\nu}T_\phi^{\mu\nu}$, where $T^{\mu\nu}$ is the energy momentum tensor of the scalar field. It can be seen from
\begin{equation}
    \delta S_\phi = -\frac12\int\dx\metdens T_\phi^{\mu\nu}\delta g_{\mu\nu},
\end{equation}
that the energy-momentum tensor must have the same dimensions as the Lagrangian. A factor of $\hbar$ must thus be inserted in the combination $\hbar^\alpha\kappa h_{\mu\nu}$ to ensure it be dimensionless. This comes out to
\begin{equation}
    [\hbar^\alpha\kappa h_{\mu\nu}] = [\hbar]^{\alpha - \frac{3-d}{2}},
\end{equation}
implying immediately that every occurrence of $\kappa$ must be accompanied by $\hbar^\frac{3-d}{2}$, which becomes $\hbar^{-\frac12}$ in four dimensions. Another way to see this is from the dimensions of the $d$-dimensional Newton's constant: when Planck's constant is turned on,
\begin{equation}
    [G_d] = [\text{L}]^{d-3}[\text{M}]^{-1}.
\end{equation}
These dimensions can be deduced from the $d$-dimensional Poisson's equation for the gravitational field \cite{Zwiebach:2004tj}. They imply that one must make the replacement
\begin{equation}
    G_d \to \hbar^{d-3}G_d
\end{equation}
(to keep the dimensions the same) in agreement with what was stated before.

Now knowing where all the $\hbar$'s are supposed to go, one can proceed in the usual way. Compute the quantity of interest in natural units with unitarity, Feynman diagrams or other methods. Then, make the replacements
\begin{equation}
    \kappa \to \frac{\kappa}{\hbar^\frac{d-3}{2}}, \qqquad p^\mu \to \hbar\bar p^\mu,
\end{equation}
and take $\hbar \to 0$. There is a final subtle point that must be emphasized, and this is the difference in the classical nature of massless particles and massive particles. In an effective field theory, one is always operating in a low-energy limit. This means massless particles become waves with well-defined wavenumber $\bar p^\mu$ in the classical limit, while massive particles become particles with well-defined momentum $p^\mu$. This implies that one must make the replacement
\begin{equation}
    p^\mu \to \hbar\bar p^\mu
\end{equation}
only for massless momenta. In the massive case, the wavenumber diverges exactly opposite to the manner $\hbar$ goes to zero, such that the expression $\hbar\bar p^\mu$ stays finite. For more details on the classical limit, see \cite{Kosower:2018adc}.

A very important consequence of working in a joint low-energy and classical limit is that the higher-order derivative operators in eqs. \eqref{eq:greft} and \eqref{eq:scalareft} in general can be ignored, since they contain higher powers in the graviton momentum, which scales as $\hbar$. Another important consequence is that graviton loops never contribute to classical physics, as is discussed in, e.g., \cite{Bern:2019crd}.
\chapter{Worldline quantum field theory}\label{ch:worldline formalism}
In the preceding chapter, I described the effective field theory approach to gravitational dynamics, where one describes compact bodies as excitations of scalar fields. Classical observables can then be extracted from this theory by taking the classical limit of its scattering amplitudes. In worldline quantum field theory, the classical limit is taken at the level of the action \cite{Mogull:2020sak}. This makes taking the classical limit in a concrete computation very easy: one simply includes only tree-level diagrams in the diagrammatic expansion. In section \ref{sec:wftformalism}, I provide an overview of the worldline formalism. Then, in section \ref{sec:feynruleswfwqft}, I derive the Feynman rules in the weak-field expansion, after which I turn to the Feynman rules in the Schwarzschild--Tangherlini background in section \ref{sec:wqftfeynrulesschwarzschil}. Lastly, in section \ref{sec:relationsbetween}, I discuss the relations between these two sets of Feynman rules.

\section{Description of the formalism}\label{sec:wftformalism}
Worldline quantum field theory has its origins in worldline effective field theory, which was developed to describe the binary problem in general relativity. It is described at the fundamental level by the action
\begin{equation}
    S[g,\{x_i\}] = S_\text{G}[g] + \sum_{i=1}^2S_{\text{wl},i}[x_i,g],
\label{eq:wlaction2particles}\end{equation}
where $S_\text{G}$ is the gauge-fixed gravitational action from eq. \eqref{eq:gravitationalaction}, and
\begin{equation}
\begin{aligned}
    S_{\text{wl},i}[x_i,g] = &-\frac{M}{2}\int\dd\tau\,\big(g_{\mu\nu}(x_i)\dot x_i^\mu\dot x_i^\nu + 1\big) \\
    &+ c_R\int\dd\tau\,R(x_i) + c_V\int\dd\tau\,R_{\mu\nu}(x_i)\dot x_i^\mu\dot x_i^\nu + \cdots
\end{aligned}
\end{equation}
is the point particle action I introduced in sec. \ref{sec:action for a point particle}, now augmented with the appropriate higher-derivative operators representing tidal effects in the spirit of effective field theory \cite{Bini:2012gu}. The $c_R$ and $c_V$ terms are in fact unphysical, as they can be absorbed in a field redefinition of $h_{\mu\nu}$ \cite{Goldberger:2004jt}. Due to working in a low-energy expansion, I will not concern myself with the higher-order operators. In the literature with which I am familiar, the background metric is always taken to be the Minkowski metric. In worldline effective field theory, one has traditionally performed a non-relativistic joint expansion in Newton's constant and the relative velocity $v^\mu$ of the bodies \cite{Goldberger:2004jt,Goldberger:2006bd,Goldberger:2009qd}. This is known as the non-relativistic general relativity formalism. Here, a decomposition of the graviton into potential and radiation modes is performed and the path integral over the field configurations of the potential modes is computed perturbatively, while the radiation modes and the trajectories $x_i^\mu$ are kept classical---with the trajectories in a sense acting as sources for the potential graviton. This results in an effective action for the radiation graviton and the trajectories $x_i^\mu$ from which one can derive equations of motion. The non-relativistic formalism is especially well-suited for describing the bound system, since the virial theorem implies that
\begin{equation}
    v^2 \sim \frac{Gm}{r},
\end{equation}
where $m$ is a mass characteristic of the system. More recently, the use of the post-Min\-kowskian expansion, where one expands solely in Newton's constant, has been utilized to great effect in the context of worldline effective field theory \cite{Kalin:2020mvi,Kalin:2020fhe}.

In worldline quantum field theory, in contrast to worldline effective field theory, both the graviton and the trajectories $x_i^\mu$ are quantized. Observables such as the change in the momentum of one of the point particles over a scattering event or the radiation waveform are then related to one-point functions in the theory, and these can be readily computed in the post-Minkowskian expansion. This is quite advantageous, as one circumvents entirely the need for solving the equations of motion obtained from the aforementioned effective action; the one-point functions already solve the equations of motion \cite{Boulware:1968zz}.

Explorations of what quantities of interest other than one-point functions can be calculated in worldline quantum field theory and the possibility of taking the background metric to be curved has remained relatively obscure, however. I am familiar only with the analyses in \cite{Wang:2022ntx,Comberiati:2022ldk,Kosmopoulos:2023bwc, Akpinar:2025huz}. An example is exactly the topic of this thesis: the classical Compton amplitude, which I compute in the following chapter at first and 2PM. For this computation, the relevant system contains a single black hole, and my action will thus be the one in eq. \eqref{eq:wlaction2particles} with only one point particle action. If a coordinate system is placed, such that the black hole crosses the origin when its proper time $\tau = 0$, then it is physically obvious that its free motion is given by $v^\mu\tau$, with $v^\mu$ being its velocity. It is hence sensible to perform a background field expansion of the trajectory
\begin{equation}
    x^\mu(\tau) = v^\mu\tau + z^\mu(\tau),
\label{eq:inertial-exp}\end{equation}
and work with the deflection $z^\mu(\tau)$ from the free motion. I will refer to this as the inertial expansion. Once quantized, one should think of the deflection field as a quantum field living on the worldline of the black hole, which is exactly the faint dotted line that accompanied the vertices in the section \ref{sec:feynmanrulesgeneralback}.

The worldline quantum field theory I will be working with is thus defined by the generating functional
\begin{equation}
    Z[J,f] = \int\mathscr{D}h\mathscr{D}z\,\ee^{\ii S[g,x] + \ii\int\dx h_{\mu\nu}J^{\mu\nu} + \ii\int\dd\tau\,z^\mu f_\mu},
\label{eq:wqftpathintegral}\end{equation}
where the background metric is either the Minkowski one or the Schwarzschild--Tangherlini one. Note that it is reasonable to label the source of the deflection field as $f_\mu$, as it can be seen as a force. In the following sections, I will first derive the Feynman rules for this theory in flat space, after which I move to the curved case.

It is shown explicitly in \cite{Mogull:2020sak} that eq. \eqref{eq:wqftpathintegral} is the classical limit of a theory of a real massive scalar coupled to Einstein gravity, in the sense that expectation values in the worldline quantum field theory equal classical amplitudes computed in the full theory. I will not pursue a full derivation of this fact, as it is lengthy. A very schematic upshot of the process is the following: One can introduce a worldline representation for the dressed propagator of the scalar field and subsequently use this to ``partially'' integrate out the scalar from path integrals computing expectation values bilinear in the scalar. What is left is exactly the worldline action plus certain ghost fields (which do not contribute classically). Then, putting the external scalar legs on-shell by LSZ reduction, one obtains the desired result.\footnote{For an alternative approach to the reduction, see \cite{Bonezzi:2025iza}.}

\section{Feynman rules in the weak-field expansion}\label{sec:feynruleswfwqft}
The derivation of the worldline Feynman rules in the weak-field expansion was discussed comprehensively in \cite{Mogull:2020sak}. The starting point for the derivation is the worldline action
\begin{equation}
    S_\text{wl}[g,x] = -\frac{M}{2}\int\dd\tau\,\big(g_{\mu\nu}(x_i)\dot x_i^\mu\dot x_i^\nu + 1\big).
\end{equation}
The constant term in the action has no role to play in the context of Feynman rules, so I can drop it in the following steps. The first order of business is deriving the propagator for the deflection field. Inserting the weak-field expansion, the linearity in the metric of this action implies that
\begin{equation}
    S_\text{wl}[\eta + \kappa h,x] = S_\text{wl}[\eta,x] + \kappa S_\text{wl}[h,x].
\label{eq:wl-action-decomp}\end{equation}
It is from the first term that the propagator is obtained. Inserting the inertial expansion from eq. \eqref{eq:inertial-exp}, I get
\begin{equation}
    S_\text{wl}[\eta, v\tau + z] = -\frac{M}{2}\int\mathrm{d}\tau\,\eta_{\rho\sigma}\dot z^\rho\dot z^\sigma - M\int\mathrm{d}\tau\, v\cdot\dot z.
\end{equation}
The second term is total derivative so it can be dropped. The first I can integrate by parts to get
\begin{equation}
    -\frac{M}{2}\int\mathrm{d}\tau\,\eta_{\rho\sigma}\dot z^\rho\dot z^\sigma = \frac{M}{2}\int\mathrm{d}\tau\, z^\rho\eta_{\rho\sigma}\frac{\mathrm d^2}{\mathrm d\tau^2}{}z^\sigma.
\end{equation}
By introducing the Fourier transformed deflection
\begin{equation}
    z^\rho(\tau) = \int_\omega\ee^{-\ii \omega\tau}z^\rho(\omega), \qqquad \int_\omega = \int\frac{\dd\omega}{2\pi},
\label{eq:deflectionfourier}\end{equation}
the propagator can be seen to be
\begin{equation}
    \begin{tikzpicture}[baseline={(current bounding box.center)}]
        \coordinate (in) at (-1,0);
        \coordinate (out) at (2,0);
        \coordinate (x) at (-.3,0);
        \coordinate (y) at (1.3,0);
        \draw [zUndirected] (x) -- (y) node [midway, below] {$\omega$};
        \draw [dotted, thick] (in) -- (x);
        \draw [dotted, thick] (y) -- (out);
        \draw [fill] (x) node [above] {$z^\rho(\omega)$};
        \draw [fill] (y) node [above] {$z^\sigma(\omega)$};
    \end{tikzpicture} = \frac{-\ii\eta^{\rho\sigma}}{M\omega^2},
\end{equation}
The $\ii0$ prescription of the deflection is immaterial for the computations in this thesis, so I have omitted it. The second term of eq. \eqref{eq:wl-action-decomp} gives rise to the interactions between the graviton living in the bulk and the deflection living on the worldline. There is an infinite tower of interaction vertices on the worldline, all of which possess a single graviton leg but any amount of deflection legs, including none. This tower arises due to the dependence of the graviton on the worldline trajectory, which may be Fourier transformed away;
\begin{equation}
    h_{\mu\nu}(x(\tau)) = \int_k\ee^{\ii k\cdot x(\tau)}h_{\mu\nu}(-k).
\end{equation}
The expansion of the exponential after inserting the inertial expansion then produces
\begin{align}
    h_{\mu\nu}(x(\tau)) &= \int_k\ee^{\ii \tau k\cdot v}\Big[\sum_{j=0}^\infty\frac{\ii^j(k\cdot z(\tau))^j}{j!}\Big]h_{\mu\nu}(-k) \notag\\
    &= \sum_{j=0}^\infty\frac{\ii^j}{j!}\int_{k,\omega_1,\ldots,\omega_j}\ee^{\ii \tau k\cdot v + \ii\tau\omega_{1\cdots j}}h_{\mu\nu}(-k)\prod_{l=1}^jk\cdot z(-\omega_l)
\label{eq:gravitononwlfourier}\end{align}
with
\begin{equation}
    \omega_{1\cdots j} = \sum_{i=1}^j\omega_i.
\end{equation}
To obtain a compact expression from which the vertex can be extracted easily, it is most convenient to treat separately each term from the second term of eq. \eqref{eq:wl-action-decomp},
\begin{equation}
    \kappa S_\text{wl}[h,v\tau + z] = -M\kappa\int\dd\tau\,h_{\mu\nu}(x(\tau))\Big[\frac12v^\mu v^\nu + v^{(\mu}\dot z^{\nu)}(\tau) + \frac12\dot z^\mu(\tau)\dot z^\nu(\tau)\Big].
\end{equation}
The goal will be to express the above such that the sum over $j$ from eq. \eqref{eq:gravitononwlfourier} can be pulled out of the integrals. The first term requires no processing. The second term is
\begin{align}
    &h_{\mu\nu}(x(\tau))v^{(\mu}\dot z^{\nu)}(\tau) \notag\\ &= \sum_{j=0}^\infty\frac{\ii^j}{j!}\int_{k,\omega_1,\ldots,\omega_{j+1}}\ee^{\ii \tau k\cdot v + \ii\tau\omega_{1\cdots(j+1)}}h_{\mu\nu}(-k)\prod_{l=1}^jk\cdot z(-\omega_l)\ii\omega_{j+1}v^{(\mu}z^{\nu)}(-\omega_{j+1}),
\end{align}
where I Fourier transformed the remaining deflection and labeled its energy $\omega_{j+1}$. By splitting the dot products of the type $k\cdot z$, the above can be rewritten as
\begin{equation}
    \sum_{j=0}^\infty\frac{\ii^{j+1}}{j!}\int_{k,\omega_1,\ldots,\omega_{j+1}}\ee^{\ii \tau k\cdot v + \ii\tau\omega_{1\cdots(j+1)}}h_{\mu\nu}(-k)\prod_{l=1}^{j+1}z^{\rho_{l}}(-\omega_l)\prod_{m=1}^jk_{\rho_m}v^{(\mu}\delta^{\nu)}_{\rho_{j+1}}\omega_{j+1}
\end{equation}
The above expression allows a large freedom to relabel, meaning it is invariant under the symmetrization
\begin{equation}
    \prod_{m=1}^jk_{\rho_m}v^{(\mu}\delta^{\nu)}_{\rho_{j+1}}\omega_{j+1} \to \frac{1}{j+1}\sum_{n=1}^{j+1}\prod_{m\neq n}^{j+1}\omega_nk_{\rho_m}^{(\mu}\delta^{\nu)}_{\rho_n}.
\end{equation}
Making this replacement and reindexing the sum by taking $j+1\to j$, I get
\begin{equation}
    \sum_{j=1}^\infty\frac{\ii^j}{j!}\int_{k,\omega_1,\ldots,\omega_j}\ee^{\ii \tau k\cdot v + \ii\tau\omega_{1\cdots j}}h_{\mu\nu}(-k)\prod_{l=1}^jz^{\rho_{l}}(-\omega_l)\sum_{n=1}^j\prod_{m\neq n}^j\omega_nk_{\rho_m}^{(\mu}\delta^{\nu)}_{\rho_n}.
\label{eq:secondtermwlaction}\end{equation}
A story, similar in spirit to the above, can be told about the third term. I will not go through it in detail; suffice it to say that the result is
\begin{align}
    &\frac12 h_{\mu\nu}(x(\tau))\dot z^\mu(\tau)\dot z^\nu(\tau) \notag\\ &= \sum_{j=2}^\infty\frac{\ii^j}{j!}\int_{k,\omega_1,\ldots,\omega_j}\ee^{\ii \tau k\cdot v + \ii\tau\omega_{1\cdots(j+1)}}h_{\mu\nu}(-k)\prod_{l=1}^jz^{\rho_l}(-\omega_l)\sum_{n<m}^j\prod_{s\neq n,m}^j\omega_n\omega_mk_{\rho_s}
    \delta^{(\mu}_{\rho_n}\delta^{\nu)}_{\rho_m}.
\label{eq:thirdtermwlaction}\end{align}
The summations over $j$ in eqs. \eqref{eq:secondtermwlaction} and \eqref{eq:thirdtermwlaction} can be extended downward to $j=0$ with the understanding that the newly included terms vanish. Thus, when all is said and done, the integration over $\tau$ can be performed to give an overall delta function enforcing energy conservation on the worldline, and, all told, the interaction action becomes
\begin{equation}
\begin{aligned}
    \kappa S_\text{wl}[h,x] &= -\kappa M\sum_{j=0}^\infty\frac{\ii^j}{j!}
    \int_{k,\omega_1,\ldots,\omega_j}\hat\delta(k\cdot v + \omega_{1\cdots j})
    h_{\mu\nu}(-k)\prod_{l=1}^jz^{\rho_l}(-\omega_l) \\
    &\times\Bigg[\frac12\prod_{n=1}^jk_{\rho_n}v^\mu v^\nu+
    \sum_{n=1}^j\prod_{m\neq n}^j\omega_nk_{\rho_m}\!
    v^{(\mu}\delta^{\nu)}_{\rho_n}+
    \sum_{n<m}^j\prod_{s\neq n,m}^j\omega_n\omega_mk_{\rho_s}
    \delta^{(\mu}_{\rho_n}\delta^{\nu)}_{\rho_m}\Bigg],
\end{aligned}
\label{eq:wl-fouriered}\end{equation}
which was originally derived in \cite{Mogull:2020sak}. In this form, it is very easy to apply eq. \eqref{eq:vertexrule} to derive the Feynman rules. For the computations in this thesis, only the $j=0$ and $j=1$ pieces are required. The zeroth order term is,
\begin{equation}
    \kappa S_\text{wl}[h,x]\vert_{z^0} = -\frac{\kappa M}{2}\int_k\hat\delta(k\cdot v)v^\mu v^\nu h_{\mu\nu}(-k).
\label{eq:wl-source}\end{equation}
This piece of the action is in fact related to the energy-momentum tensor of the Schwarz\-schild--Tangherlini metric, which I shall make use of in the next section. It gives rise to a vertex where the worldline sources a graviton,
\begin{equation}
    \begin{tikzpicture}[baseline={(current bounding box.center)}]
        \coordinate (in) at (-1,0);
        \coordinate (out) at (1,0);
        \coordinate (x) at (0,0);
        \node (k) at (0,-1.3) {$h_{\mu\nu}(k)$};
        \draw [dotted, thick] (in) -- (x);
        \draw [dotted, thick] (x) -- (out);
        \draw [photon] (x) -- (k);
        \draw [fill] (x) circle (.08);
    \end{tikzpicture} = -\frac{\ii\kappa M}{2}\hat\delta(k\cdot v)v^\mu v^\nu.
\label{eq:wl-source-diag}\end{equation}
Even though I draw the worldline as a faint dotted line in these diagrams, I emphasize that it plays no mathematical significance exactly as was the case in section \ref{sec:feynmanrulesgeneralback}. The terms of the action linear in the deflection are
\begin{equation}
    \kappa S_\text{wl}[h,x]\vert_{z^1} = -\frac{\ii\kappa M}{2}\int_k\hat\delta(k\cdot v + \omega)(v^\mu v^\nu k_\rho + 2\omega v\indices{^{(\mu}}\delta\indices*{^{\nu)}_\rho})z^\rho(-\omega)h_{\mu\nu}(-k),
\end{equation}
from which I derive the Feynman rule
\begin{align}
    \begin{tikzpicture}[baseline={(current bounding box.center)}]
        \coordinate (in) at (-1,0);
        \coordinate (out) at (1,0);
        \coordinate (x) at (0,0);
        \node (k) at (0,-1.3) {$h_{\mu\nu}(k)$};
        \draw (out) node [right] {$z^\rho(\omega)$};
        \draw [dotted, thick] (in) -- (x);
        \draw [zUndirected] (x) -- (out);
        \draw [photon] (x) -- (k);
        \draw [fill] (x) circle (.08);
    \end{tikzpicture} &= \frac{\kappa M}{2}\hat\delta(k\cdot v + \omega)V\indices{^{\mu\nu}_\rho}(k,\omega) \notag \\
    &= \frac{\kappa M}{2}\hat\delta(k\cdot v + \omega)(v^\mu v^\nu k_\rho + 2\omega v\indices{^{(\mu}}\delta\indices*{^{\nu)}_\rho}).
\label{eq:lindeflecvertex}\end{align}
I will use these rules to compute the Compton amplitude at first and second post-Min\-kowskian order in the next chapter. First, I must focus on deriving the Feynman rules in the curved expansion.

Before proceeding, it is worth dwelling on one interesting aspect about these Feynman rules. The rules involving the deflection contain solely energy-conserving delta functions, as the deflection is defined only on the worldline. This is quite alien if one is familiar only with quantum field theories defined in what one might call the bulk spacetime. In the bulk, all Feynman rules are accompanied by momentum-conserving delta functions coming from the ubiquitous presence of $\int\mathrm{d}^dx$ in the action. This is not the case here, and it means that one obtains loop-like integrals in the tree-level diagrammatic expansion due to the leftover momenta not killed by momentum conservation.

\section{Feynman rules with Schwarzschild--Tangherlini background}\label{sec:wqftfeynrulesschwarzschil}
Once the background is chosen to be the Schwarzschild--Tangherlini metric, there are two important things that merit a discussion. The first is the fact that the worldline action contains the metric evaluated on the worldline, which is formally divergent for the metric under consideration,
\begin{equation}
    \bar g_{\mu\nu}(x(\tau)) = \infty.
\end{equation}
It will turn out that these infinities disappear in dimensional regularization, leaving only the flat space rules from the previous section. The second important thing is the cancellation that occurs between the part of the Einstein--Hilbert action that is linear in $h_{\mu\nu}$ and the part of the worldline action that is independent of the deflection,
\begin{equation}
    S_\text{EH}[g]\vert_{h^1} + S_\text{wl}[g,x]\vert_{z^0h^1} = 0.
\end{equation}
This means that the Feynman rule in eq. \eqref{eq:wl-source-diag} vanishes in this expansion of the metric. The implications of this will be discussed in section \ref{sec:relationsbetween}.

\subsection{Invariance of the rules under choice of background}
I will demonstrate in this section that the worldline Feynman rules in a Schwarzschild--Tan\-gherlini background are exactly the same as in the weak-field expansion. In fact, the worldline Feynman rules in any background $g_{\mu\nu}$ that admits an expansion of the type
\begin{equation}
    g_{\mu\nu}(x) = \eta_{\mu\nu} + \sum_n\frac{g_{\mu\nu}^{(n)}}{\abs{x_\perp}^n}.
\end{equation}
will be in exact correspondence with the weak-field ones, as the argument relies only on this expansion. The worldline action contains the metric evaluated on the worldline, which means that any piece of the metric going as $\abs{x_\perp}^{-n}$ will be formally divergent at the point of evaluation. This corresponds to a divergent self-energy of the black hole. However, as I am working in dimensional regularization, this will translate to purely scaleless integrals in momentum space, such that any piece of the metric deviating from flat space contributes nothing to the Feynman rules. If one were using another regularization procedure such as imposing a cutoff on the momentum integrals, these pieces would not vanish. They would, however, be purely divergent and could be absorbed in a renormalization of the mass \cite{Cheung:2024byb, Kosmopoulos:2023bwc}.

It is useful for this discussion to decompose the background metric into its flat and curved parts
\begin{equation}
    \bar g_{\mu\nu}(x) = \eta_{\mu\nu} + \bar\gamma_{\mu\nu}(x).
\end{equation}
Then, the linearity of the worldline action in the metric implies that
\begin{equation}
    S_\text{wl}[\bar g + \kappa h,x] = S_\text{wl}[\eta,x] + \kappa S_\text{wl}[h,x] + S_\text{wl}[\bar\gamma,x].
\label{eq:wl-action-decomp-curved}\end{equation}
This coincides with what was found in eq. \eqref{eq:wl-action-decomp} plus the extra piece $S_\text{wl}[\bar\gamma,x]$ which describes the self-force of the worldline, as it contains the curved part of the metric evaluated on the worldline. To prove the scalelessness, I start by manipulating $S_\text{wl}[\bar\gamma,x]$ analogously to the way $S_\text{wl}[h,x]$ was manipulated in the previous section. First, $\bar\gamma_{\mu\nu}$ is Fourier transformed,
\begin{equation}
    \bar\gamma_{\mu\nu}(x(t)) = \int_{\ell_1}\ee^{\ii\ell_1\cdot x(\tau)}\bar\gamma_{\mu\nu}(-\ell_1).
\label{eq:curved-piece-fourier}\end{equation}
Expanding the exponential in eq. \eqref{eq:curved-piece-fourier} and inserting yields an expression identical to what we obtained before, i.e.,
\begin{equation}
\begin{aligned}
    S_\text{wl}[\bar\gamma,x] &= - M\sum_{j=0}^\infty\frac{\ii^j}{j!}
    \int_{\ell_1,\omega_1,\ldots,\omega_j}\hat\delta(\ell_1\cdot v + \omega_{1\cdots j})
    \bar\gamma_{\mu\nu}(-\ell_1)\prod_{l=1}^jz^{\rho_l}(-\omega_l) \\
    &\times\Bigg[\frac12\prod_{n=1}^jk_{\rho_n}v^\mu v^\nu+
    \sum_{n=1}^j\prod_{m\neq n}^j\omega_nk_{\rho_m}\!
    v^{(\mu}\delta^{\nu)}_{\rho_n}+
    \sum_{n<m}^j\prod_{s\neq n,m}^j\omega_n\omega_mk_{\rho_s}
    \delta^{(\mu}_{\rho_n}\delta^{\nu)}_{\rho_m}\Bigg].
\end{aligned}
\label{eq:scaleless}\end{equation}
The $z^j$ order piece of this can be written
\begin{equation}
    S_\text{wl}[\bar\gamma,x]\vert_{z^j} \propto \int_{\ell_1,\omega_1,\ldots,\omega_j}\prod_{l=1}^jz^{\rho_l}(-\omega_l)N^{\mu\nu}_{\rho_1\cdots\rho_j}[\ell_1,\{\omega_i\}_{i=1..j}]\hat\delta(\ell_1\cdot v + \omega_{1\cdots j})\gamma_{\mu\nu}(-\ell_1),
\end{equation}
where the factor $N^{\mu\nu}_{\rho_1\cdots\rho_j}[\ell_1,\{\omega_i\}_{i=1..j}]$ contains the (complicated) numerator. As was shown above, $\bar\gamma_{\mu\nu}(-\ell_1)$ is given at order $\kappa^{2n}$ by the fan-type $(n-1)$-loop integral in eq. \eqref{eq:nth-order-fan-integral}. Thus, extracting the $\kappa^{2n}$-order piece of the above leads to
\begin{equation}
\begin{aligned}
    S_\text{wl}[\bar\gamma,x]\vert_{z^j\kappa^{2n}} \propto& \int_{\ell_1,\ldots,\ell_n,\omega_1,\ldots,\omega_j}\prod_{l=1}^jz^{\rho_l}(-\omega_l)N^{\mu\nu}_{\rho_1\cdots\rho_j}[\ell_1,\{\omega_i\}_{i=1..j}] \\
    &\hspace{1em}\times\frac{\hat\delta(\ell_1\cdot v + \omega_{1\cdots j})\hat\delta(\ell_1\cdot v)\hat\delta(\ell_2\cdot v)\cdots\hat\delta(\ell_{n}\cdot v)}{\ell_2^2\cdots\ell_n^2(\ell_1+\ell_{2\cdots n})^2}.
\end{aligned}
\end{equation}
Using $\hat\delta(\ell_1\cdot v)$ and substituting $\ell_1 \to \ell_1 - \ell_{2\cdots n}$ the expression becomes
\begin{equation}
\begin{aligned}
    \!S_\text{wl}[\bar\gamma,x]\vert_{z^j\kappa^{2n}} &\propto \int_{\omega_1,\ldots,\omega_j}\prod_{l=1}^jz^{\rho_l}(-\omega_l)\hat\delta(\omega_{1\cdots j}) \\
    &\times\int_{\ell_1,\ldots,\ell_n}\frac{\hat\delta(\ell_1\cdot v)\hat\delta(\ell_2\cdot v)\cdots\hat\delta(\ell_{n}\cdot v)N^{\mu\nu}_{\rho_1\cdots\rho_j}[\ell_1 - \ell_{2\cdots n},\{\omega_i\}_{i=1..j}]}{\ell_1^2\ell_2^2\cdots\ell_n^2}.
\end{aligned}
\end{equation}
When expanded, each piece of the above factorizes into manifestly scaleless integrals. To see this, notice from eq. \eqref{eq:scaleless} that each term in the numerator $N^{\mu\nu}_{\rho_1\cdots\rho_j}$ from the above expression will be proportional to products of the form
\begin{equation}
    \prod_{i=1}^\iota(\ell_1 - \ell_{2\cdots n})_{\rho_{s_i}},
\label{eq:generic-numerator}\end{equation}
where $\iota = j-2, j-1, j$ corresponds to terms coming from the first, second, and third term of eq. \eqref{eq:scaleless}, and $s_i = 1,\ldots,j$. When expanded, a term in eq. \eqref{eq:generic-numerator} will in general contain $1\leq R\leq\iota$ different loop momenta, meaning that it can be written
\begin{equation}
    \prod_{i=1}^R\prod_{s_i}\ell_{r_i\rho_{s_i}},
\end{equation}
where $\rho_{s_i}$ denotes the indices that $\ell_{r_i}$ has in the term. With this knowledge, one can deduce that any term in $S_\text{wl}[\bar\gamma,x]\vert_{z^j\kappa^{2n}}$ will be proportional to a product of integrals of the form
\begin{equation}
    \Bigg[\int_{\ell}\frac{\hat\delta(\ell\cdot v)}{\ell^2}\Bigg]^{n-R}\prod_{i=1}^R\int_{\ell_{r_i}}\frac{\hat\delta(\ell_{r_i}\cdot v)\prod_{s_i}\ell_{r_i\rho_{s_i}}}{\ell_{r_i}^2}.
\end{equation}
These integrals are all scaleless, so $S_\text{wl}[\bar\gamma,x]$ vanishes in dimensional regularization. Diagrammatically, this can be understood as the vanishing of all diagrams where the worldline emits and absorbs a fat graviton, e.g.,
\begin{equation}
    \begin{tikzpicture}[baseline={(current bounding box.center)}]
        \coordinate (in) at (-1.5,0);
        \coordinate (out) at (1.5,0);
        \coordinate (x1) at (-.8,0);
        \coordinate (x2) at (.8,0);
        \draw [dotted, thick] (in) -- (out);
        \draw [white] (x1) to [bend left=80] (x2);
        \draw [photon2] (x1) to [bend right=80] (x2);
        \draw [fill] (x1) circle (.08);
        \draw [fill] (x2) circle (.08);
    \end{tikzpicture}\,, \quad \begin{tikzpicture}[baseline={(current bounding box.center)}]
        \coordinate (in) at (-1.5,0);
        \coordinate (out) at (1.5,0);
        \coordinate (x1) at (-.8,0);
        \coordinate (x2) at (.8,0);
        \draw [dotted, thick] (in) -- (x1);
        \draw [dotted, thick] (x2) -- (out);
        \draw [zUndirected] (x1) -- (x2);
        \draw [white] (x1) to [bend left=80] (x2);
        \draw [photon2] (x1) to [bend right=80] (x2);
        \draw [fill] (x1) circle (.08);
        \draw [fill] (x2) circle (.08);
    \end{tikzpicture}\,, \quad \begin{tikzpicture}[baseline={(current bounding box.center)}]
        \coordinate (in) at (-1.5,0);
        \coordinate (out) at (1.5,0);
        \coordinate (x1) at (-.8,0);
        \coordinate (x2) at (.8,0);
        \draw [dotted, thick] (in) -- (x1);
        \draw [dotted, thick] (x2) -- (out);
        \draw [zUndirected] (x1) -- (x2);
        \draw [zUndirected] (x1) to [bend left=80] (x2);
        \draw [photon2] (x1) to [bend right=80] (x2);
        \draw [fill] (x1) circle (.08);
        \draw [fill] (x2) circle (.08);
    \end{tikzpicture}\,, \quad \text{etc.}
\end{equation}
Finally, after the dust has settled, one ends up simply left with the same action as in the weak-field expansion,
\begin{equation}
    S_\text{wl}[\bar g + \kappa h,x] = S_\text{wl}[\eta + \kappa h,x],
\end{equation}
meaning the Feynman rules are identical.

\subsection{Cancellation of source rule}
In this section, I discuss the aforementioned cancellation occurring between the Einstein--Hilbert and worldline actions. The simplest of the flat space rules was the source rule derived from eq. \eqref{eq:wl-source}. This equation can be rewritten in terms of the energy-momentum tensor of the background,
\begin{align}
    \bar T^{\mu\nu}(x) &= \frac{-2}{\bmetdens}\frac{\delta S_\text{wl}[\bar g,v\tau]}{\delta\bar g_{\mu\nu}(x)} \notag\\
    &= \frac{M}{\bmetdens}\int\mathrm{d}\tau\,\delta^{(d)}(x - v\tau)v^\mu v^\nu.
\end{align}
To wit,
\begin{equation}
    \kappa S_\text{wl}[h,x]\vert_{z^0} = -\frac{\kappa}{2}\int\mathrm{d}^dx\, \bmetdens \bar T^{\mu\nu}(x)h_{\mu\nu}(x).
\end{equation}
When this term is added to the linear part of the Einstein-Hilbert action from eq. \eqref{eq:eh-variation},
\begin{equation}
     S_\text{EH}[\bar g + \kappa h]\vert_{h^1} = \frac{2}{\kappa}\int\dx\bmetdens\bar G^{\mu\nu}h_{\mu\nu},
\end{equation}
where $\bar G^{\mu\nu}$ is the background Einstein tensor, Einstein's equation for the background is obtained
\begin{equation}
    S_\text{EH}[\bar g + \kappa h]\vert_{h^1} + \kappa S_\text{wl}[h,x]\vert_{z^0} = \frac{2}{\kappa}\int\mathrm{d}^dx\,\bmetdens\Big[\bar G^{\mu\nu}(x) -\frac{\kappa^2}{4}\bar T^{\mu\nu}(x)\Big]h_{\mu\nu}(x),
\end{equation}
which exactly vanishes as $\bar g$ is the Schwarzschild--Tangherlini metric. It is worth noting that the absence of a source rule in the curved expansion means it is not possible to draw diagrams containing self-energy pieces such as
\begin{equation}
    \begin{tikzpicture}[baseline={(current bounding box.center)}]
        \coordinate (in) at (-2,0);
        \coordinate (out) at (2,0);
        \coordinate (x1) at (-1.3,0);
        \coordinate (x2) at (0,0);
        \coordinate (x3) at (1.3,0);
        \coordinate (gout) at (1.3,-1);
        \draw [dotted, thick] (in) -- (out);
        \draw [zUndirected] (x2) -- (x3);
        \draw [photon] (x1) to [bend right=80] (x2);
        \draw [photon] (x3) -- (gout);
        \draw [fill] (x1) circle (.08);
        \draw [fill] (x2) circle (.08);
        \draw [fill] (x3) circle (.08);
    \end{tikzpicture}.
\end{equation}
In the weak-field expansion, one generically encounters them and has to exclude them manually.

\section{Relation between the two expansions}\label{sec:relationsbetween}
The source term in eq. \eqref{eq:wl-source} is exactly the Feynman rule describing the graviton's interaction with the background geometry. Its elimination in the expansion around the Schwarzschild--Tangherlini background in fact eliminates the class of source-containing diagrams from the expansion. One might wonder with this vertex gone how the graviton feels the background; in other words, what object takes the source vertex's place? The answer is that the graviton vertices derived from the Einstein--Hilbert action now contain this information. In the previous chapter, I derived the two-point vertex in eq. \eqref{eq:exacttwopointvertex}, however the same pattern holds true for higher-point vertices. Diagrammatically, this can be expressed as
\begin{equation}
    \begin{tikzpicture}[baseline={(current bounding box.center)}]
        \coordinate (in) at (-1,0);
        \coordinate (out) at (1,0);
        \coordinate (x) at (0,0);
        \coordinate (g1) at (-1,-1.2);
        \coordinate (g2) at (-.7,-1.2-.7);
        \coordinate (gn) at (1,-1.2);
        \coordinate (v) at (0,-1.2);
        \draw [dotted, thick] (in) -- (x);
        \draw [dotted, thick] (x) -- (out);
        \draw [photon2] (x) -- (v);
        \draw [photon] (v) -- (g1);
        \draw [photon] (v) -- (g2);
        \draw [photon] (v) -- (gn);
        \draw [loosely dotted, thick] ([shift=(-110:.7)]0,-1.2) arc (-110:-25:.7);
        \draw [fill] (v) circle (.08);
        \draw [fill] (x) circle (.08);
    \end{tikzpicture} = \begin{tikzpicture}[baseline={(current bounding box.center)}]
        \coordinate (in) at (-1,0);
        \coordinate (out) at (1,0);
        \coordinate (x) at (0,0);
        \coordinate (g1) at (-1,-1.2);
        \coordinate (g2) at (-.7,-1.2-.7);
        \coordinate (gn) at (1,-1.2);
        \coordinate (v) at (0,-1.2);
        \draw [dotted, thick] (in) -- (x);
        \draw [photon] (x) -- (v);
        \draw [dotted, thick] (x) -- (out);
        \draw [photon] (v) -- (g1);
        \draw [photon] (v) -- (g2);
        \draw [photon] (v) -- (gn);
        \draw [loosely dotted, thick] ([shift=(-110:.7)]0,-1.2) arc (-110:-25:.7);
        \draw [fill] (v) circle (.04);
        \draw [fill] (x) circle (.08);
    \end{tikzpicture} + \frac{1}{2!}\hspace{2pt}\begin{tikzpicture}[baseline={(current bounding box.center)}]
        \coordinate (in) at (-1,0);
        \coordinate (out) at (1,0);
        \coordinate (x1) at (-.6,0);
        \coordinate (x2) at (.6,0);
        \coordinate (g1) at (-1,-1.2);
        \coordinate (g2) at (-.7,-1.2-.7);
        \coordinate (gn) at (1,-1.2);
        \coordinate (v1) at (0,-.5);
        \coordinate (v2) at (0,-1.2);
        \draw [dotted, thick] (in) -- (out);
        \draw [photon] (x1) -- (v1);
        \draw [photon] (x2) -- (v1);
        \draw [photon] (v1) -- (v2);
        \draw [photon] (v2) -- (g1);
        \draw [photon] (v2) -- (g2);
        \draw [photon] (v2) -- (gn);
        \draw [loosely dotted, thick] ([shift=(-110:.7)]0,-1.2) arc (-110:-25:.7);
        \draw [fill] (v1) circle (.04);
        \draw [fill] (v2) circle (.04);
        \draw [fill] (x1) circle (.08);
        \draw [fill] (x2) circle (.08);
    \end{tikzpicture} + \frac{1}{2!}\hspace{2pt}\begin{tikzpicture}[baseline={(current bounding box.center)}]
        \coordinate (in) at (-1,0);
        \coordinate (out) at (1,0);
        \coordinate (x1) at (-.6,0);
        \coordinate (x2) at (.6,0);
        \coordinate (g1) at (-1,-1.2);
        \coordinate (g2) at (-.7,-1.2-.7);
        \coordinate (gn) at (1,-1.2);
        \coordinate (v) at (0,-1.2);
        \draw [dotted, thick] (in) -- (out);
        \draw [photon] (x1) -- (v);
        \draw [photon] (x2) -- (v);
        \draw [photon] (v) -- (g1);
        \draw [photon] (v) -- (g2);
        \draw [photon] (v) -- (gn);
        \draw [loosely dotted, thick] ([shift=(-110:.7)]0,-1.2) arc (-110:-25:.7);
        \draw [fill] (v) circle (.04);
        \draw [fill] (x1) circle (.08);
        \draw [fill] (x2) circle (.08);
    \end{tikzpicture} + \cdots.
\label{eq:se-wfe-correspondence}\end{equation}
On the left-hand side, I have put the all-order-in-$\kappa^2$ $n$-point graviton vertex coming from the gravitational action in the curved expansion, which is represented by a fat graviton connecting to the worldline to suggest the interaction with the background. On the right-hand side one has all the source-containing diagrams coming from the weak-field expansion. One could label these diagrams potential diagrams, since they represent exactly the gravitational field generated by the black hole. This also tracks with the delta functions appearing in the curved vertices, which force the graviton momenta to be spacelike (see eqs. \eqref{eq:2-pt-kappa2} and \eqref{eq:2-pt-kappa4}). Notice that due to the appearance of loop-like integrals in tree-level diagrams in worldline quantum field theory, the last two diagrams on the right-hand contain what is effectively a one-loop integral. These diagrams are of order $\kappa^4$. On the curved side, this order comes from the vertex in eq. \eqref{eq:2-pt-kappa4}, which is exactly a one-loop integral.

It is interesting to dwell briefly on the origin of the factors of a half. From the curved side, they arise simply from the Taylor expansions of the vertex. On the weak-field side, however, they are symmetry factors necessary due to the invariance of the diagram under interchange of the source vertices. In the following chapter where I compute the 2PM Compton amplitude, these factors of a half are crucial for ensuring the gauge invariance of the amplitude.

An attractive feature of this resummation is that the curved expansion graviton vertex, if connected to external momenta, will only ever contain two loop momenta in the numerator, while the corresponding $\kappa^{2n}$ weak-field expansion diagram, in general, will contain $2(n-1)$, making this effective tensor reduction of the integrand quite considerable at higher post-Minkowskian orders.
\chapter{Compton amplitude}\label{ch:compton amplitude}
This chapter contains a detailed discussion of the computation of the 1PM and 2PM classical, gravitational Compton amplitude, with the second order amplitude being the main result of the thesis. Gravitational Compton scattering is the process where a graviton scatters off a compact object, for instance a black hole. The name of the process is a reference to Compton scattering in quantum electrodynamics, which is the process where a photon scatters off an electron. The computations are carried out in worldline quantum field theory using both a flat and a Schwarzschild--Tangherlini background. For the curved calculations, I employ the expanded vertices in eq. \eqref{eq:2-pt-kappa2} and \eqref{eq:2-pt-kappa4}. In section \ref{sec:prelims}, I begin by describing the kinematics of the scattering situation and the diagrammatic expansion. Then, in section \ref{sec:1pmamp}, I discuss the curved and flat computation of the first-order Compton amplitude. Following this, in section \ref{sec:2pmintegrand}, I describe the curved and flat computation of the integrand of the second-order Compton amplitude. Section \ref{sec:technical} details how integration-by-parts identities allow a magnificent simplification of the integrand, and how one computes the so-called master integrals that remain to be integrated. Finally, in section \ref{sec:2pmamp}, I present the result and perform a couple of non-trivial consistency checks on the result.

All computations in this chapter were carried out using \texttt{xTensor} \cite{xAct}. Note also that all calculations take place in a post-Minkowskian context, so indices are raised and lowered with the Minkowski metric.

\section{Preliminary remarks}\label{sec:prelims}
The situation under study is the scattering of graviton off a black hole modeled as a point particle. The kinematics can be visualized as
\begin{equation}
    \begin{tikzpicture}[baseline={(current bounding box.center)}]
        \coordinate (in) at (-1,0);
        \coordinate (middle) at (0,0);
        \coordinate (out) at (1,0);
        \coordinate (gin) at (-1,-.93);
        \coordinate (gout) at (1,-.93);
        \draw [zUndirected] (middle) -- (in) node [left] {$Mv - \displaystyle\frac{q}{2}$} node [midway,above,sloped] {$\rightarrow$};
        \draw [zUndirected] (middle) -- (out) node [right] {$Mv + \displaystyle\frac{q}{2}$} node [midway,above,sloped] {$\rightarrow$};
        \draw [photon] (-.25,-.45) -- (gin) node [left] {$p_1$} node [midway,below,sloped] {$\rightarrow$};
        \draw [photon] (.25,-.45) -- (gout) node [right] {$p_2$} node [midway,below,sloped] {$\rightarrow$};
        \fill [darkgray] (0,-.25) circle (.4);
    \end{tikzpicture}
\end{equation}
The gray blob represents the exact Compton amplitude. The momentum of the incoming graviton is $p_1^\mu$, while the momentum of the outgoing graviton is $p_2^\mu$. The momentum of the incoming black hole is $Mv^\mu - q^\mu/2$ while the outgoing momentum is $Mv^\mu + q^\mu/2$, implying that the momentum transfer is
\begin{equation}
    q^\mu = p^\mu_2 - p^\mu_1.
\end{equation}
From the on-shell condition of the outgoing graviton,
\begin{equation}
    (p_1 + q)^2 = 0,
\end{equation}
one can derive that
\begin{equation}
    p_1\cdot q = -\frac{q^2}{2}.
\end{equation}
Meanwhile, from the on-shell condition of the incoming and outgoing black hole momenta,
\begin{equation}
    \Big(Mv - \frac{q}{2}\Big)^2 = \Big(Mv + \frac{q}{2}\Big)^2,
\end{equation}
one can see that
\begin{equation}
    v\cdot q = 0.
\end{equation}
The energy of the gravitons
\begin{equation}
    \omega = p_1\cdot v = p_2\cdot v
\end{equation}
is conserved in the process due to the conservation of energy on the worldline. The kinematics can thus be parameterized by the energy $\omega$ and the dimensionless ratio
\begin{equation}
    y = \frac{-q^2}{4\omega^2},
\label{eq:kinpardef}\end{equation}
which controls whether or not one is in the \emph{forward-scattering limit}. To see this, consider
\begin{equation}
    q^2 = -2p_{1}\cdot p_{2} = -2\omega^2(1 - \cos\theta),
\end{equation}
where $\theta$ is the angle between $p_{1\perp}^\mu$ and $p_{2\perp}^\mu$.\footnote{Recall that $V_\perp^\mu$ is the projection of $V^\mu$ onto the spacelike subspace orthogonal to $v^\mu$.} From this, the physical region is seen to be $y \in [0,1]$, with $y \to 0$ being exactly the forward-scattering limit. The external gravitons have physical polarization tensors $\varepsilon_{1\mu\nu}$ and $\varepsilon_{2\mu\nu}$, which satisfy
\begin{equation}
    \varepsilon_{i\mu\nu} = \varepsilon_{i\nu\mu}, \qquad p_i^\mu\varepsilon_{i\mu\nu} = 0, \qquad \varepsilon\indices{_{i\mu}^\mu} = 0.
\end{equation}
They can each be factorized into two polarization vectors
\begin{equation}
    \varepsilon_{i\mu\nu} = \varepsilon_{i\mu}\varepsilon_{i\nu},
\end{equation}
In worldline quantum field theory, the velocity that appears in the Feynman rules is the average of the one at the asymptotic past and future. By parameterizing the momenta of the black hole as I have done, this average velocity is simply $v^\mu$. This fact comes from the i0 prescription for the deflection propagator (for more details, see \cite{Mogull:2020sak}).

In the curved expansion, the diagrammatic expansion of the Compton amplitude takes the structured form
\begin{equation}\label{eq:ComptonGen}
\begin{aligned}
    \begin{tikzpicture}[baseline={(current bounding box.center)}]
        \coordinate (in) at (-1,0);
        \coordinate (out) at (1,0);
        \coordinate (gin) at (-1,-.93);
        \coordinate (gout) at (1,-.93);
        \draw [dotted, thick] (in) -- (out);
        \draw [photon] (gin) -- (-.25,-.45);
        \draw [photon] (gout) -- (.25,-.45);
        \fill [darkgray] (0,-.25) circle (.4);
    \end{tikzpicture} &= \begin{tikzpicture}[baseline={(current bounding box.center)}]
        \coordinate (in) at (0,0);
        \coordinate (out) at (1.2,0);
        \coordinate (x) at (.6,0);
        \coordinate (gin) at (0,-.8);
        \coordinate (gout) at (1.2,-.8);
        \coordinate (v) at (.6,-.8);
        \draw [dotted, thick] (in) -- (x);
        \draw [dotted, thick] (x) -- (out);
        \draw [photon2] (x) -- (v);
        \draw [photon] (v) -- (gin);
        \draw [photon] (v) -- (gout);
        \draw [fill] (v) circle (.08);
        \draw [fill] (x) circle (.08);
    \end{tikzpicture} + \begin{tikzpicture}[baseline={(current bounding box.center)}]
        \coordinate (in) at (0,0);
        \coordinate (out) at (1.8,0);
        \coordinate (x1) at (0.6,0);
        \coordinate (x2) at (1.2,0);
        \coordinate (gin) at (0,-.8);
        \coordinate (gout) at (1.8,-.8);
        \coordinate (v1) at (0.6,-.8);
        \coordinate (v2) at (1.2,-.8);
        \draw [dotted, thick] (in) -- (out);
        \draw [photon2] (x1) -- (v1);
        \draw [photon2] (x2) -- (v2);
        \draw [photon] (v1) -- (gin);
        \draw [photon] (v1) -- (v2);
        \draw [photon] (v2) -- (gout);
        \draw [fill] (v1) circle (.08);
        \draw [fill] (x1) circle (.08);
        \draw [fill] (v2) circle (.08);
        \draw [fill] (x2) circle (.08);
    \end{tikzpicture} + \cdots \\
    &+ \begin{tikzpicture}[baseline={(current bounding box.center)}]
        \coordinate (in) at (0,0);
        \coordinate (out) at (1.2,0);
        \coordinate (x1) at (0.4,0);
        \coordinate (x2) at (0.8,0);
        \coordinate (gin) at (0,-.8);
        \coordinate (gout) at (1.2,-.8);
        \draw [dotted, thick] (in) -- (x1);
        \draw [dotted, thick] (x2) -- (out);
        \draw [zUndirected] (x1) -- (x2);
        \draw [photon] (x1) -- (gin);
        \draw [photon] (x2) -- (gout);
        \draw [fill] (x1) circle (.08);
        \draw [fill] (x2) circle (.08);
    \end{tikzpicture} + \Bigg(\begin{tikzpicture}[baseline={(current bounding box.center)}]
        \coordinate (in) at (0,0);
        \coordinate (out) at (1.8,0);
        \coordinate (x1) at (0.4,0);
        \coordinate (x2) at (0.8,0);
        \coordinate (x3) at (1.2,0);
        \coordinate (v2) at (1.2,-.8);
        \coordinate (gin) at (0,-.8);
        \coordinate (gout) at (1.8,-.8);
        \draw [dotted, thick] (in) -- (x1);
        \draw [dotted, thick] (x2) -- (out);
        \draw [zUndirected] (x1) -- (x2);
        \draw [photon] (x1) -- (gin);
        \draw [photon] (x2) -- (v2);
        \draw [photon2] (x3) -- (v2);
        \draw [photon] (v2) -- (gout);
        \draw [fill] (x1) circle (.08);
        \draw [fill] (x2) circle (.08);
        \draw [fill] (v2) circle (.08);
        \draw [fill] (x3) circle (.08);
    \end{tikzpicture} + \text{perms}\Bigg) + \cdots \\
    &+ \begin{tikzpicture}[baseline={(current bounding box.center)}]
        \coordinate (in) at (0,0);
        \coordinate (out) at (2.2,0);
        \coordinate (x1) at (0.4,0);
        \coordinate (x2) at (0.8,0);
        \coordinate (x3) at (1.4,0);
        \coordinate (x4) at (1.8,0);
        \coordinate (gin) at (0,-.8);
        \coordinate (gout) at (2.2,-.8);
        \draw [dotted, thick] (in) -- (x1);
        \draw [dotted, thick] (x2) -- (out);
        \draw [zUndirected] (x1) -- (x2);
        \draw [photon] (x1) -- (gin);
        \draw [photon] (x2) to [bend right=60] (x3);
        \draw [zUndirected] (x3) -- (x4);
        \draw [photon] (x4) -- (gout);
        \draw [fill] (x1) circle (.08);
        \draw [fill] (x2) circle (.08);
        \draw [fill] (x3) circle (.08);
        \draw [fill] (x4) circle (.08);
    \end{tikzpicture} + \Bigg(\begin{tikzpicture}[baseline={(current bounding box.center)}]
        \coordinate (in) at (0,0);
        \coordinate (out) at (2.8,0);
        \coordinate (x1) at (0.4,0);
        \coordinate (x2) at (0.8,0);
        \coordinate (x3) at (1.4,0);
        \coordinate (x4) at (1.8,0);
        \coordinate (gin) at (0,-.8);
        \coordinate (v) at (2.2,-.8);
        \coordinate (x5) at (2.2,0);
        \coordinate (gout) at (2.8,-.8);
        \draw [dotted, thick] (in) -- (x1);
        \draw [dotted, thick] (x2) -- (out);
        \draw [zUndirected] (x1) -- (x2);
        \draw [photon] (x1) -- (gin);
        \draw [photon] (x2) to [bend right=60] (x3);
        \draw [zUndirected] (x3) -- (x4);
        \draw [photon] (x4) -- (v);
        \draw [photon2] (x5) -- (v);
        \draw [fill] (v) circle (.08);
        \draw [photon] (v) -- (gout);
        \draw [fill] (x1) circle (.08);
        \draw [fill] (x2) circle (.08);
        \draw [fill] (x3) circle (.08);
        \draw [fill] (x4) circle (.08);
        \draw [fill] (x5) circle (.08);
    \end{tikzpicture} + \text{perms}\Bigg) + \cdots \\
    &+ \cdots.
\end{aligned}
\end{equation}
The gray blob again represents the exact Compton amplitude, and `perms' denotes all possible permutations of background injections and deflections. In the weak-field expansion, the diagrammatics is more involved due to the presence of the arbitrary-multiplicity vertices. Expanding the Compton amplitude one gets
\begin{equation}
\begin{aligned}
    \begin{tikzpicture}[baseline={(current bounding box.center)}]
        \coordinate (in) at (-1,0);
        \coordinate (out) at (1,0);
        \coordinate (gin) at (-1,-.93);
        \coordinate (gout) at (1,-.93);
        \draw [dotted, thick] (in) -- (out);
        \draw [photon] (gin) -- (-.25,-.45);
        \draw [photon] (gout) -- (.25,-.45);
        \fill [darkgray] (0,-.25) circle (.4);
    \end{tikzpicture} &= \begin{tikzpicture}[baseline={(current bounding box.center)}]
        \coordinate (in) at (0,0);
        \coordinate (out) at (1.2,0);
        \coordinate (x) at (.6,0);
        \coordinate (gin) at (0,-.8);
        \coordinate (gout) at (1.2,-.8);
        \coordinate (v) at (.6,-.8);
        \draw [dotted, thick] (in) -- (x);
        \draw [dotted, thick] (x) -- (out);
        \draw [photon] (x) -- (v);
        \draw [photon] (v) -- (gin);
        \draw [photon] (v) -- (gout);
        \draw [fill] (v) circle (.04);
        \draw [fill] (x) circle (.08);
    \end{tikzpicture} + \begin{tikzpicture}[baseline={(current bounding box.center)}]
        \coordinate (in) at (0,0);
        \coordinate (out) at (1.2,0);
        \coordinate (x1) at (0.4,0);
        \coordinate (x2) at (0.8,0);
        \coordinate (gin) at (0,-.8);
        \coordinate (gout) at (1.2,-.8);
        \draw [dotted, thick] (in) -- (x1);
        \draw [dotted, thick] (x2) -- (out);
        \draw [zUndirected] (x1) -- (x2);
        \draw [photon] (x1) -- (gin);
        \draw [photon] (x2) -- (gout);
        \draw [fill] (x1) circle (.08);
        \draw [fill] (x2) circle (.08);
    \end{tikzpicture} \\
    &+ \begin{tikzpicture}[baseline={(current bounding box.center)}]
        \coordinate (in) at (0,0);
        \coordinate (out) at (2.2,0);
        \coordinate (x1) at (0.4,0);
        \coordinate (x2) at (0.8,0);
        \coordinate (x3) at (1.4,0);
        \coordinate (x4) at (1.8,0);
        \coordinate (gin) at (0,-.8);
        \coordinate (gout) at (2.2,-.8);
        \draw [dotted, thick] (in) -- (x1);
        \draw [dotted, thick] (x2) -- (out);
        \draw [zUndirected] (x1) -- (x2);
        \draw [photon] (x1) -- (gin);
        \draw [photon] (x2) to [bend right=60] (x3);
        \draw [zUndirected] (x3) -- (x4);
        \draw [photon] (x4) -- (gout);
        \draw [fill] (x1) circle (.08);
        \draw [fill] (x2) circle (.08);
        \draw [fill] (x3) circle (.08);
        \draw [fill] (x4) circle (.08);
    \end{tikzpicture} + \Bigg(\begin{tikzpicture}[baseline={(current bounding box.center)}]
        \coordinate (in) at (0,0);
        \coordinate (out) at (1.8,0);
        \coordinate (x1) at (0.4,0);
        \coordinate (x2) at (0.8,0);
        \coordinate (x3) at (1.2,0);
        \coordinate (v2) at (1.2,-.8);
        \coordinate (gin) at (0,-.8);
        \coordinate (gout) at (1.8,-.8);
        \draw [dotted, thick] (in) -- (x1);
        \draw [dotted, thick] (x2) -- (out);
        \draw [zUndirected] (x1) -- (x2);
        \draw [photon] (x1) -- (gin);
        \draw [photon] (x2) -- (v2);
        \draw [photon] (x3) -- (v2);
        \draw [photon] (v2) -- (gout);
        \draw [fill] (x1) circle (.08);
        \draw [fill] (x2) circle (.08);
        \draw [fill] (v2) circle (.04);
        \draw [fill] (x3) circle (.08);
    \end{tikzpicture} + \text{perms}\Bigg) \\
    &+ \frac12\Bigg(\begin{tikzpicture}[baseline={(current bounding box.center)}]
        \coordinate (in) at (0,0);
        \coordinate (out) at (1.8,0);
        \coordinate (x1) at (0.6,0);
        \coordinate (x2) at (1.2,0);
        \coordinate (gin) at (0,-.8);
        \coordinate (gout) at (1.8,-.8);
        \coordinate (v1) at (.9,-.4);
        \coordinate (v2) at (.9,-.8);
        \draw [dotted, thick] (in) -- (out);
        \draw [photon] (x1) -- (v1);
        \draw [photon] (x2) -- (v1);
        \draw [photon] (v1) -- (v2);
        \draw [photon] (v2) -- (gin);
        \draw [photon] (v2) -- (gout);
        \draw [fill] (v1) circle (.04);
        \draw [fill] (v2) circle (.04);
        \draw [fill] (x1) circle (.08);
        \draw [fill] (x2) circle (.08);
    \end{tikzpicture} + \begin{tikzpicture}[baseline={(current bounding box.center)}]
        \coordinate (in) at (0,0);
        \coordinate (out) at (1.8,0);
        \coordinate (x1) at (0.6,0);
        \coordinate (x2) at (1.2,0);
        \coordinate (gin) at (0,-.8);
        \coordinate (gout) at (1.8,-.8);
        \coordinate (v) at (.9,-.8);
        \draw [dotted, thick] (in) -- (out);
        \draw [photon] (x1) -- (v);
        \draw [photon] (x2) -- (v);
        \draw [photon] (v2) -- (gin);
        \draw [photon] (v2) -- (gout);
        \draw [fill] (v) circle (.04);
        \draw [fill] (x1) circle (.08);
        \draw [fill] (x2) circle (.08);
    \end{tikzpicture}\Bigg) + \begin{tikzpicture}[baseline={(current bounding box.center)}]
        \coordinate (in) at (0,0);
        \coordinate (out) at (1.8,0);
        \coordinate (x1) at (0.6,0);
        \coordinate (x2) at (1.2,0);
        \coordinate (gin) at (0,-.8);
        \coordinate (gout) at (1.8,-.8);
        \coordinate (v1) at (0.6,-.8);
        \coordinate (v2) at (1.2,-.8);
        \draw [dotted, thick] (in) -- (out);
        \draw [photon] (x1) -- (v1);
        \draw [photon] (x2) -- (v2);
        \draw [photon] (v1) -- (gin);
        \draw [photon] (v1) -- (v2);
        \draw [photon] (v2) -- (gout);
        \draw [fill] (v1) circle (.04);
        \draw [fill] (x1) circle (.08);
        \draw [fill] (v2) circle (.04);
        \draw [fill] (x2) circle (.08);
    \end{tikzpicture} \\
    &+ \cdots.
\end{aligned}
\label{eq:flat2pm}\end{equation}
Here, each diagram has a well-defined power of $\kappa^2$, with the first line being order $\kappa^2$ and the remaining depicted diagrams being order $\kappa^4$. As I stressed in the previous chapter, the symmetry factor of two comes from the two identical sources connecting to the same vertex.

\section{First-order amplitude}\label{sec:1pmamp}
In this section I discuss the computation of the gravitational Compton amplitude at first post-Minkowskian order. First, the computation is completed using the curved expansion. Afterwards, I compute it with the weak-field rules. In the end, the result ends up being the same, as one would expect. The curved version of this computation was first reported in \cite{Kosmopoulos:2023bwc}. At this low order, both computations involve two diagrams of similar complexity.

\subsection{Curved computation}\label{sec:curvedcomp}
The first diagram involves the vertex from eq. \eqref{eq:2-pt-kappa2} and looks like
\begin{equation}
    \begin{tikzpicture}[baseline={(current bounding box.center)}]
        \coordinate (in) at (-1,0);
        \coordinate (out) at (1,0);
        \coordinate (x) at (0,0);
        \coordinate (gin) at (-1,-1.2);
        \coordinate (gout) at (1,-1.2);
        \coordinate (v) at (0,-1.2);
        \draw [dotted, thick] (in) -- (x);
        \draw [dotted, thick] (x) -- (out);
        \draw [photon2] (x) -- (v) node [midway, left, black] {$q\!\downarrow$};
        \draw [photon] (v) -- (gin) node [left] {$p_1$} node [midway, below=.3em] {$\rightarrow$};
        \draw [photon] (v) -- (gout) node [right] {$p_2$} node [midway, below=.3em] {$\rightarrow$};
        \draw [fill] (v) circle (.14) node {\color{white}\tiny$1$};
        \draw [fill] (x) circle (.08);
    \end{tikzpicture} = \ii\hat\delta(q\cdot v)\frac{\kappa^2M}{q^2}\varepsilon_{1\mu_1\nu_1}\varepsilon_{2\mu_2\nu_2}\bar N_{(1)}^{\mu_1\nu_1\,\mu_2\nu_2}.
\label{eq:1pm-2pt}\end{equation}
Recall that $N_{(1)}^{\mu_1\nu_1\,\mu_2\nu_2}$ was the numerator factor that appeared when expanding the background metric in the Einstein--Hilbert action. The second diagram involves a deflection mode traveling on the worldline,
\begin{equation}
\begin{aligned}
    &\begin{tikzpicture}[baseline={(current bounding box.center)}]
        \coordinate (in) at (-1.3,0);
        \coordinate (out) at (1.3,0);
        \coordinate (x1) at (-.6,0);
        \coordinate (x2) at (.6,0);
        \coordinate (gin) at (-.6,-1);
        \coordinate (gout) at (.6,-1);
        \draw [dotted, thick] (in) -- (x1);
        \draw [dotted, thick] (x2) -- (out);
        \draw [zUndirected] (x1) -- (x2) node [midway, above] {$\stackrel{\displaystyle\omega}{\rightarrow}$};
        \draw [photon] (x1) -- (gin) node [below] {$p_1$} node [midway, left=.3em] {$\uparrow$};
        \draw [photon] (x2) -- (gout) node [below] {$p_2$} node [midway, right=.3em] {$\downarrow$};;
        \draw [fill] (x1) circle (.08);
        \draw [fill] (x2) circle (.08);
    \end{tikzpicture} \\
    &= \ii\hat\delta(q\cdot v)\frac{\kappa^2M}{4\omega^2}\varepsilon_{1\mu_1\nu_1}\varepsilon_{2\mu_2\nu_2}(v^{\mu_1}v^{\nu_1}p_1^\rho - 2\omega v\indices{^{(\mu_1}}\eta^{\nu_1)\rho})(v^{\mu_2}v^{\nu_2}p_{2\rho} - 2\omega v\indices{^{(\mu_2}}\delta\indices*{^{\nu_2)}_\rho}).
\end{aligned}
\label{eq:1pm-deflection}\end{equation}
Here, I used the integration over the deflection energy and the delta function from one of the vertices to fix the energy flowing through the worldline to $\omega = p_1\cdot v$. In summing these contributions, all dependence on $d$ cancels automatically, meaning the result is independent of the dimension. If the polarization tensors are factorized, the result simplifies to,
\begin{equation}
    \ii\hat\delta(q\cdot v)\mathcal{M}^\text{1PM}(p_1, p_2) = -\ii\hat\delta(q\cdot v)\frac{\kappa^2M}{2}\frac{(v\cdot f_1\cdot f_2\cdot v)^2}{q^2\omega^2},
\label{eq:1pm-compton}\end{equation}
which is expressed in terms of the momentum space field strength tensor
\begin{equation}
    f_{i\mu\nu} = p_{i\mu}\varepsilon_{i\nu} - p_{i\nu}\varepsilon_{i\mu}
\label{eq:fieldstrengthtensor}\end{equation}
of the incoming and outgoing graviton and the notation $(V\cdot f_i)^\mu = V^\nu f\indices{_{i\nu}^\mu}$. As was noted in \cite{Kosmopoulos:2023bwc}, the normalization of the amplitude in eq. \eqref{eq:1pm-compton} differs by a factor of $2M$ from the conventional result
\begin{equation}
    \widetilde{\mathcal{M}}^\text{1PM}(p_1,p_2) = -\kappa^2M^2\frac{(v\cdot f_1\cdot f_2\cdot v)^2}{q^2\omega^2}
\end{equation}
(see e.g. \cite{Brandhuber:2021eyq}). This is a purely definitional issue arising from the relation
\begin{equation}
    \ii\hat\delta(2q\cdot Mv)\widetilde{\mathcal{M}}^\text{1PM}(p_1,p_2) = \ii\hat\delta(q\cdot v)\mathcal{M}^\text{1PM}(p_1, p_2).
\end{equation}
In other words, it is a question of the normalization of the argument of the momentum-conserving delta function.

\subsection{Flat computation}
Using the weak-field rules, there is instead of the diagram in eq. \eqref{eq:1pm-2pt} a diagram involving the cubic vertex. It is
\begin{equation}
    \begin{tikzpicture}[baseline={(current bounding box.center)}]
        \coordinate (in) at (-1,0);
        \coordinate (out) at (1,0);
        \coordinate (x) at (0,0);
        \coordinate (gin) at (-1,-1.2);
        \coordinate (gout) at (1,-1.2);
        \coordinate (v) at (0,-1.2);
        \draw [dotted, thick] (in) -- (x);
        \draw [dotted, thick] (x) -- (out);
        \draw [photon] (x) -- (v) node [midway, left] {$q\!\downarrow$};
        \draw [photon] (v) -- (gin) node [left] {$p_1$} node [midway, below=.3em] {$\rightarrow$};
        \draw [photon] (v) -- (gout) node [right] {$p_2$} node [midway, below=.3em] {$\rightarrow$};;
        \draw [fill] (v) circle (.04);
        \draw [fill] (x) circle (.08);
    \end{tikzpicture} = \ii\hat\delta(q\cdot v)\frac{\kappa^2M}{2}\varepsilon_{1\mu_1\nu_1}\varepsilon_{2\mu_2\nu_2}V^{\mu_1\nu_1\,\rho\sigma\,\mu_2\nu_2}(-p_1,-q, p_2)\frac{P_{\rho\sigma\,\gamma\delta}}{q^2}v^\gamma v^\delta,
\label{eq:1pm-3pt}\end{equation}
where the integral over the momentum emitted from the worldline and the delta function from the three-point vertex was used to fix the injected momentum $q = p_2 - p_1$. Then, as the deflection diagram is the same in both expansions, the weak-field calculation the Compton amplitude is
\begin{equation}
    \ii\hat\delta(q\cdot v)\mathcal{M}^\text{1PM}(p_1, p_2) = \eqref{eq:1pm-3pt} + \eqref{eq:1pm-deflection},
\end{equation}
which, as indicated, coincides with the result obtained from the curved expansion. For this to be true the diagram in eq. \eqref{eq:1pm-deflection} and in eq. \eqref{eq:1pm-3pt} must be equal, and I find indeed that this is the case. This equality is the two-point version of the correspondence depicted in eq. \eqref{eq:se-wfe-correspondence} at order $\kappa^2$.

\section{Second-order integrand}\label{sec:2pmintegrand}
In this section, the computation of the classical Compton amplitude at second post-Minkowskian order is discussed. This is the lowest order that experiences a resummation of potential graviton diagrams, as two flat space diagrams combine into one curved,
\begin{equation}
    \begin{tikzpicture}[baseline={(current bounding box.center)}]
        \coordinate (in) at (-1,0);
        \coordinate (out) at (1,0);
        \coordinate (x1) at (-.6,0);
        \coordinate (x2) at (.6,0);
        \coordinate (g1) at (-1,-1.2);
        \coordinate (g2) at (1,-1.2);
        \coordinate (v1) at (0,-.5);
        \coordinate (v2) at (0,-1.2);
        \draw [dotted, thick] (in) -- (out);
        \draw [photon] (x1) -- (v1);
        \draw [photon] (x2) -- (v1);
        \draw [photon] (v1) -- (v2);
        \draw [photon] (v2) -- (g1);
        \draw [photon] (v2) -- (g2);
        \draw [fill] (v1) circle (.04);
        \draw [fill] (v2) circle (.04);
        \draw [fill] (x1) circle (.08);
        \draw [fill] (x2) circle (.08);
    \end{tikzpicture} \quad\text{and}\quad \begin{tikzpicture}[baseline={(current bounding box.center)}]
        \coordinate (in) at (-1,0);
        \coordinate (out) at (1,0);
        \coordinate (x1) at (-.6,0);
        \coordinate (x2) at (.6,0);
        \coordinate (g1) at (-1,-1.2);
        \coordinate (g2) at (1,-1.2);
        \coordinate (v) at (0,-1.2);
        \draw [dotted, thick] (in) -- (out);
        \draw [photon] (x1) -- (v);
        \draw [photon] (x2) -- (v);
        \draw [photon] (v2) -- (g1);
        \draw [photon] (v2) -- (g2);
        \draw [fill] (v) circle (.04);
        \draw [fill] (x1) circle (.08);
        \draw [fill] (x2) circle (.08);
    \end{tikzpicture} \qquad\rightarrow\qquad \begin{tikzpicture}[baseline={(current bounding box.center)}]
        \coordinate (in) at (-1,0);
        \coordinate (out) at (1,0);
        \coordinate (x) at (0,0);
        \coordinate (gin) at (-1,-1.2);
        \coordinate (gout) at (1,-1.2);
        \coordinate (v) at (0,-1.2);
        \draw [dotted, thick] (in) -- (x);
        \draw [dotted, thick] (x) -- (out);
        \draw [photon2] (x) -- (v);
        \draw [photon] (v) -- (gin);
        \draw [photon] (v) -- (gout);
        \draw [fill] (v) circle (.14) node {\color{white}\tiny$2$};
        \draw [fill] (x) circle (.08);
    \end{tikzpicture}.
\end{equation}
I obtain the integrand using the curved expansion first, after which I obtain it with flat space diagrams.

\subsection{Curved computation}
The curved diagrammatic expansion of the Compton amplitude was given in eq. \eqref{eq:ComptonGen}. Similarly to the 1PM case, one must use the expanded vertex rules from eqs. \eqref{eq:2-pt-kappa2} and \eqref{eq:2-pt-kappa4} to obtain the 2PM order piece of this. Five diagrams emerge through this procedure that I label with (i)--(v). These contain no non-trivial integrations over the deflection energy, as it is always forced to be the graviton energy $\omega$. In addition, the diagrams always contain a delta function $\hat\delta(\ell\cdot v)$ forcing the energy of the loop momentum to vanish. I will carefully go through the manipulations needed to see this for the first diagram, which is of a radiation-reaction type:
\begin{align}
    &\begin{tikzpicture}[baseline={(current bounding box.center)}]
        \coordinate (in) at (-2,0);
        \coordinate (out) at (2,0);
        \coordinate (x1) at (-1.5,0);
        \coordinate (x2) at (-.5,0);
        \coordinate (x3) at (.5,0);
        \coordinate (x4) at (1.5,0);
        \coordinate (gin) at (-1.5,-1);
        \coordinate (gout) at (1.5,-1);
        \draw [dotted, thick] (in) -- (x1);
        \draw [dotted, thick] (x2) -- (x3);
        \draw [dotted, thick] (x4) -- (out);
        \draw [zUndirected] (x1) -- (x2) node [midway, above] {$\stackrel{\displaystyle\lambda_1}{\rightarrow}$};
        \draw [zUndirected] (x3) -- (x4) node [midway, above] {$\stackrel{\displaystyle\lambda_2}{\rightarrow}$};
        \draw [photon] (x1) -- (gin) node [below] {$p_1$} node [midway, left=.3em] {$\uparrow$};
        \draw [photon] (x2) to [bend right=80] (x3) node [midway, below=.8em] {$\stackrel{\displaystyle\rightarrow}{p_1 + \ell}$};
        \draw [photon] (x4) -- (gout) node [below] {$p_2$} node [midway, right=.3em] {$\downarrow$};
        \draw [fill] (x1) circle (.08);
        \draw [fill] (x2) circle (.08);
        \draw [fill] (x3) circle (.08);
        \draw [fill] (x4) circle (.08);
    \end{tikzpicture} \notag\\
    &= -\ii\frac{\kappa^4M^2}{16}\int_{\ell,\lambda_1,\lambda_2}\hat\delta(\lambda_1-p_1\cdot v)\hat\delta(p_1\cdot v + \ell\cdot v - \lambda_1)\hat\delta(\lambda_2 - p_1\cdot v - \ell\cdot v)\hat\delta(p_2\cdot v - \lambda_2) \notag\\
    &\hspace{3.3cm}\times\varepsilon_{1\mu_1\nu_1}V\indices{^{\mu_1\nu_1}_\rho_1}(-p_1,\lambda_1)\frac{\eta^{\rho\sigma}}{\lambda_1^2}V\indices{^{\alpha\beta}_\sigma}(p_1+\ell,-\lambda_1)\frac{P_{\alpha\beta\alpha'\beta'}}{(p_1 + \ell)^2 + \ii0} \notag\\
    &\hspace{3.3cm}\times V\indices{^{\alpha'\beta'}_{\rho'}}(-p_1-\ell,\lambda_2)\frac{\eta^{\rho'\sigma'}}{\lambda_2^2}V\indices{^{\mu_2\nu_2}_{\sigma'}}(p_2,-\lambda_2)\varepsilon_{2\mu_2\nu_2}.
\end{align}
Here $V\indices{^{\mu\nu}_{\rho}}(k,\lambda)$ is the worldline vertex from eq. \eqref{eq:lindeflecvertex}. The momentum flowing through the graviton line is parameterized as $p_1 + \ell$ such that the remaining integral conforms to what is obtained from the other diagrams. I labeled the energies flowing through the two deflection lines $\lambda_1$ and $\lambda_2$; the integral over $\lambda_1$ fixes the energy of the first deflection mode to be $p_1\cdot v$, while the one over $\lambda_2$ fixes the second deflection mode to have energy $p_2\cdot v$. Doing these trivial integrals simplifies the diagram to
\begin{align}
    &\begin{tikzpicture}[baseline={(current bounding box.center)}]
        \coordinate (in) at (-2,0);
        \coordinate (out) at (2,0);
        \coordinate (x1) at (-1.5,0);
        \coordinate (x2) at (-.5,0);
        \coordinate (x3) at (.5,0);
        \coordinate (x4) at (1.5,0);
        \coordinate (gin) at (-1.5,-1);
        \coordinate (gout) at (1.5,-1);
        \draw [dotted, thick] (in) -- (x1);
        \draw [dotted, thick] (x2) -- (x3);
        \draw [dotted, thick] (x4) -- (out);
        \draw [zUndirected] (x1) -- (x2) node [midway, above] {$\stackrel{\displaystyle p_1\cdot v}{\rightarrow}$};
        \draw [zUndirected] (x3) -- (x4) node [midway, above] {$\stackrel{\displaystyle p_2\cdot v}{\rightarrow}$};
        \draw [photon] (x1) -- (gin) node [below] {$p_1$} node [midway, left=.3em] {$\uparrow$};
        \draw [photon] (x2) to [bend right=80] (x3) node [midway, below=.8em] {$\stackrel{\displaystyle\rightarrow}{p_1 + \ell}$};
        \draw [photon] (x4) -- (gout) node [below] {$p_2$} node [midway, right=.3em] {$\downarrow$};
        \draw [fill] (x1) circle (.08);
        \draw [fill] (x2) circle (.08);
        \draw [fill] (x3) circle (.08);
        \draw [fill] (x4) circle (.08);
    \end{tikzpicture} \notag\\
    &= -\ii\hat\delta(p_2\cdot v - p_1\cdot v)\frac{\kappa^4M^2}{16}\int_\ell\hat\delta(\ell\cdot v)\varepsilon_{1\mu_1\nu_1}V\indices{^{\mu_1\nu_1}_\rho_1}(-p_1,p_1\cdot v)\frac{\eta^{\rho\sigma}}{(p_1\cdot v)^2} \notag\\
    &\hspace{5.2cm}\times V\indices{^{\alpha\beta}_\sigma}(p_1+\ell,-p_1\cdot v)\frac{P_{\alpha\beta\alpha'\beta'}}{(p_1 + \ell)^2 + \ii0}V\indices{^{\alpha'\beta'}_{\rho'}}(-p_1-\ell,p_2\cdot v) \notag\\
    &\hspace{5.2cm}\times \frac{\eta^{\rho'\sigma'}}{(p_2\cdot v)^2}V\indices{^{\mu_2\nu_2}_{\sigma'}}(p_2,-p_2\cdot v)\varepsilon_{2\mu_2\nu_2}.
\end{align}
I pulled out the delta function that ensures overall energy conservation. Using the definition of the momentum transfer,
\begin{equation}
    \hat\delta(p_2\cdot v - p_1\cdot v) = \hat\delta(q\cdot v).
\end{equation}
The Feynman rules thus ensure that $p_1\cdot v = p_2\cdot v = \omega$ as is expected. The pattern of fixed energy $\omega$ flowing through the deflection mode and the overall factor $\hat\delta(q\cdot v)$ repeats itself for all the diagrams, so I can label all the energies of the deflection lines with $\omega$ from the outset. Defining the diagram numerator as
\begin{align}
    N_\text{i}(\varepsilon_1,\varepsilon_2,p_1,p_2,v,\ell) &= -\frac{1}{16}\varepsilon_{1\mu_1\nu_1}V\indices{^{\mu_1\nu_1}_\rho_1}(-p_1,p_1\cdot v)\eta^{\rho\sigma}V\indices{^{\alpha\beta}_\sigma}(p_1+\ell,-p_1\cdot v)P_{\alpha\beta\alpha'\beta'} \notag\\
    &~~~~\times V\indices{^{\alpha'\beta'}_{\rho'}}(-p_1-\ell,p_2\cdot v)\eta^{\rho'\sigma'}V\indices{^{\mu_2\nu_2}_{\sigma'}}(p_2,-p_2\cdot v)\varepsilon_{2\mu_2\nu_2},
\end{align}
the diagram can be reduced to the simple form
\begin{equation}
    \begin{tikzpicture}[baseline={(current bounding box.center)}]
        \coordinate (in) at (-2,0);
        \coordinate (out) at (2,0);
        \coordinate (x1) at (-1.5,0);
        \coordinate (x2) at (-.5,0);
        \coordinate (x3) at (.5,0);
        \coordinate (x4) at (1.5,0);
        \coordinate (gin) at (-1.5,-1);
        \coordinate (gout) at (1.5,-1);
        \draw [dotted, thick] (in) -- (x1);
        \draw [dotted, thick] (x2) -- (x3);
        \draw [dotted, thick] (x4) -- (out);
        \draw [zUndirected] (x1) -- (x2) node [midway, above] {$\stackrel{\displaystyle\omega}{\rightarrow}$};
        \draw [zUndirected] (x3) -- (x4) node [midway, above] {$\stackrel{\displaystyle\omega}{\rightarrow}$};
        \draw [photon] (x1) -- (gin) node [below] {$p_1$} node [midway, left=.3em] {$\uparrow$};
        \draw [photon] (x2) to [bend right=80] (x3) node [midway, below=.8em] {$\stackrel{\displaystyle\rightarrow}{p_1 + \ell}$};
        \draw [photon] (x4) -- (gout) node [below] {$p_2$} node [midway, right=.3em] {$\downarrow$};
        \draw [fill] (x1) circle (.08);
        \draw [fill] (x2) circle (.08);
        \draw [fill] (x3) circle (.08);
        \draw [fill] (x4) circle (.08);
    \end{tikzpicture} = \ii\hat\delta(q\cdot v)\frac{\kappa^4M^2}{\omega^4}\int_\ell\frac{\hat\delta(\ell\cdot v)N_\text{i}(\varepsilon_1,\varepsilon_2,p_1,p_2,v,\ell)}{(p_1 + \ell)^2 + \ii0}.
\end{equation}
Following similar manipulations, the next diagram---which involves the Feynman rule from eq. \eqref{eq:2-pt-kappa2}---can be brought into the form
\begin{equation}
    \begin{tikzpicture}[baseline={(current bounding box.center)}]
        \coordinate (in) at (-1.5,0);
        \coordinate (out) at (2.5,0);
        \coordinate (x1) at (-1,0);
        \coordinate (y1) at (-.25,-.8);
        \coordinate (x2) at (.5,0);
        \coordinate (x3) at (2,0);
        \coordinate (gin) at (-.25,-1.4);
        \coordinate (gout) at (2,-1.4);
        \draw [dotted, thick] (in) -- (x2);
        \draw [dotted, thick] (x4) -- (out);
        \draw [zUndirected] (x2) -- (x3) node [midway, above] {$\stackrel{\displaystyle\omega}{\rightarrow}$};
        \draw [photon] (y1) -- (gin) node [below] {$p_1$} node [midway, left=.3em] {$\uparrow$};
        \draw [photon2] (x1) -- (y1) node [midway, left, black] {$\ell\searrow$};
        \draw [photon] (y1) -- (x2) node [midway, right] {$\nearrow p_1+\ell$};
        \draw [photon] (x3) -- (gout) node [below] {$p_2$} node [midway, right=.3em] {$\downarrow$};
        \draw [fill] (x1) circle (.08);
        \draw [fill] (x2) circle (.08);
        \draw [fill] (x3) circle (.08);
        \draw [fill] (y1) circle (.14) node {\color{white}\tiny$1$};
    \end{tikzpicture} = \ii\hat\delta(q\cdot v)\frac{\kappa^4M^2}{\omega^2}\int_\ell\frac{\hat\delta(\ell\cdot v)N_\text{ii}(\varepsilon_1,\varepsilon_2,p_1,p_2,v,\ell)}{\ell^2[(p_1 + \ell)^2+\ii0]}.
\end{equation}
A mirrored version of diagram (ii) is also included. The mirrored diagram is labeled (iii), and it can be easily obtained from the diagram (ii) by exchanging $(p_1,\varepsilon_1)$ with $(p_2,\varepsilon_2)$ and changing the sign of $\omega$. Next, there is a diagram coming from the second-order piece of the curved vertex from eq. \eqref{eq:2-pt-kappa4}. It is
\begin{equation}
    \begin{tikzpicture}[baseline={(current bounding box.center)}]
        \coordinate (in) at (-1,0);
        \coordinate (out) at (1,0);
        \coordinate (x) at (0,0);
        \coordinate (gin) at (-.65,-1.6);
        \coordinate (gout) at (.65,-1.6);
        \coordinate (v) at (0,-1);
        \draw [dotted, thick] (in) -- (x);
        \draw [dotted, thick] (x) -- (out);
        \draw [photon2] (x) -- (v) node [midway, left, black] {$q\!\downarrow$};
        \draw [photon] (v) -- (gin) node [left] {$p_1$} node [midway,above,sloped] {$\rightarrow$};
        \draw [photon] (v) -- (gout) node [right] {$p_2$} node [midway,above,sloped] {$\rightarrow$};
        \draw [fill] (v) circle (.14) node {\color{white}\tiny$2$};
        \draw [fill] (x) circle (.08);
    \end{tikzpicture} = \ii\hat\delta(q\cdot v)\kappa^4M^2\int_\ell\frac{\hat\delta(\ell\cdot v)N_\text{iv}(\varepsilon_1,\varepsilon_2,p_1,p_2,v,\ell)}{\ell^2(q - \ell)^2}.
\label{eq:resummeddiagram}\end{equation}
Lastly, there is a diagram with two occurrences of the first order vertex which has the energy flowing through a graviton line. It takes the form
\begin{equation}
    \begin{tikzpicture}[baseline={(current bounding box.center)}]
        \coordinate (in) at (-1.25,0);
        \coordinate (out) at (1.25,0);
        \coordinate (x1) at (-.6,0);
        \coordinate (x2) at (.6,0);
        \coordinate (y1) at (-.6,-1);
        \coordinate (y2) at (.6,-1);
        \coordinate (g1) at (-1.25,-1.6);
        \coordinate (g2) at (1.25,-1.6);
        \coordinate (v) at (0,-.8);
        \draw [dotted, thick] (in) -- (out);
        \draw [photon2] (x1) -- (y1) node [midway, left, black] {$\ell\downarrow$};
        \draw [photon2] (x2) -- (y2) node [midway, right, black] {$\downarrow q-\ell$};
        \draw [photon] (y1) -- (y2) node [midway, below] {$\stackrel{\displaystyle\rightarrow}{p_1+\ell}$};
        \draw [photon] (y1) -- (g1) node [left] {$p_1$} node [midway,above,sloped] {$\rightarrow$};
        \draw [photon] (y2) -- (g2) node [right] {$p_2$} node [midway,above,sloped] {$\rightarrow$};
        \draw [fill] (y1) circle (.14) node {\color{white}\tiny$1$};
        \draw [fill] (y2) circle (.14) node {\color{white}\tiny$1$};
        \draw [fill] (x1) circle (.08);
        \draw [fill] (x2) circle (.08);
    \end{tikzpicture} = \ii\hat\delta(q\cdot v)\kappa^4M^2\int_\ell\frac{\hat\delta(\ell\cdot v)N_\text{v}(\varepsilon_1,\varepsilon_2,p_1,p_2,v,\ell)}{\ell^2[(p_1+\ell)^2+\ii0](q - \ell)^2}.
\end{equation}
All the diagrams have unit symmetry factor. Upon summing them, I find that the amplitude
\begin{equation}
    \ii\hat\delta(q\cdot v)\mathcal{M}^\text{2PM} = (\text{i}) + \cdots + (\text{v}),
\end{equation}
consists exclusively of integrals from the integral family
\begin{equation}
    K_{\nu_1,\nu_2,\nu_3,\lambda_1,\lambda_2} = \tilde\mu^{4-d}\int_\ell\frac{\hat\delta(\ell\cdot v)(\varepsilon_1\cdot\ell)^{\lambda_1}(\varepsilon_2\cdot\ell)^{\lambda_2}}{[\ell^2]^{\nu_1}[(p_1 + \ell)^2 + \ii0]^{\nu_2}[(q - \ell)^2]^{\nu_3}}.
\label{eq:intfam}\end{equation}
I exposed here the $\tilde\mu^2$ which was part of my definition of the $d$-dimensional Newton's constant in eq. \eqref{eq:ddimnewtonconstant}. The $\ii0$ prescription is made explicit for the middle propagator, since it is the only one that can go on-shell. The other two are forced to be potential (spacelike) by the delta function and the requirement that $q\cdot v = 0$.

\subsection{Flat computation}
The weak-field diagrammatic expansion of the Compton amplitude was given in eq. \eqref{eq:flat2pm}. The second post-Minkowskian piece of it comprises six diagrams which I label (i$'$)--(vi$'$). Owing to the equivalence between the flat and curved worldline Feynman rules, the first diagram is identical to the curved space one
\begin{equation}
    \begin{tikzpicture}[baseline={(current bounding box.center)}]
        \coordinate (in) at (-2,0);
        \coordinate (out) at (2,0);
        \coordinate (x1) at (-1.5,0);
        \coordinate (x2) at (-.5,0);
        \coordinate (x3) at (.5,0);
        \coordinate (x4) at (1.5,0);
        \coordinate (gin) at (-1.5,-1);
        \coordinate (gout) at (1.5,-1);
        \draw [dotted, thick] (in) -- (x1);
        \draw [dotted, thick] (x2) -- (x3);
        \draw [dotted, thick] (x4) -- (out);
        \draw [zUndirected] (x1) -- (x2) node [midway, above] {$\stackrel{\displaystyle\omega}{\rightarrow}$};
        \draw [zUndirected] (x3) -- (x4) node [midway, above] {$\stackrel{\displaystyle\omega}{\rightarrow}$};
        \draw [photon] (x1) -- (gin) node [below] {$p_1$} node [midway, left=.3em] {$\uparrow$};
        \draw [photon] (x2) to [bend right=80] (x3) node [midway, below=.8em] {$\stackrel{\displaystyle\rightarrow}{p_1 + \ell}$};
        \draw [photon] (x4) -- (gout) node [below=-.3em] {$p_2$} node [midway, right=.3em] {$\downarrow$};
        \draw [fill] (x1) circle (.08);
        \draw [fill] (x2) circle (.08);
        \draw [fill] (x3) circle (.08);
        \draw [fill] (x4) circle (.08);
    \end{tikzpicture} = \ii\hat\delta(q\cdot v)\frac{\kappa^4M^2}{\omega^4}\int_\ell\frac{\hat\delta(\ell\cdot v)N_{\text{i}'}(\varepsilon_1,\varepsilon_2,p_1,p_2,v,\ell)}{(p_1 + \ell)^2 + \ii0}.
\end{equation}
For coherence of notation, I labeled the numerator $N_{\text{i}'}$ even though it is obviously equal to $N_\text{i}$. The next diagram involves the cubic graviton vertex and looks like
\begin{equation}
    \begin{tikzpicture}[baseline={(current bounding box.center)}]
        \coordinate (in) at (-1.5,0);
        \coordinate (out) at (2.5,0);
        \coordinate (x1) at (-1,0);
        \coordinate (y1) at (-.25,-.8);
        \coordinate (x2) at (.5,0);
        \coordinate (x3) at (2,0);
        \coordinate (gin) at (-.25,-1.4);
        \coordinate (gout) at (2,-1.4);
        \draw [dotted, thick] (in) -- (x2);
        \draw [dotted, thick] (x4) -- (out);
        \draw [zUndirected] (x2) -- (x3) node [midway, above] {$\stackrel{\displaystyle\omega}{\rightarrow}$};
        \draw [photon] (y1) -- (gin) node [below] {$p_1$} node [midway, left=.3em] {$\uparrow$};
        \draw [photon] (x1) -- (y1) node [midway, left, black] {$\ell\searrow$};
        \draw [photon] (y1) -- (x2) node [midway, right] {$\nearrow p_1+\ell$};
        \draw [photon] (x3) -- (gout) node [below] {$p_2$} node [midway, right=.3em] {$\downarrow$};
        \draw [fill] (x1) circle (.08);
        \draw [fill] (x2) circle (.08);
        \draw [fill] (x3) circle (.08);
        \draw [fill] (y1) circle (.04);
    \end{tikzpicture} = \ii\hat\delta(q\cdot v)\frac{\kappa^4M^2}{\omega^2}\int_\ell\frac{\hat\delta(\ell\cdot v)N_{\text{ii}'}(\varepsilon_1,\varepsilon_2,p_1,p_2,v,\ell)}{\ell^2[(p_1 + \ell)^2+\ii0]}.
\end{equation}
Exactly as before, a mirrored version of diagram (ii$'$) labeled (iii$'$) needs to be included, and it is again found by exchanging $(p_1,\varepsilon_1)$ with $(p_2,\varepsilon_2)$ and changing the sign of $\omega$. Now follow the two potential diagrams which resum into eq. \eqref{eq:resummeddiagram}. They are
\begin{equation}
    \begin{tikzpicture}[baseline={(current bounding box.center)}]
        \coordinate (in) at (-1.5,0);
        \coordinate (out) at (1,0);
        \coordinate (x1) at (-1,0);
        \coordinate (v1) at (-.25,-.8);
        \coordinate (x2) at (.5,0);
        \coordinate (v2) at (-.25,-1.6);
        \coordinate (gin) at (-1,-2.4);
        \coordinate (gout) at (.5,-2.4);
        \draw [dotted, thick] (in) -- (out);
        \draw [photon] (v1) -- (v2) node [midway, left, black] {$q\!\downarrow$};
        \draw [photon] (x1) -- (v1) node [midway, left, black] {$\ell\searrow$};
        \draw [photon] (v1) -- (x2) node [midway, right] {$\swarrow q-\ell$};
        \draw [photon] (v2) -- (gout) node [right] {$p_2$} node [midway,above,sloped] {$\rightarrow$};
        \draw [photon] (v2) -- (gin) node [left] {$p_1$} node [midway,above,sloped] {$\rightarrow$};
        \draw [fill] (x1) circle (.08);
        \draw [fill] (x2) circle (.08);
        \draw [fill] (v1) circle (.04);
        \draw [fill] (v2) circle (.04);
    \end{tikzpicture} = \ii\hat\delta(q\cdot v)\kappa^4M^2\int_\ell\frac{\hat\delta(\ell\cdot v)N_{\text{iv}'}(\varepsilon_1,\varepsilon_2,p_1,p_2,v,\ell)}{\ell^2(q - \ell)^2},
\end{equation}
and
\begin{equation}
    \begin{tikzpicture}[baseline={(current bounding box.center)}]
        \coordinate (in) at (-1.5,0);
        \coordinate (out) at (1,0);
        \coordinate (x1) at (-1,0);
        \coordinate (v1) at (-.25,-.8);
        \coordinate (x2) at (.5,0);
        \coordinate (gin) at (-1,-1.6);
        \coordinate (gout) at (.5,-1.6);
        \draw [dotted, thick] (in) -- (out);
        \draw [photon] (x1) -- (v1) node [midway, left, black] {$\ell\searrow$};
        \draw [photon] (v1) -- (x2) node [midway, right] {$\swarrow q-\ell$};
        \draw [photon] (v1) -- (gout) node [right] {$p_2$} node [midway,above,sloped] {$\rightarrow$};
        \draw [photon] (v1) -- (gin) node [left] {$p_1$} node [midway,above,sloped] {$\rightarrow$};
        \draw [fill] (x1) circle (.08);
        \draw [fill] (x2) circle (.08);
        \draw [fill] (v1) circle (.04);
    \end{tikzpicture} = \ii\hat\delta(q\cdot v)\kappa^4M^2\int_\ell\frac{\hat\delta(\ell\cdot v)N_{\text{v}'}(\varepsilon_1,\varepsilon_2,p_1,p_2,v,\ell)}{\ell^2(q - \ell)^2}.
\end{equation}
Finally, there is the diagram with two cubic graviton vertices,
\begin{equation}
    \begin{tikzpicture}[baseline={(current bounding box.center)}]
        \coordinate (in) at (-1.25,0);
        \coordinate (out) at (1.25,0);
        \coordinate (x1) at (-.6,0);
        \coordinate (x2) at (.6,0);
        \coordinate (y1) at (-.6,-1);
        \coordinate (y2) at (.6,-1);
        \coordinate (g1) at (-1.25,-1.6);
        \coordinate (g2) at (1.25,-1.6);
        \coordinate (v) at (0,-.8);
        \draw [dotted, thick] (in) -- (out);
        \draw [photon] (x1) -- (y1) node [midway, left, black] {$\ell\downarrow$};
        \draw [photon] (x2) -- (y2) node [midway, right, black] {$\downarrow q-\ell$};
        \draw [photon] (y1) -- (y2) node [midway, below] {$\stackrel{\displaystyle\rightarrow}{p_1+\ell}$};
        \draw [photon] (y1) -- (g1) node [left] {$p_1$} node [midway,above,sloped] {$\rightarrow$};
        \draw [photon] (y2) -- (g2) node [right] {$p_2$} node [midway,above,sloped] {$\rightarrow$};
        \draw [fill] (y1) circle (.04);
        \draw [fill] (y2) circle (.04);
        \draw [fill] (x1) circle (.08);
        \draw [fill] (x2) circle (.08);
    \end{tikzpicture} = \ii\hat\delta(q\cdot v)\kappa^4M^2\int_\ell\frac{\hat\delta(\ell\cdot v)N_{\text{vi}'}(\varepsilon_1,\varepsilon_2,p_1,p_2,v,\ell)}{\ell^2[(p_1+\ell)^2+\ii0](q - \ell)^2}.
\end{equation}
Upon summing these diagrams, I find again that the amplitude
\begin{equation}
    \ii\hat\delta(q\cdot v){\mathcal{M}'}^\text{2PM} = (\text{i}') + \cdots + (\text{vi}'),
\end{equation}
can be expressed as a sum of integrals from the family in eq. \eqref{eq:intfam}. Even though the diagrams in the flat and curved computations look very similar, they have quite different origins. The pieces from the flat diagrams where the external graviton line is connected to the worldline by a graviton involves a combination of pieces from the worldline and Einstein--Hilbert action. In the curved computation, however, these pieces emerge solely from the quadratic part of the Einstein--Hilbert actions. The way I have chosen to draw the curved diagrams is therefore a matter of convention, but I would argue that the similarity between the two sets of diagrams highlight the naturalness of this choice.

\section{Technical interlude}\label{sec:technical}
It is not easy to establish the equality between the amplitudes $\mathcal{M}^\text{2PM}$ and ${\mathcal{M}'}^\text{2PM}$ resulting from, respectively, the curved and flat computations without further processing, which is why I have distinguished them from each other. They also contain many different integrals (i.e., many different values of $\nu_1$, $\nu_2$, $\nu_3$, $\lambda_1$ and $\lambda_2$ cf. eq. \eqref{eq:intfam}), making the problem of integrating the amplitudes quite challenging. Fortunately, there exists linear relations between these integrals, known as integration-by-parts identities, which allow one to reduce the integration problem to solving a few master integrals. In this section, I therefore introduce these identities and explain how to integrate the master integrals that appear.

\subsection{Integration-by-parts identities}
Integration-by-parts identities are so named because they follow from the vanishing of integrals of total derivatives in dimensional regularization,
\begin{equation}
    \int_\ell \frac{\partial}{\partial \ell^\mu}\big[V^\mu f(\ell, P)\big] = 0.
\label{eq:totaldervanishes}\end{equation}
Here, $P^\mu$ represents any external vector in the problem. In the case at hand, it could be $p_i^\mu$, $\varepsilon_i^\mu$, or $v^\mu$ or any linear combination of them such as $q^\mu$. The vector $V^\mu$ represents $\ell^\mu$ or $P^\mu$. The relevant function $f$ for the case at hand is of course
\begin{equation}
    f(\ell, P) = \tilde\mu^{4-d}\frac{\hat\delta(\ell\cdot v)(\varepsilon_1\cdot\ell)^{\lambda_1}(\varepsilon_2\cdot\ell)^{\lambda_2}}{[\ell^2]^{\nu_1}[(p_1 + \ell)^2 + \ii0]^{\nu_2}[(q - \ell)^2]^{\nu_3}},
\end{equation}
however, it can in general be any function of all possible dot products made of external vectors and the loop momentum.

The identity in eq. \eqref{eq:totaldervanishes} is valid at any number of loops, but only the one-loop case is relevant here. Following \cite{Abreu:2022mfk, Lee:2008tj}, it is quite simple to prove it by demanding that the integral
\begin{equation}
    I = \int_\ell f(\ell,P)
\end{equation}
behave ordinarily under linear variable substitutions $\ell^\mu \to A\ell^\mu + B^\mu$, where $A$ is a number and $B^\mu$ is a vector independent of $\ell^\mu$. Consider the transformation $\ell^\mu \to \ell^\mu + \alpha B^\mu$ with $\alpha$ infinitesimal. Under such translations, $I$ must be invariant:
\begin{align}
    I &= \int_\ell f(\ell + \alpha B, P) \notag\\
    &= \int_\ell f(\ell, P) + \alpha\int_\ell\frac{\partial}{\partial\ell^\mu}\Big[B^\mu f(\ell, P)\Big] + \mathcal{O}(\alpha^2).
\end{align}
This implies
\begin{equation}
    \int_\ell\frac{\partial}{\partial\ell^\mu}\Big[B^\mu f(\ell, P)\Big] = 0,
\end{equation}
which is exactly eq. \eqref{eq:totaldervanishes} for $V^\mu$ independent of $\ell^\mu$. To establish the case with $V^\mu = \ell^\mu$, consider the transformation $\ell^\mu \to \ee^{\alpha}\ell^\mu$ with $\alpha$ infinitesimal. Under such rescalings,
\begin{subequations}
\begin{align}
    \dd^d\ell &\to \ee^{d\alpha}\dd^d\ell = \dd^d\ell + d\alpha\dd^d\ell + \mathcal{O}(\alpha^2), \\
    f(\ell, P) &\to f(\ee^\alpha\ell, P) = f(\ell,P) + \alpha\ell^\mu\frac{\partial}{\partial\ell^\mu}f(\ell,P) + \mathcal{O}(\alpha^2),
\end{align}
\end{subequations}
so
\begin{align}
    I &= \ee^{d\alpha}\int_\ell f(\ee^\alpha\ell,P) \notag\\
    &= \int_\ell f(\ell, P) + \alpha\bigg[d\int_\ell f(\ell,P) + \int_\ell\ell^\mu\frac{\partial}{\partial\ell^\mu}f(\ell,P)\bigg] + \mathcal{O}(\alpha^2).
\end{align}
By noticing that
\begin{equation}
    \frac{\partial\ell^\mu}{\partial\ell^\mu} = \delta^\mu_\mu = d,
\end{equation}
I can rewrite the above as
\begin{equation}
    I = I + \alpha\int_\ell\frac{\partial}{\partial\ell^\mu}\big[\ell^\mu f(\ell,P)\big] + \mathcal{O}(\alpha^2),
\end{equation}
which immediately implies the vanishing of the second term and therefore establishes the sought-after identity.

The integral family under study contains the delta function $\hat\delta(\ell\cdot v)$, so it is not immediately clear how to apply the machinery of integration-by-parts. In fact, as the $\ii0$ prescription is immaterial for integration-by-parts \cite{Abreu:2022mfk}, the delta functions can be treated on equal footing with the propagators due to the Sokhotski--Plemelj formula
\begin{equation}
    \ii\hat\delta(x) = \frac{1}{x - \ii0} + \frac{1}{x + \ii0}
\label{eq:sokhotski}\end{equation}
and its generalization
\begin{equation}
    \frac{\ii}{(-1)^{n+1}n!}\hat\delta^{(n)}(x) = \frac{1}{(x - \ii0)^{n+1}} + \frac{1}{(x + \ii0)^{n+1}},
\label{eq:sokhotskigeneralized}\end{equation}
where the superscript on the delta function denotes differentiation. The generalization follows from simple differentiation of eq. \eqref{eq:sokhotski}. The use of this identity in the context of Feynman integrals is sometimes known as \emph{reverse unitarity} \cite{Herrmann:2021lqe,Brunello:2024ibk,Brandhuber:2021eyq}. Knowing this, it is clear that it is useful generalize the integral family to
\begin{equation}
    K_{\nu_0,\nu_1,\nu_2,\nu_3,\lambda_1,\lambda_2} = \tilde\mu^{4-d}\int_\ell\frac{\hat\delta^{(\nu_0-1)}(\ell\cdot v)(\varepsilon_1\cdot\ell)^{\lambda_1}(\varepsilon_2\cdot\ell)^{\lambda_2}}{[\ell^2]^{\nu_1}[(p_1 + \ell)^2 + \ii0]^{\nu_2}[(q - \ell)^2]^{\nu_3}},
\end{equation}
and to consider the delta function to be a propagator through eq. \eqref{eq:sokhotskigeneralized} for the purposes of integration-by-parts.\footnote{An alternative to this approach would have been to \emph{do} the integral along $v^\mu$, eliminating the delta function. This would have left a family of $(d-1)$-dimensional integrals to be reduced.}

The next step is then to derive the linear relations among the integrals with differing indices.\footnote{The set $(\nu_0,\nu_1,\nu_2,\nu_3,\lambda_1,\lambda_2)$ constitutes the indices of a Feynman integral.} I will give an example of how to derive such relations, after which it will be clear why this task is best left to a computer. From eq. \eqref{eq:totaldervanishes},
\begin{equation}
    \tilde\mu^{4-d}\int_\ell v^\mu\frac{\partial}{\partial\ell^\mu}\frac{\hat\delta^{(\nu_0-1)}(\ell\cdot v)(\varepsilon_1\cdot\ell)^{\lambda_1}(\varepsilon_2\cdot\ell)^{\lambda_2}}{[\ell^2]^{\nu_1}[(p_1 + \ell)^2 + \ii0]^{\nu_2}[(q - \ell)^2]^{\nu_3}} = 0.
\end{equation}
Working out the derivative results in
\begin{align}
    0 &= K_{\nu_0+1,\nu_1,\nu_2,\nu_3,\lambda_1,\lambda_2} + \lambda_1(v\cdot\varepsilon_1)K_{\nu_0,\nu_1,\nu_2,\nu_3,\lambda_1-1,\lambda_2} + \lambda_2(v\cdot\varepsilon_2)K_{\nu_0,\nu_1,\nu_2,\nu_3,\lambda_1,\lambda_2-1} \notag\\
    &+ 2\nu_1(\nu_0-1)K_{\nu_0-1,\nu_1+1,\nu_2,\nu_3,\lambda_1,\lambda_2} + 2\nu_2(\nu_0-1)K_{\nu_0-1,\nu_1,\nu_2+1,\nu_3,\lambda_1,\lambda_2} \notag\\
    &- 2\nu_2\omega K_{\nu_0,\nu_1,\nu_2+1,\nu_3,\lambda_1,\lambda_2} - 2\nu_3(\nu_0-1)K_{\nu_0-1,\nu_1,\nu_2,\nu_3+1,\lambda_1,\lambda_2}.
\label{eq:bigmess}\end{align}
The idea is then to compute relations like these for all choices of $V^\mu$ (cf. eq. \eqref{eq:totaldervanishes}), and from them derive relations that allow one to reduce any integral to a linear combination of a finite basis of master integrals. This is obviously an error-prone and laborious process---fortunately, there exists software packages such as \texttt{LiteRed} for \texttt{Mathematica} \cite{Lee:2012cn, Lee:2013mka}, which by employing an algorithm due to Laporta \cite{Laporta:2000dsw} can do this automatically.\footnote{\texttt{LiteRed} can also handle cut propagators through the \texttt{CutDs} option, which immediately sets to zero any integral containing the cut propagator in the numerator.}

Applying this to the integral family associated with the 2PM Compton amplitude, I find that both the amplitude from the curved and flat calculation admit the same expansion
\begin{equation}
    \mathcal{M}^\text{2PM} = {\mathcal{M}'}^\text{2PM} = \kappa^4M^2\sum_{i=1}^3c_i\mathcal{K}_i,
\end{equation}
where the three master integrals are
\begin{equation}
    \mathcal{K}_1 = K_{0,1,0,0,0}, \qquad \mathcal{K}_2 = K_{1,0,1,0,0}, \qquad \mathcal{K}_3 = K_{1,1,1,0,0}.
\end{equation}
The expansion coefficients from the curved and flat calculation are in complete agreement, which demonstrates that the curved and flat calculation give the same answer. The master integral coefficients $c_i$ are polynomials in dot products of the momenta, the velocity, and the polarizations, naturally restricted to be quadratic in each polarization:
\begin{equation}
    c_i = \varepsilon_{1\mu_1}\varepsilon_{1\nu_1}\varepsilon_{2\mu_2}\varepsilon_{2\nu_2}c^{\mu_1\nu_1\,\mu_2\nu_2}_i(p_1,p_2,v).
\end{equation}
Due to the gauge symmetry discussed in chapter \ref{ch:perturbative quantization of gravity}, these coefficients must satisfy a Ward identity
\begin{equation}
    p_{1\mu_1}c^{\mu_1\nu_1\,\mu_2\nu_2}_i(p_1,p_2,v) = p_{2\mu_2}c^{\mu_1\nu_1\,\mu_2\nu_2}_i(p_1,p_2,v) = 0.
\end{equation}
I find that this is indeed the case. Expressions for the coefficients are presented in section \ref{sec:resultddims} for $d$ dimensions and in section \ref{sec:resultfourdims} for four-dimensions. First, it is essential to discuss in detail how the master integrals are evaluated. It is noteworthy that the integration-by-parts reduction also takes care of all the tensor reduction. Were this not the case, one would have to resort to Passarino-Veltman reduction \cite{Passarino:1978jh}.

\subsection{Master integrals}
\begin{figure}[t]
    \centering
    \begin{tikzpicture}[baseline={(current bounding box.center)}]
        \coordinate (v1) at (-1.3,0);
        \coordinate (v2) at (1.3,0);
        \coordinate (v3) at (-1.3,-1.5);
        \coordinate (v4) at (1.3,-1.5);
        \coordinate (v5) at (0,.1);
        \coordinate (v6) at (0,-.5);
        \draw [zUndirected,looseness=2] (v1) to[in=180+60,out=-60] (v2);
        \draw [zUndirected,looseness=2] (v3) to[in=180-60,out=60] (v4);
        \draw [zUndirected,red] (v5) -- (v6);
    \end{tikzpicture}\hspace{1cm}
    \begin{tikzpicture}[baseline={(current bounding box.center)}]
        \coordinate (v1) at (-1.3,0);
        \coordinate (v2) at (1.3,0);
        \coordinate (v3) at (-.6,-2);
        \coordinate (v4) at (.6,-2);
        \coordinate (v5) at (-.9,-.41);
        \coordinate (v6) at (.9,-.41);
        \coordinate (v7) at (0,-.1);
        \coordinate (v8) at (0,-.7);
        \draw [zUndirected] (v1) -- (v4);
        \draw[zUndirected] (v2) -- (v3);
        \draw[zUndirected] (v5) -- (v6);
        \draw[zUndirected,red] (v7) -- (v8);
    \end{tikzpicture}\hspace{1cm}
    \begin{tikzpicture}[baseline={(current bounding box.center)}]
        \coordinate (out1) at (-1.3,0);
        \coordinate (out2) at (1.3,0);
        \coordinate (out3) at (-1.3,-2);
        \coordinate (out4) at (1.3,-2);
        \coordinate (in1) at (-.8,-.5);
        \coordinate (in2) at (.8,-.5);
        \coordinate (in3) at (-.8,-2+.5);
        \coordinate (in4) at (.8,-2+.5);
        \coordinate (v1) at (0,-.2);
        \coordinate (v2) at (0,-.8);
        \draw[zUndirected] (out1) -- (in1);
        \draw[zUndirected] (out2) -- (in2);
        \draw[zUndirected] (out3) -- (in3);
        \draw[zUndirected] (out4) -- (in4);
        \draw[zUndirected] (in1) -- (in2);
        \draw[zUndirected] (in2) -- (in4);
        \draw[zUndirected] (in3) -- (in4);
        \draw[zUndirected] (in3) -- (in1);
        \draw[zUndirected,red] (v1) -- (v2);
    \end{tikzpicture}
    \caption{The cut bubble $\mathcal{K}_1$, the cut triangle $\mathcal{K}_2$, and the cut box $\mathcal{K}_3$. The red line signifies the propagator that is cut.}
    \label{fig:mis}
\end{figure}
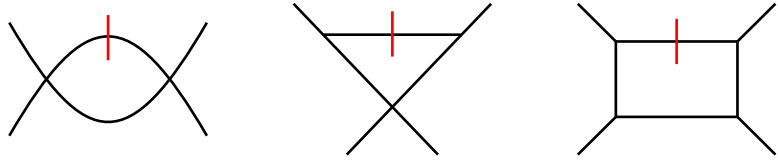

In this section, I compute the three master integrals that span the 2PM Compton amplitude. Writing them out, they are
\begin{subequations}
\begin{align}
    \mathcal{K}_1 &= \tilde\mu^{4-d}\int_\ell\frac{\hat\delta(\ell\cdot v)}{(p_1 + \ell)^2 + \ii0}, \\
    \mathcal{K}_2 &= \tilde\mu^{4-d}\int_\ell\frac{\hat\delta(\ell\cdot v)}{\ell^2(q - \ell)^2}, \\
    \mathcal{K}_3 &= \tilde\mu^{4-d}\int_\ell\frac{\hat\delta(\ell\cdot v)}{\ell^2[(p_1 + \ell)^2 + \ii0](q - \ell)^2}.
\end{align}
\end{subequations}
By inspecting the topologies, they can be recognized as a cut bubble, a cut triangle, and a cut box, which is depicted graphically in figure \ref{fig:mis}.\footnote{A propagator is ``cut'' if it appears in a delta function instead of in the denominator.} The cut bubble and the cut triangle are quite simple to compute, while the cut box is somewhat trickier.

\paragraph{Cut bubble.} I start with the cut bubble by decomposing the loop momentum into its parallel and perpendicular parts
\begin{equation}
    \ell^\mu = \ell_\parallel^\mu + \ell_\perp^\mu, \qqquad \dd^d\ell = \dd^{d-1}\ell_\perp\dd\ell_\parallel,
\end{equation}
in analogy with eq. \eqref{eq:decomppos}. This allows using the delta function to do the integral along $v^\mu$,
\begin{equation}
    \mathcal{K}_1 = \tilde\mu^{4-d}\int\frac{\dd^{d-1}\ell_\perp}{(2\pi)^{d-1}}\frac{1}{(p_{1\perp} + \ell)^2 + \omega^2 + \ii0}.
\end{equation}
As the integral is now over a purely spacelike subspace, it is convenient to change metric signature; I write $\mathbf{V} = V_\perp$ and denote positive-definite dot products by $\mathbf{V}^2 = -V_\perp^2$. This yields
\begin{align}
    \mathcal{K}_1 = -\tilde\mu^{4-d}\int\frac{\dd^{d-1}\ell_\perp}{(2\pi)^{d-1}}\frac{1}{\boldsymbol\ell^2 - \omega^2 - \ii0},
\end{align}
where I shifted the loop momentum by $\boldsymbol\ell \to \boldsymbol\ell - \mathbf{p}_1$. This integral can be evaluated by an analytic continuation. Consider the Euclidean integral
\begin{equation}
    I = \int\mathrm{d}^{d-1}\boldsymbol\ell\,\frac{1}{\boldsymbol\ell^2 + A}
\end{equation}
with $A>0$. The spherical symmetry implies
\begin{equation}
    I = \int\mathrm{d}\Omega_{d-1}\int_0^\infty\mathrm{d}\abs{\boldsymbol\ell}\,\frac{\abs{\boldsymbol\ell}^{d-2}}{\abs{\boldsymbol\ell}^2+A},
\end{equation}
whereupon the substitution $t = \abs{\boldsymbol\ell}^2/A$ may be performed and the integral over the solid angle can be done. This gives
\begin{equation}
    \frac{\pi^{\frac{d-1}{2}}A^{\frac{d-3}{2}}}{\Gamma\big(\frac{d-1}{2}\big)}\int_0^\infty\mathrm{d}t\,\frac{t^{\frac{d-3}{2}}}{t+1}.
\end{equation}
This is recognized to be Euler's beta function $B(\nu_1,\nu_2)$ in the representation
\begin{equation}
    B(n_1,n_2) = \int_0^\infty\mathrm{d}t\,t^{n_1-1}(1+t)^{-n_1-n_2}.
\label{eq:eulerbeta}\end{equation}
Using this and the fact that
\begin{equation}
    B(n_1, n_2) = \frac{\Gamma(n_1)\Gamma(n_2)}{\Gamma(n_1 +n_2)},
\end{equation}
the integral evaluates to
\begin{equation}
    I = \pi^{\frac{d-1}{2}}A^{\frac{d-3}{2}}\Gamma\Big(\frac{3-d}{2}\Big).
\end{equation}
Now, this result may be continued to negative values of $A$ as the $\ii0$ prescription takes care of the choice of branch cut. Hence, setting $A = -\omega^2 - \ii0$ and taking into account the differing normalizations of the integrals, I obtain
\begin{equation}
    \mathcal{K}_1 = -\frac{\tilde\mu^{4-d}(-\omega^2-\ii0)^\frac{d-3}{2}\Gamma\big(\frac{3-d}{2}\big)}{(4\pi)^\frac{d-1}{2}}.
\end{equation}

\paragraph{Cut triangle.} The integration along the velocity can be performed for the second master integral in the same way as for the first, yielding
\begin{equation}
    \mathcal{K}_2 = \tilde\mu^{4-d}\int\frac{\dd^{d-1}\boldsymbol\ell}{(2\pi)^{d-1}}\,\frac{1}{\boldsymbol\ell^2(\mathbf{q} - \boldsymbol\ell)^2}.
\end{equation}
This integral is easiest to evaluate via Feynman parameters \cite{Weinzierl:2022eaz, Smirnov:2006ry}.\footnote{While the Feynman parameterization of $\mathcal{K}_2$ is not hard to derive by hand, it is convenient to use software packages such as \texttt{FeynCalc} \cite{Shtabovenko:2016sxi, Shtabovenko:2020gxv, Shtabovenko:2023idz, Mertig:1990an}.} The Feynman parameterization of the cut triangle is
\begin{equation}
    \mathcal{K}_2 = \frac{\Gamma\big(\frac{5-d}{2}\big)\tilde\mu^{4-d}}{(4\pi)^\frac{d-1}{2}(-q^2)^\frac{5-d}{2}}\int_{\alpha_i\geq0}\mathrm{d}^2\alpha\,\delta(1 - \alpha_1 - \alpha_2)\frac{(\alpha_1 + \alpha_2)^{3-d}}{(\alpha_1\alpha_2)^\frac{5-d}{2}}.
\end{equation}
I used that $q^2 = q_\perp^2 = -\mathbf{q}^2$. By the Cheng-Wu theorem \cite{Cheng:1969eh}, Feynman integrals are invariant under the choice of non-empty subsets of Feynman parameters $\alpha_i$ one includes in the delta function. Using this to exclude $\alpha_1$ from the delta function, the integration over $\alpha_2$ simply forces this parameter to be unity,
\begin{equation}
    \mathcal{K}_2 = \frac{\Gamma\big(\frac{5-d}{2}\big)\tilde\mu^{4-d}}{(4\pi)^\frac{d-1}{2}(-q^2)^\frac{5-d}{2}}\int_0^\infty\mathrm{d}\alpha_1\,\frac{(\alpha_1 + 1)^{3-d}}{\alpha_1^\frac{5-d}{2}}.
\end{equation}
Again, it is Euler's beta function in the representation from eq. \eqref{eq:eulerbeta} that emerges, meaning the integral evaluates to
\begin{equation}
    \mathcal{K}_2 = \frac{\Gamma\big(\frac{5-d}{2}\big)\Gamma\big(\frac{d-3}{2}\big)^2\tilde\mu^{4-d}}{(4\pi)^\frac{d-1}{2}\Gamma(d-3)(-q^2)^\frac{5-d}{2}}.
\end{equation}

\paragraph{Cut box.} After integrating out the delta function in the same manner as above, the cut box becomes
\begin{equation}
    \mathcal{K}_3 = -\tilde\mu^{4-d}\int\frac{\mathrm{d}^{d-1}\boldsymbol\ell}{(2\pi)^{d-1}}\frac{1}{\boldsymbol\ell^2[(\mathbf{p}_1 + \boldsymbol\ell)^2 - \omega^2 - \ii0](\mathbf{q} - \boldsymbol\ell)^2}.
\end{equation}
The Feynman parameterization of this integral is
\begin{equation}
    \mathcal{K}_3(y) = -\frac{\Gamma\big(\frac{7-d}{2}\big)}{(4\pi)^\frac{d-1}{2}}\int_{\alpha_i\geq0}\mathrm{d}^3\alpha\,\delta\bigg(1-\sum_{i\in S}\alpha_i\bigg)\frac{(\alpha_1 + \alpha_2 + \alpha_3)^{4-d}}{(-q^2\alpha_1\alpha_3 -\omega^2\alpha_2^2 - \ii0)^\frac{7-d}{2}},
\label{eq:cutboxfeyns}\end{equation}
where $S$ is any non-empty subset of the Feynman parameters. This is not an easy integral to compute directly, so I shall tackle it with the method of differential equations \cite{Kotikov:1990kg,gehrmann2000differential,Smirnov:2008iw,larsen2016integration,lee2015reducing}. This method is intimately related with the existence of integration-by-parts identities that I discussed in the previous section. Namely, the idea of the method is to differentiate the integral with respect to some kinematic parameter, which will result in some linear combination of integrals from the integral family not dissimilar to the expression in eq. \eqref{eq:bigmess}. Using the integration-by-parts identities, every integral in this sum is again reduced to a linear combination of master integrals, yielding a closed system of differential equations for the master integrals. In this case, two of the three master integrals were already computed, so the system will be one-dimensional. What kinematic parameter should be chosen? Naïvely, the integral can depend independently on the energy $\omega$ and, by Lorentz invariance, the squared momentum transfer $q^2$. However, by scaling out the mass dimension from all the momenta,
\begin{equation}
    \mathbf{P} \to \frac{\mathbf{P}}{\omega}, \quad\text{where}\quad \mathbf{P} = \boldsymbol\ell, \mathbf{p}_1, \mathbf{q},
\label{eq:mass-rescaling}\end{equation}
the energy $\omega$ can be extracted as a prefactor, implying that the integral depends only on the kinematic ratio $y$ defined in eq. \eqref{eq:kinpardef}:
\begin{equation}
    \mathcal{K}_3(\omega,q^2) = \omega^{d-7}\tilde\mu^{4-d}\mathsf{K}_3(y), \qquad y = \frac{-q^2}{4\omega^2}.
\label{eq:dimlessintegral}\end{equation}
An easy prescription to go from the calligraphic to the sans-serif integrals is to simply set $\omega\to1$, $q^2 \to -4y$, and $\tilde\mu^2 \to 1$. The integral $\mathsf{K}_3(y)$ is a dimensionless function only of $y$, so an ordinary differential equation in $y$ can be derived for it. This is in practice most easily accomplished with the built-in functionality of \texttt{LiteRed}---I will briefly discuss, however, the most conceptually important point, which is deriving a form of $\dd/\dd y$ that can be easily applied to a Feynman integral. This can be done conveniently by making the ansatz \cite{Henn:2014qga}
\begin{equation}
    \frac{\dd}{\dd y} = (\xi_vv^\mu + \xi_qq^\mu + \xi_{p_1}p_1^\mu)\frac{\partial}{\partial p_1^\mu},
\end{equation}
and demanding that the derivative be properly normalized and commute with the on-shell conditions,
\begin{equation}
    \frac{\dd y}{\dd y} = 1, \qqquad \frac{\dd p_1^2}{\dd y} = 0, \qqquad \frac{\dd (p_1 + q)^2}{\dd y} = 0.
\end{equation}
The choice of $\partial/\partial p_1^\mu$ is arbitrary. One could just as well have chosen $\partial/\partial v^\mu$ or $\partial/\partial q^\mu$. The solution of the system is
\begin{equation}
    \xi_v = \frac{-y}{2y(y-1)}, \qqquad \xi_q = \frac{1}{4y(y-1)}, \qqquad \xi_{p_1} = \frac{1}{2y(y-1)}.
\end{equation}
Using \texttt{LiteRed}, I obtain the differential equation
\begin{equation}
    \frac{\mathrm{d}}{\mathrm{d}y}\mathsf{K}_3(y) = \frac{1}{y(y - 1)}\bigg(\frac{d-3}{8}\mathsf{K}_1(y) - \frac{d-4}{4}\mathsf{K}_2(y) - \frac{(d-6+2y)}{2}\mathsf{K}_3(y)\bigg),
\end{equation}
where $\mathsf{K}_1$ and $\mathsf{K}_2$ are the cut bubble and triangle with their $\omega$ dependence properly extracted using the prescription below eq. \eqref{eq:dimlessintegral}:
\begin{subequations}
\begin{alignat}{2}
    \mathsf{K}_1(y) &= A_{\mathsf{K}_1}(d) &&= -\frac{(-1-\ii0)^\frac{d-3}{2}\Gamma\big(\frac{3-d}{2}\big)}{(4\pi)^\frac{d-1}{2}}, \\
    \mathsf{K}_2(y) &= \frac{A_{\mathsf{K}_2}(d)}{y^\frac{5-d}{2}} &&= \frac{\Gamma\big(\frac{5-d}{2}\big)\Gamma\big(\frac{d-3}{2}\big)^2}{2^{5-d}(4\pi)^\frac{d-1}{2}\Gamma(d-3)}\frac{1}{y^\frac{5-d}{2}}.
\end{alignat}
\end{subequations}
Observe that the differential equation contains the factor $y(y-1)$ from the derivative. The solution is in terms of hypergeometric functions,
\begin{equation}
\begin{aligned}
    \mathsf{K}_3(y) &= (1 - y)^\frac{4-d}{2}\Bigg(\frac{C(d)}{y^\frac{6-d}{2}} + A_{\mathsf{K}_1}(d)\,{}_2F_1\Big[\begin{gathered}\scriptstyle \frac{6-d}{2},\,\frac{6-d}{2} \\[-.5em]\scriptstyle\frac{8-d}{2}\end{gathered}; y\Big] + \frac{A_{\mathsf{K}_2}(d)}{y^\frac{5-d}{2}}\,{}_2F_1\Big[\begin{gathered}\scriptstyle \frac12,\,\frac{6-d}{2} \\[-.5em] \scriptstyle\frac32\end{gathered}; y\Big]\Bigg),
\end{aligned}
\end{equation}
which is easily obtained with \texttt{Mathematica}. There is an integration constant $C(d)$ which must be determined from a boundary condition. To that end, notice that the asymptotic expansion of the solution in the forward-scattering limit is
\begin{equation}
    \mathsf{K}_3(y) \sim \frac{C(d)}{y^\frac{6-d}{2}} \quad\text{as}\quad y\to0.
\label{eq:mi3-asym}\end{equation}
This follows from
\begin{equation}
    {}_2F_1\Big[\begin{gathered} a,\,c \\[-.5em] c\end{gathered}; 0\Big] = 0,
\end{equation}
which can be seen to hold for any $a$, $b$, and $c$ directly from the series definition of the hypergeometric function \cite{NIST:DLMF}. It is hence necessary to obtain the asymptotic expansion of $\mathsf{K}_3(y)$ in the forward-scattering limit. This is made possible by the so‑called method of regions, which provides a systematic way of decomposing the integral into contributions from distinct kinematic regions (each defined by a particular scaling of loop momenta or parameters), expanding the integrand in each region before integration, and then recombining these pieces to obtain the full asymptotic behavior \cite{Becher:2015,Smirnov:1994tg}. In \cite{Pak:2010pt,Semenova:2018cwy}, an especially attractive approach was proposed, allowing the expansion to be performed at the level of Feynman parameters. For a pedagogical outline of this method, see also appendix A in \cite{Brandhuber:2021eyq}. By applying the rescaling prescription to the Feynman parameterization from eq. \eqref{eq:cutboxfeyns}, I obtain
\begin{align}
    \mathsf{K}_3(y) &= -\frac{\Gamma\big(\frac{7-d}{2}\big)}{(4\pi)^\frac{d-1}{2}}\int_{\alpha_i\geq0}\mathrm{d}^3\alpha\,\delta\bigg(1-\sum_{i\in S}\alpha_i\bigg)\frac{(\alpha_1 + \alpha_2 + \alpha_3)^{4-d}}{(4y\alpha_1\alpha_3 - \alpha_2^2 - \ii0)^\frac{7-d}{2}} \notag\\
    &= -\frac{\Gamma\big(\frac{7-d}{2}\big)}{(4\pi)^\frac{d-1}{2}}\int_{\alpha_i\geq0}\mathrm{d}^2\alpha\,\frac{(1 + \alpha_1 + \alpha_3)^{4-d}}{(4y\alpha_1\alpha_3 - 1 - \ii0)^\frac{7-d}{2}},
\label{eq:dimlessfeynparamsbox}\end{align}
where the Cheng-Wu theorem again allowed choosing $S = \{2\}$. The method involves finding all variable substitutions $\alpha_i\to y^{v_i}\alpha_i$ that result in a non-scaleless integral with $y$-homogeneous Symanzik polynomials after expanding the integrand in $y$. Three definitions are needed for the previous sentence to make sense. Firstly, the first and second Symanzik polynomials are respectively the numerator and denominator of eq. \eqref{eq:dimlessfeynparamsbox} \cite{Weinzierl:2022eaz}. Secondly, a Feynman parameterized integral is scaleless if it receives an overall scaling under a rescaling of a proper subset of the Feynman parameters \cite{Pak:2010pt}. Thirdly, by each Symanzik polynomial being $y$-homogeneous, I mean that each term in the polynomial contains the same power of $y$.\footnote{There is a beautiful geometric reinterpretation of the procedure outlined here in terms of convex hulls and their normal vectors. One considers each monomial in each Symanzik polynomial as a vector in the space of the exponents of each Feynman parameter plus the kinematic parameter one is expanding in. The set of monomials will have a convex hull, and each facet of the hull with an upward-point normal vector represents a region that contributes in the expansion. A proper treatment of this is beyond the scope of this thesis, but see \cite{Pak:2010pt} for more details.} Using the algorithm in \cite{Brandhuber:2021eyq}, I find the only such substitution is $\alpha_i \to y^{-\frac12}\alpha_i$, which leads to an integral that is readily calculable:
\begin{equation}
\begin{array}{cc}
    \begin{aligned}
        \mathsf{K}_3(y) &\sim -\frac{1}{y^\frac{6-d}{2}}\frac{\Gamma\big(\frac{7-d}{2}\big)}{(4\pi)^\frac{d-1}{2}}\int_{\alpha_i\geq0}\mathrm{d}^2\alpha\,\frac{(\alpha_1 + \alpha_3)^{4-d}}{(4\alpha_1\alpha_3 - 1 - \ii0)^\frac{7-d}{2}} \\
        &= (-1-\ii0)^\frac{d+1}{2}\frac{\pi}{4}\frac{\big(\ii+\cot\frac{d\pi}{2}\big)\Gamma\big(\frac{5-d}{2}\big)\cos\frac{d\pi}{2}}{(4\pi)^\frac{d-1}{2}}\frac{1}{y^\frac{6-d}{2}}
    \end{aligned} & \qquad\text{as}\quad y\to0.
\end{array}
\end{equation}
Comparison with eq. \eqref{eq:mi3-asym} then yields
\begin{equation}
    C(d) = (-1-\ii0)^\frac{d+1}{2}\frac{\pi}{4}\frac{\big(\ii+\cot\frac{d\pi}{2}\big)\Gamma\big(\frac{5-d}{2}\big)\cos\frac{d\pi}{2}}{(4\pi)^\frac{d-1}{2}},
\end{equation}
which finally determines the cut box to be
\begin{equation}
\begin{aligned}
    \mathcal{K}_3 = \frac{\omega^{d-7}\tilde\mu^{4-d}}{(1-y)^\frac{d-4}{2}(4\pi)^\frac{d-1}{2}}\Bigg(&(-1-\ii0)^\frac{d+1}{2}\frac{\pi}{4}\frac{\big(\ii+\cot\frac{d\pi}{2}\big)\Gamma\big(\frac{5-d}{2}\big)\cos\frac{d\pi}{2}}{y^\frac{6-d}{2}} \\
    &~- (-1-\ii0)^\frac{d-3}{2}\Gamma\big(\frac{3-d}{2}\big)\,{}_2F_1\Big[\begin{gathered}\scriptstyle \frac{6-d}{2},\,\frac{6-d}{2} \\[-.5em]\scriptstyle\frac{8-d}{2}\end{gathered}; y\Big] \\
    &~+ \frac{\Gamma\big(\frac{5-d}{2}\big)\Gamma\big(\frac{d-3}{2}\big)^2}{2^{5-d}\Gamma(d-3)}\frac{1}{y^\frac{5-d}{2}}\,{}_2F_1\Big[\begin{gathered}\scriptstyle \frac12,\,\frac{6-d}{2} \\[-.5em] \scriptstyle\frac32\end{gathered}; y\Big]\Bigg).
\end{aligned}
\end{equation}
For the four-dimensional result, the Laurent expansion of the integrals around $d = 4-2\epsilon$ will be necessary. These are
\begin{subequations}
\begin{align}
    \mathcal{K}_1 &= -\frac{\ii\omega}{4\pi}\Bigg[1 + \epsilon\bigg(\ii\pi + 2 - \log\frac{4\omega^2}{\mu^2}\bigg)\Bigg] + \mathcal{O}(\epsilon), \label{eq:cutbubble4dim}\\
    \mathcal{K}_2 &= \frac{1}{8\abs{q}} + \mathcal{O}(\epsilon), \label{eq:cuttriangle4dim}\\
    \mathcal{K}_3 &= \frac{-\ii}{8\pi q^2\omega}\Bigg[\frac{1}{\epsilon} - \log\frac{-q^2}{\mu^2}\Bigg] + \mathcal{O}(\epsilon). \label{eq:cutbox4dim}
\end{align}
\end{subequations}
Notice that these are written in terms of $\mu$, not $\tilde\mu$. The extra factors in $\tilde\mu$ (cf. eq. \eqref{eq:ddimnewtonconstant}) were used to cancel some factors in the integral. In the expansion I also used that
\begin{equation}
    \sqrt{-1-\ii0} = -\ii, \qqquad \log(-1-\ii0) = -\ii\pi.
\end{equation}

\section{Second-order amplitude}\label{sec:2pmamp}
Having computed the master integrals, I can now present the result. In the first section, I present the $d$-dimensional master integral coefficients $c_i$ after having set up some compact notation. After this, I then specify to four dimensions and provide a discussion of the result in this case including various cross checks.

\subsection{Result in \textit{d} dimensions}\label{sec:resultddims}
By integration-by-parts reduction, I found that the amplitude admitted the expansion
\begin{equation}
    \mathcal{M}^\text{2PM} = \kappa^4M^2\sum_{i=1}^3c_i\mathcal{K}_i,
\end{equation}
with gauge-invariant master integral coefficients $c_i$. Having now computed the master integrals, it is time to deal with the coefficients. To represent them in a compact manner, it is very convenient to exploit the gauge-invariance that each coefficient satisfies. It implies that each coefficient is an element of a gauge-invariant vector space, meaning it must admit an expansion in products involving the field strength tensors $f_1$ and $f_2$ defined in eq. \eqref{eq:fieldstrengthtensor}. A practical way to find the basis for this space is to examine products that are quadratic in each field strength and from these form a linearly independent set that spans each coefficient. As proven in \cite{Feng:2020jck}, it is sufficient to consider the products that can be expressed through the so-called fundamental $f$-terms
\begin{equation}
    V\cdot f_i\cdot f_j\cdot V', \qqquad V\cdot f_i\cdot V',
\end{equation}
where the notation defined below eq. \eqref{eq:fieldstrengthtensor} was used, and $V^\mu$ and $V'^\mu$ represents either $p_1^\mu$, $p_2^\mu$ or $v^\mu$. The reason this is enough is essentially due to no new dot products being generated by including more field strengths in the string, so any string with more than two field strengths is expressible as a linear combination of products of fundamental $f$-terms. Nevertheless, the basis formed from all the products of the fundamental $f$-terms is overdetermined. In fact, it can easily be seen from the relations
\begin{subequations}
\begin{align}
    (V\cdot f_i\cdot V')(v\cdot p_i) &= (V\cdot f_i\cdot v)(V'\cdot p_i) + (v\cdot f_i\cdot V')(V\cdot p_i), \\
    (V\cdot f_i\cdot f_j\cdot V')(v\cdot p_i) &= (V\cdot f_i\cdot v)(p_i\cdot f_j\cdot V') + (v\cdot f_i\cdot f_j\cdot V')(V\cdot p_i),
\end{align}
\end{subequations}
along with the other relations
\begin{equation}
    V\cdot f_i\cdot f_j\cdot V' = (V\cdot f_j\cdot f_i\cdot V'), \qquad v\cdot f_i\cdot v = 0, \qquad v\cdot f_i\cdot V = -V\cdot f_i\cdot v,
\end{equation}
\begin{table}[t]
    \centering
    \addtolength{\tabcolsep}{-0.2em}
    \begin{tabular}{cc} \toprule
        $c_1\vert_{\mathsf{F}_1^2}$  & $\scriptstyle-\frac{-d^3+6 d^2+(d ((d-7) d+20)-22) y-15 d+16}{8 (d-4) (d-2)^2 (d-1) (y-1)^2}$ \\ \midrule
        
        $c_1\vert_{\mathsf{F}_2^2}$  & $\scriptstyle\frac{(d-3) d (d (d+6 y-6)-20 y+16)}{256 (d-4) (d-2)^2 (d-1) (y-1)^4}$ \\ \midrule
        
        $c_1\vert_{\mathsf{F}_1\mathsf{F}_2}$  & $\scriptstyle-\frac{-d^4+7 d^3-18 d^2+(d (d ((d-9) d+28)-24)-16) y+12 d+16}{32 (d-4) (d-2)^2 (d-1) (y-1)^3}$ \\ \midrule
        
        $c_2\vert_{\mathsf{F}_1^2}$  & $\scriptstyle-\frac{\omega ^2 (-4 d^4+33 d^3-(d-5) (d (6 d^2-39 d+88)-67) y^2-99 d^2+(d-5) (d-3)^3 y^3+(d-1) (d (d (9
   d-79)+223)-209) y+127 d-57)}{32 (d-3) (d-2)^2 (d-1) (y-1)^2}$ \\ \midrule
   
        $c_2\vert_{\mathsf{F}_2^2}$  & $\scriptstyle-\frac{\omega ^2 (-4 d^3+(d (6 d^2-62 d+181)-155) y+30 d^2+(d-5) (d-3)^2 y^3-(d-5) (d (4 d-23)+29) y^2-72
   d+55)}{128 (d-2)^2 (d-1) (y-1)^4}$ \\ \midrule
        
        $c_2\vert_{\mathsf{F}_1\mathsf{F}_2}$  & $\scriptstyle-\frac{\omega ^2 (d (d (9 d^2-96 d+364)-586) y+(d-5) (d-3)^3 y^3-(d-5) (d (d (5 d-37)+93)-77) y^2+d (d ((44-5
   d) d-138)+182)+333 y-83)}{32 (d-3) (d-2)^2 (d-1) (y-1)^3}$ \\ \midrule
   
        $c_3\vert_{\mathsf{F}_1^2}$  & $\scriptstyle-\frac{\omega ^4 (d^3-6 d^2+(d ((d-8) d+19)-16) y^2-2 (d-4) (d-2) (d-1) y+11 d-6)}{4 (d-3) (d-2) (d-1) (y-1)^2}$ \\ \midrule
        
        $c_3\vert_{\mathsf{F}_2^2}$  & $\scriptstyle-\frac{\omega ^4 (d^2+8 (d-2) y-6 d+8 y^2+8)}{64 (d-2) (d-1) (y-1)^4}$ \\ \midrule
        
        $c_3\vert_{\mathsf{F}_1\mathsf{F}_2}$  & $\scriptstyle\frac{\omega ^4 (-d^3+7 d^2+2 ((d-5) d+8) y^2+(d-2) ((d-9) d+12) y-14 d+8)}{8 (d-3) (d-2) (d-1) (y-1)^3}$ \\ \bottomrule
    \end{tabular}
    \caption{The master integral coefficients for the second post-Minkowskian order Compton amplitude in $d$ dimensions.}
    \label{tab:ddimcoeffs}
\end{table}
that the set of fundamental $f$-terms can be reduced to
\begin{equation}
    \Big\{v\cdot f_1\cdot f_2\cdot v, \quad v\cdot f_1\cdot p_2, \quad v\cdot f_2\cdot p_1\Big\}.
\end{equation}
This motivates the definition of the dimensionless quantities
\begin{subequations}
\begin{align}
    \mathsf{F}_1 &= \frac{v\cdot f_1\cdot f_2\cdot v}{\omega^2}, \label{eq:f1def}\\
    \mathsf{F}_2 &= \frac{(v\cdot f_1\cdot p_2)(v\cdot f_2\cdot p_1)}{\omega^4}. \label{eq:f2def}
\end{align}
\end{subequations}
To see that these are independent, consider that only $\mathsf{F}_1$ contains the product $\varepsilon_1\cdot\varepsilon_2$, while only $\mathsf{F}_2$ contains the product $(p_1\cdot\varepsilon_2)(p_2\cdot\varepsilon_1)$. The end result is that the set
\begin{equation}
    \Big\{\mathsf{F}_1^2, \quad \mathsf{F}_2^2, \quad \mathsf{F}_1\mathsf{F}_2\Big\},
\end{equation} 
should form a gauge-invariant basis for the coefficients, which I find they do. The expansion coefficients in this gauge-invariant basis of the $d$-dimensional master integral coefficients can be found in table \ref{tab:ddimcoeffs}.
\begin{table}
    \centering
    \begin{tabular}{cc} \toprule
       $c_1^{(-1)}$  & $\frac{1}{48}\bigg[\frac{5y - 6}{2(y-1)^2}\mathsf{F}_1^2 - \frac{2 + y}{8(y-1)^4}\mathsf{F}_2^2 + \frac{y - 2}{(y-1)^3}\mathsf{F}_1\mathsf{F}_2\bigg]$  \\ \midrule
         $ c_1^{(0)} $ & $\frac{1}{288}\bigg[\frac{4y - 3}{(y-1)^2}\mathsf{F}_1^2 + \frac{4 + 17y}{8(y-1)^4}\mathsf{F}_2^2 + \frac{7 - 2y}{(y-1)^3}\mathsf{F}_1\mathsf{F}_2\bigg]$ \\ \midrule
        $ c_2^{(0)} $ & $\frac{\omega^2}{384}\bigg[\frac{45 + 15y - 45y^2 + y^3}{(y-1)^2}\mathsf{F}_1^2+ \frac{9+39y - y^2 + y^3}{4(y-1)^4}\mathsf{F}_2^2 + \frac{27 + 27y - 23y^2 + y^3}{(y-1)^3}\mathsf{F}_1\mathsf{F}_2\bigg]$\\ \midrule
        $c_3^{(0)}$ & $\frac{\omega^4}{6}\bigg[\frac{2y^2 - 3}{2(y-1)^2}\mathsf{F}_1^2 - \frac{2y + y^2}{8(y-1)^4}\mathsf{F}_2^2 + \frac{y^2 - 2y}{(y-1)^3}\mathsf{F}_1\mathsf{F}_2\bigg]$\\ \midrule
        $c_3^{(1)}$ & $\frac{\omega^4}{36}\bigg[\frac{31y^2 - 36y}{(y-1)^2}\mathsf{F}_1^2 + \frac{3 - 8y - 10y^2}{8(y-1)^4}\mathsf{F}_2^2 + \frac{9 - 29y + 13y^2}{(y-1)^3}\mathsf{F}_1\mathsf{F}_2\bigg]$ \\ \bottomrule
    \end{tabular}
    \caption{The master integral coefficients for the second post-Minkowskian order Compton amplitude. By $c_i^{(n)}$, the coefficient of $\epsilon^n$ in the expansion of the master integral coefficients is meant.}
    \label{tab:4dimcoeffs}
\end{table}

\subsection{Result in four dimensions}\label{sec:resultfourdims}
Having presented the $d$-dimensional result, I now specify to four dimensions. As can be seen from table \ref{tab:ddimcoeffs}, the coefficient $c_1$ here experiences a divergence. Indeed, the expansion of the cut box around four dimensions in eq. \eqref{eq:cutbox4dim} also contains an $\epsilon$-pole. When expanded about $d = 4-2\epsilon$, the coefficients admit an expansion in the regulator $\epsilon$,
\begin{equation}
    c_i = \sum_{n=n_\text{min}}^\infty c_i^{(n)}\epsilon^n.
\end{equation}
The first coefficient $c_1$ has $n_\text{min} = -1$, while the second and third, $c_2$ and $c_3$, have $n_\text{min} = 0$. This explains why I expanded $\mathcal{K}_1$ to first-order in $\epsilon$ in eq. \eqref{eq:cutbubble4dim}; the first-order piece is needed to obtain the full finite part of the amplitude in four dimensions. The expansion coefficients of the master integral coefficients $c_i^{(n)}$ are given in table \ref{tab:4dimcoeffs}.

By inserting the explicit expressions of the master integrals and the master integral coefficients, one arrives at an expression for the amplitude involving the functions appearing in the master integrals,
\begin{equation}
    \mathcal{M}^\text{2PM} = \frac{d_\text{IR}}{\epsilon} + d_1\log\frac{4\omega^2}{\mu^2} + d_2\log\frac{-q^2}{\mu^2} + d_\text{Im} \ii + \frac{d_q}{\abs{q}} + R.
\label{eq:result}\end{equation}
It contains an $\epsilon$-pole part given by
\begin{align}
    \frac{d_\text{IR}}{\epsilon} &= \epsilon^{-1}c_1^{(-1)}\mathcal{K}_1\vert_{\epsilon^{0}} + c_3^{(0)}\mathcal{K}_1\vert_{\epsilon^{-1}} \notag\\
    &= \frac{1}{\epsilon}\frac{\ii\kappa^4M^2\omega^3\mathsf{F}_1^2}{32\pi q^2}.
\end{align}
Notice that it is proportional to $\mathsf{F}_1^2$. The logarithmic terms are
\begin{align}
    d_1\log\frac{4\omega^2}{\mu^2} &= \epsilon^{-1}c_1^{(-1)}\mathcal{K}_1\vert_{\log{}} \notag\\
    &= \frac{\ii\kappa^4M^2\omega}{3\times2^6\pi}\bigg[\frac{5y-6}{2(y - 1)^2}\mathsf{F}_1^2 - \frac{2+y}{8(y - 1)^4}\mathsf{F}_2^2 - \frac{2 - y}{(y - 1)^3}\mathsf{F}_1\mathsf{F}_2\bigg]\log\frac{4\omega^2}{\mu^2},
\end{align}
and
\begin{align}
    d_2\log\frac{-q^2}{\mu^2} &= c_3^{(0)}\mathcal{K}_3\vert_{\log{}} \notag\\
    &= \frac{\ii\kappa^4M^2\omega}{3\times2^6\pi}\bigg[\frac{3 - 2y^2}{2(y - 1)^2y}\mathsf{F}_1^2 + \frac{2+y}{8(y - 1)^4}\mathsf{F}_2^2 + \frac{2 - y}{(y - 1)^3}\mathsf{F}_1\mathsf{F}_2\bigg]\log\frac{-q^2}{\mu^2}.
\end{align}
From these expressions, it is clear that
\begin{equation}
    d_1 + d_2 = -d_\text{IR}.
\label{eq:d1d2relation}\end{equation}
It is noteworthy that $d_\text{IR}$, $d_1$, and $d_2$ are imaginary. The remaining imaginary part of the amplitude lives in the term $d_\text{Im}\ii$, which is
\begin{align}
    d_\text{Im}\ii &= \epsilon^{-1}c_1^{(-1)}\mathcal{K}_1\vert_{\epsilon^1,2} + c_1^{(0)}\mathcal{K}_1\vert_{\epsilon^0} + \epsilon c_3^{(1)}\mathcal{K}_3\vert_{\epsilon^{-1}} \notag\\
    &= \frac{\ii\kappa^4M^2\omega}{3\times2^6\pi}\bigg[\frac{-1}{2(y - 1)}\mathsf{F}_1^2 - \frac{1 + 5y}{16(y - 1)^3y}\mathsf{F}_2^2 - \frac{3 - y}{2(y - 1)^2y}\mathsf{F}_1\mathsf{F}_2\bigg]
\end{align}
To be clear, $\mathcal{K}_1\vert_{\epsilon^1,2}$ refers to the term containing the $2$ from the first-order in $\epsilon$ from eq. \eqref{eq:cutbubble4dim}.
Concerning the real parts of the amplitude, there are two. A rational part and a part containing the non-analytic $\abs{q}^{-1}$. The rational part is
\begin{align}
    R &= \epsilon^{-1}c_1^{(-1)}\mathcal{K}_1\vert_{\epsilon^1,\ii\pi} \notag\\ 
    &= \frac{\kappa^4M^2\omega}{3\times2^6}\bigg[\frac{5y - 6}{2(y - 1)^2}\mathsf{F}_1^2 - \frac{2 + y}{8(y - 1)^4}\mathsf{F}_2^2 - \frac{2 - y}{(y - 1)^3}\mathsf{F}_1\mathsf{F}_2\bigg],
\end{align}
where again, to be clear, $\mathcal{K}_1\vert_{\epsilon^1,\ii\pi}$ refers to the term containing the $\ii\pi$ from the first-order in $\epsilon$ from eq. \eqref{eq:cutbubble4dim}. The non-analytic part is 
\begin{align}
    \frac{d_q}{\abs{q}} &= c_2\mathcal{K}_2 \notag \\
    &= \frac{\kappa^4M^2\omega^2}{3\times2^6}\bigg[\frac{45 + 15y - 45y^2 + y^3}{16(y - 1)^2}\mathsf{F}_1^2 \notag\\ &\hspace{2.5em}+ \frac{9 + 39y - y^2 + y^3}{64(y - 1)^4}\mathsf{F}_2^2 + \frac{27 + 27y - 23y^2 + y^3}{16(y - 1)^3}\mathsf{F}_1\mathsf{F}_2\bigg]\frac{1}{\abs{q}}.
\end{align}

\subsection{Infrared divergence}\label{sec.infrared-div}
By using the definition of $\mathsf{F}_1$ in eq. \eqref{eq:f1def} and the first-order amplitude from eq. \eqref{eq:1pm-compton}, $d_\text{IR}$ can be written
\begin{equation}
    d_\text{IR} = -\frac{\ii \kappa^2M\omega}{16\pi}\mathcal{M}^\text{1PM}.
\end{equation}
The divergent part of the second-order amplitude is thus proportional to the first-order amplitude. It should come as no surprise that an $\epsilon$-pole shows up in the final result; the Compton amplitude has, after all, two external massless gravitons, and amplitudes with massless final external states are known to be divergent in the infrared \cite{Strominger:2017zoo}. In fact, Weinberg, building on the work \cite{Bloch:1937fer,Jauch:1954rnc}, proved in \cite{Weinberg:1965nx} that the infrared divergences associated with external photons and gravitons must resum into an exponential in the exact $S$-matrix. Weinberg used cutoff regularization in his derivation, while I have used dimensional regularization. The translation between the two schemes was provided in \cite{Brandhuber:2023hhy}, where the classical limit was also discussed. The upshot is that one should expect the exact Compton amplitude $\mathcal{M}$ to be related to a finite Compton amplitude $\mathcal{M}_\text{finite}$ by the relation
\begin{equation}
    \mathcal{M} = \ee^{\frac{\ii W}{\epsilon}}\mathcal{M}_\text{finite},
\end{equation}
where $W$ is a factor---linear in $\kappa^2$---that can be determined from the formulas provided in \cite{Weinberg:1965nx}. Expanding both sides in $\kappa^2$, I obtain
\begin{align}
    \sum_{n=1}^\infty\mathcal{M}^{n\text{PM}} &= \Big(1 + \frac{\ii W}{\epsilon} + \frac{-W^2}{2\epsilon^2} + \cdots\Big)\big(\mathcal{M}_\text{finite}^\text{1PM} + \mathcal{M}_\text{finite}^\text{2PM} + \mathcal{M}_\text{finite}^\text{3PM} + \cdots\big) \notag\\
    &= \mathcal{M}_\text{finite}^\text{1PM} + \Big[\frac{\ii W}{\epsilon}\mathcal{M}_\text{finite}^\text{1PM} + \mathcal{M}_\text{finite}^\text{2PM}\Big] + \Big[-\frac{W^2}{2\epsilon^2}\mathcal{M}_\text{finite}^\text{1PM} + \frac{\ii W}{\epsilon}\mathcal{M}_\text{finite}^\text{2PM} + \mathcal{M}_\text{finite}^\text{3PM}\Big] + \cdots.
\end{align}
As a first observation, the exponentiation is consistent with the first-order amplitude being finite. Indeed,
\begin{equation}
    \mathcal{M}^\text{1PM} = \mathcal{M}_\text{finite}^\text{1PM}.
\end{equation}
The expected infrared divergence at second-order is thus
\begin{equation}
    \mathcal{M}^\text{2PM}\vert_\text{IR} = \frac{\ii W}{\epsilon}\mathcal{M}^\text{1PM},
\end{equation}
which is exactly what I find provided
\begin{equation}
    W = -\frac{\kappa^2M\omega}{16\pi}.
\end{equation}
It is worth remarking that the cancellations that need to take place between the cut bubble and cut box for this to result are quite non-trivial. The consistency with infrared exponentiation thus provides a strong check on the amplitude.

In the future, it would be highly interesting to compute the 3PM Compton amplitude and verify that its infrared divergence takes the form
\begin{equation}
    \mathcal{M}^\text{3PM}\vert_\text{IR} = -\frac{W^2}{2\epsilon^2}\mathcal{M}^\text{1PM} + \frac{\ii W}{\epsilon}\mathcal{M}_\text{finite}^\text{2PM}.
\end{equation}

\subsection{Scalar bending angle}
Reassured by the correct infrared behavior of the amplitude, it is interesting to perform a different, final check on the amplitude.

In \cite{Bjerrum-Bohr:2014zsa, Bjerrum-Bohr:2016hpa}, the classical 2PM amplitude for a massless scalar scattering gravitationally off a massive scalar was computed in the low-energy and forward-scattering limit.\footnote{In the same work, the case of photons and massless fermions was also treated in addition to the scalar case, with the leading quantum correction being provided in all three cases. This shall not be relevant here, however.} Following the discussion on transmutation in \cite{Cheung:2017ems}, the Compton amplitude for a massless scalar can be extracted from the gravitational Compton amplitude by isolating the coefficient of the dot product $(\varepsilon_1\cdot\varepsilon_2)^2$, meaning it should be possible to compare (the correct limit of) the result from eq. \eqref{eq:result} with the one from \cite{Bjerrum-Bohr:2016hpa}. Extracting the coefficient of $(\varepsilon_1\cdot\varepsilon_2)^2$ is simple, as
\begin{equation}
    \mathsf{F_1}^2\vert_{(\varepsilon_1\cdot\varepsilon_2)^2} = 1, \qqquad \mathsf{F_2}^2\vert_{(\varepsilon_1\cdot\varepsilon_2)^2} = 0, \qqquad \mathsf{F_1}\mathsf{F}_2\vert_{(\varepsilon_1\cdot\varepsilon_2)^2} = 0.
\end{equation}
It is useful to use eq. \eqref{eq:d1d2relation} to rewrite the logarithms as
\begin{equation}
    d_1\log\frac{4\omega^2}{\mu^2} + d_2\log\frac{-q^2}{\mu^2} = d_\text{IR}\log\frac{\mu^2}{M^2} + d_1\log\frac{4\omega^2}{M^2} + d_2\log\frac{-q^2}{M^2}.
\end{equation}
The term $\log\mu^2/M^2$ has the same coefficient as the infrared divergence---specifically, it is proportional to the first-order amplitude, suggesting that it also exponentiates to an unphysical phase. Hence, I neglect this term below. Now, to obtain something that can be compared with \cite{Bjerrum-Bohr:2016hpa}, it is necessary to correct the normalization of the amplitude following the discussion in section \ref{sec:curvedcomp}. This results in
\begin{equation}
    2M\mathcal{M}^\text{2PM} \simeq 2M\Big[d_1\log\frac{4\omega^2}{M^2} + d_2\log\frac{-q^2}{M^2} + d_\text{Im} \ii + \frac{d_q}{\abs{q}} + R\Big],
\end{equation}
where I dropped the infrared divergence also following the discussion in \cite{Bjerrum-Bohr:2016hpa}. In the limit $\omega \ll M$ and $y \ll 1$, 
\begin{equation}
    \log\frac{4\omega^2}{M^2} \ll \log\frac{-q^2}{M^2},
\end{equation}
so the logarithm of the energy can be dropped. Additionally, the $d_q$ and $d_2$ terms will dominate the $d_\text{Im}$ and $R$ terms, as the latter contain no poles in $y$ in their $\mathsf{F}_1^2$ part. Finally, the expressions for the relevant coefficients in the limit is
\begin{subequations}
\begin{align}
    d_q\big\vert_{(\varepsilon_1\cdot\varepsilon_2)^2} &\simeq \frac{15\kappa^4M^2\omega^2}{1024}, \\
    d_2\big\vert_{(\varepsilon_1\cdot\varepsilon_2)^2} &\simeq -\frac{\ii\kappa^4M^2\omega^3}{32\pi q^2},
\end{align}
\end{subequations}
meaning I obtain
\begin{equation}
    2M\mathcal{M}^\text{2PM}\vert_{(\varepsilon_1\cdot\varepsilon_2)^2} \simeq \frac{15\kappa^4M^3\omega^2}{512\abs{q}} - \frac{\ii\kappa^4M^3\omega^3}{16\pi q^2}\log\frac{-q^2}{M^2}.
\label{eq:comparisonscalar}\end{equation}
This matches the expression obtained in \cite{Bjerrum-Bohr:2016hpa} almost exactly, the only difference being the sign in the first term and the numerical prefactor in the second term. In the reference, the sign of the first term is negative, while the numerical factor on the second term term is $1/8$.

\chapter{Conclusion and perspectives}\label{ch:conclusion}
I have treated in this thesis various aspects related to the quantization of gravity in curved backgrounds and the extraction of classical amplitudes from such a theory. Specifically, I have computed from this framework the well-known 1PM Compton amplitude and the 2PM Compton amplitude, which is a new result. For both of these amplitudes, I demonstrated that the curved result is equal to the flat space result.

Aspects of the action formulation for general relativity and the treatment of point particles have been discussed, and the derivation of the $d$-dimensional metric sourced by a point particle, the Schwarzschild--Tangherlini metric, was derived in great detail.

I provided a discussion of the analogy between gauge theories and gravity to set the stage for the quantum field-theoretic treatment of general relativity. Path integral quantization of a generic quantum field theory was discussed, after which this method was applied to the theory of a graviton living on an arbitrary curved background. In this context, I discussed the gauge symmetry associated with this theory, and showed how to resolve the apparently divergent partition function that it causes. This entailed performing the Faddeev--Popov procedure on the theory. Then, I provided an account of the Feynman rules in flat space, after which I generalized the discussion to arbitrary backgrounds. Specifying then to the Schwarzschild--Tangherlini background, I obtained the two-point interaction vertex in this theory. This vertex is not exactly calculable, so I performed a post-Minkowskian expansion of it, yielding an infinite tower of vertices.

Einstein gravity experiences divergences at loop-level, which necessitates the addition to the Einstein--Hilbert action of counterterms. I discussed this and how one reconciles the picture by recasting general relativity as an effective field theory, whose action contains an infinite set of higher-derivative operators. The low-energy quantum theory of gravity is thus a well-defined quantum field theory. In the following discussion of the classical limit, I explained how the classical limit is usually taken.

Worldline quantum field theory was then introduced as an alternative way to take the limit, and I gave a thorough discussion of the derivation of the Feynman rules of this theory, both with a flat background metric and the Schwarzschild--Tangherlini one. I demonstrated how curved space Feynman rules reorganize and, to some extent, improve Feynman rule calculations by exploiting exact-in-$G$, classical information from general relativity. Specifically, I discussed in detail how expanding around a Schwarzschild--Tangherlini background resums an infinite set of potential graviton diagrams, as shown in eq. \eqref{eq:se-wfe-correspondence}. This is exactly an example of the type of curved space resummation discussed in \cite{Kosmopoulos:2023bwc}. In addition, dimensional regularization ensured that the curved part of the metric contributes nothing to the worldline Feynman rules, and I demonstrated the cancellation between the Einstein--Hilbert and worldline actions that eliminates the class of source-containing diagrams.

Finally, I gave an account of the kinematics of the Compton scattering process, after which I computed the 1PM amplitude. The calculation of the main result of the thesis---the 2PM amplitude---required introducing integration-by-parts identities to reduce the integration problem. I then computed in detail the resulting master integrals, using in one case the method of differential equations and the method of regions. The result was found to contain an infrared divergence. This was to be expected, however, as the Compton amplitude has two external massless gravitons. Indeed, I found that the divergence is proportional to the first-order amplitude, which is consistent with Weinberg's theorem on exponentiation of infrared divergences \cite{Weinberg:1965nx}. Additionally, I extracted the scalar part of the amplitude and compared the forward-scattering and low-energy limit of this quantity with the result from \cite{Bjerrum-Bohr:2017dxw}. For the physical part (the first term of eq. \eqref{eq:comparisonscalar}), I obtain a match up to a sign. For the imaginary part (the second term of eq. \eqref{eq:comparisonscalar}), I obtain a match apart from the numerical prefactor, which differs by a factor of two.

Ultimately, what one might hope to achieve by expanding around a curved metric is some kind of resummation in Newton's constant. Although an infinite resummation of potential diagrams does occur, it might still seem that the benefits associated with the curved expansion are small and the prospect of a post-Minkowskian resummation is slim---the two-point graviton vertex in the curved expansion in eq. \eqref{eq:2-pt-expanded-diagram} is after all an infinite series of loop-diagrams. It is conceivable, however, that choosing the Kerr-Schild coordinate gauge for the background metric,
\begin{equation*}
    \bar g_{\mu\nu} = \eta_{\mu\nu} - \frac{2GM}{\abs{x_\perp}}l_\mu l_\nu, \qquad l_\mu = v_\mu - \frac{x_{\perp\mu}}{\abs{x_\perp}},
\end{equation*}
might enable progress in this direction, as the curved $n$-point graviton vertex truncates at finite order in $G$ in this gauge. The disadvantage is that the loop counting and the $G$ counting seem no longer aligned in this case. Another interesting avenue for further research would be to investigate the relation of the Compton amplitude to tidal effects. By considering the approach of \cite{Caron-Huot:2025tlq}, one might expect that progress could be made in this direction by decomposing the Compton amplitude into spherical harmonics, and that this might provide a window into tidal deformation effects.

Another natural next step is to compute and analyze the classical Compton amplitude at third and perhaps fourth post-Minkowskian order. The analysis of the master integrals that arise at these higher orders poses a highly interesting and topical problem. Additionally, as I mention in section \ref{sec.infrared-div}, it is crucial to clarify the structure of infrared divergences and their relation to lower-order amplitudes at higher post-Minkowskian orders.

Lastly, the inclusion of spin provides another interesting direction of research. It would be very interesting to derive Feynman rules from the Kerr action proposed in \cite{Bjerrum-Bohr:2025lpw}, and see if an exact-in-spin tree-level amplitude could be obtained, as the spinning, tree-level, four-point Compton is an important ingredient in modern generalized unitarity methods. Work in this direction has already been performed in \cite{Akpinar:2025huz}.
\backmatter
\bibliographystyle{JHEP}
\bibliography{KinematicAlgebra}

\end{document}